\documentclass[preprint]{aastex631}
\newcommand{\spitzer}{\textit{Spitzer} }

\renewcommand{\thefootnote}{\arabic{footnote}}

\shorttitle{Ice and Dust in Perseus and Serpens}
\shortauthors{Madden et al.}

\accepted{3/24/2022; Published in The Astrophysical Journal, 930, 2}

\begin{document}

\title{Infrared Spectroscopic Survey of the Quiescent Medium of Nearby
  Clouds: II. Ice Formation and Grain Growth in Perseus and Serpens}

\author{M.C.L. Madden}

\affiliation{Department of Astronomy, Columbia University, New York,
  NY 10027, USA}

\author{A.C.A. Boogert}

\affiliation{Institute for Astronomy, University of Hawaii, 2680
  Woodlawn Drive, Honolulu, HI 96822, USA}

\altaffiliation{Visiting Astronomer at the InfraRed Telescope Facility,
  which is operated by the University of Hawaii under contract
  80HQTR19D0030 with the National Aeronautics and Space
  Administration.}

\author{J.E. Chiar}

\affiliation{current address: Diablo Valley College, 321 Golf Club
  Road, Pleasant Hill, CA 94523, USA}

\affiliation{SETI Institute, 189 Bernardo Av., Suite 200, Mountain
  View, CA 94043, USA}

\author{C. Knez}

\affiliation{Department of Astronomy, University of Maryland, College
  Park, MD 20742, USA}

\affiliation{Johns Hopkins University Applied Physics Laboratory,
  11100 Johns Hopkins Road, Laurel, MD 20723, USA}

\altaffiliation{Visiting Astronomer at the Infrared Telescope
  Facility, which is operated by the University of Hawaii under
  contract 80HQTR19D0030 with the National Aeronautics and Space
  Administration.}

\author{Y.J. Pendleton}

\affiliation{NASA Ames Research Center, Moffett Field, CA 94035, USA}

\author{A.G.G.M. Tielens}

\affiliation{Leiden Observatory, Leiden University, P.O. Box 9513,
  2300 RA Leiden, the Netherlands}

\author{A. Yip}
\affiliation{Johns Hopkins University, Baltimore, MD 21218, USA}

%%%%%%%%%%%%%%%%%%%%%%%%%%%%%%%%%%%%%%%%%%%%%%%%
%			  ABSTRACT AND KEYWORDS
%%%%%%%%%%%%%%%%%%%%%%%%%%%%%%%%%%%%%%%%%%%%%%%%

%$abstract must be <=250 words

\begin{abstract}
 
  The properties of dust change during the transition from diffuse to
  dense clouds as a result of ice formation and dust coagulation, but
  much is still unclear about this transformation.  We present 2-20
  $\mu$m spectra of 49 field stars behind the Perseus and Serpens
  Molecular Clouds and establish relationships between the
  near-infrared continuum extinction ($A_{\rm K}$) and the depths of
  the 9.7 $\mu$m silicate ($\tau_{9.7}$) and 3.0 $\mu$m $\rm{H_2O}$
  ice ($\tau_{3.0}$) absorption bands.  The $\tau_{9.7}$/$A_{\rm K}$
  ratio varies from large, diffuse interstellar medium-like values
  ($\sim 0.55$), to much lower ratios ($\sim 0.26$). Above extinctions
  of $A_{\rm K} \sim 1.2$ ($A_{\rm V} \sim 10$; Perseus, Lupus, dense
  cores) and $\sim 2.0$ ($A_{\rm V} \sim 17$; Serpens), the
  $\tau_{9.7}$/$A_{\rm K}$ ratio is lowest. The $\tau_{9.7}$/$A_{\rm
    K}$ reduction from diffuse to dense clouds is consistent with a
  moderate degree of grain growth (sizes up to $\sim 0.5$ $\mu m$),
  increasing the near-infrared color excess (and thus $A_{\rm K}$),
  but not affecting the ice and silicate band profiles. This grain
  growth process seems to be related to the ice column densities and
  dense core formation thresholds, highlighting the importance of
  density.  After correction for Serpens foreground extinction, the
  H$_2$O ice formation threshold is in the range of $A_{\rm
    K}=0.31-0.40$ ($A_{\rm V}=2.6-3.4$) for all clouds, and thus grain
  growth takes place after the ices are formed.  Finally, abundant
  $\rm{CH_3OH}$ ice ($\sim$21\% relative to $\rm{H_2O}$) is reported
  for 2MASSJ18285266+0028242 (Serpens), a factor of $>$4 larger than
  for the other targets.
    
\end{abstract} 


%%%%%%%%%%%%%%%%%%%%%%%%%%%%%%%%%%%%%%%%%%%%%%%%
%			       BODY TEXT
%%%%%%%%%%%%%%%%%%%%%%%%%%%%%%%%%%%%%%%%%%%%%%%%

\section{Introduction}\label{sec:Introduction}

The properties of interstellar grains have long been known to be
different in dense clouds compared to diffuse clouds. The most easily
recognizable differences include the 3.4 $\mu$m aliphatic hydrocarbons
absorption feature, which is only present in diffuse clouds
\citep[e.g.,][]{Pendleton1994, Chiar1996}, and ice absorption bands,
which have only been reported toward dense clouds (e.g.,
\citealt{Boogert2015}). The absence of the 3.4 $\mu$m aliphatic
hydrocarbons absorption feature is still not well understood. The
growth of ice mantles is a consequence of extinction in the
ultraviolet (UV), reducing the effects of photodesorption, and larger
densities, increasing gas-grain interactions (e.g.,
\citealt{Hollenbach2009}). Also, the optical and infrared interstellar
extinction curve is flatter in dense clouds, with the ratio of total
to selective extinction $R_{\rm V}=A_{\rm V}/E(B-V)$ increasing from
values of $\sim$3.1 to 5.5 \citep[e.g.,][]{Indebetouw2005,
  McClure2009}. Deeper into the cloud ($A_{\rm V} \sim 20$), increased
flattening continues \citep{Cambresy2011}. This is consistent with a
reduction of the number density of the smallest grains ($<$0.1
$\mu$m), but does not necessarily trace growth of the largest grains
\citep{Weingartner2001}. Very large grains ($>$1 $\mu$m) were found to
be associated with dense clouds, however, causing `coreshine' at
wavelengths of 3.6 $\mu$m \citep{Pagani2010}. While it is
observationally and theoretically well established that ice mantles
are formed at relatively shallow dense cloud depths ($A_{\rm V} \sim
1.6$, or observationally at $A_{\rm V} \sim 3.2$ when both the front
and the back of the clouds are traced), the thickness of ice mantles
is limited to 5 nm by the available oxygen budget
\citep[e.g.,][]{Hollenbach2009}. Therefore, coagulation of small
grains, likely aided by sticky ice coated grains, must govern the
grain growth process \citep{Ormel2011}.

A sensitive indicator of the different dust properties in dense versus
diffuse clouds, is the depth of the 9.7 $\mu$m band of silicates
($\tau _{9.7}$) relative to the near-infrared continuum extinction
($A_{\rm K}$). A reduction of $\tau _{9.7}$/$A_{\rm K}$ in dense
clouds by up to a factor of 2 was observed towards a range of
sight-lines tracing nearby clouds and cores \citep{Chiar2007},
isolated dense cores \citep{Boogert2011}, and the Lupus Cloud
\citep{Boogert2013}. Models suggest that this change is likely caused
by grain growth affecting the near-infrared more than the silicate
band, because the 9.7 $\mu$m band profile shows little variation
\citep{vanBreemen2011}. Coagulation models of mixtures of ice coated
graphite and silicate grains confirm this \citep{Ormel2011}.

The $\tau _{9.7}$/$A_{\rm K}$ ratio and the 3.0 $\mu$m ice band
optical depth ($\tau_{\rm 3.0}$) are interesting observational probes
of the evolution of dust from diffuse to dense clouds. Following
\citet{Ormel2011}, it is expected that ice formation and grain growth
are correlated. Observations generally agree with that expectation,
but, perhaps, not in all lines of sight \citep{Boogert2013}. It is the
goal of this paper to further investigate the origin of the variations
of the $\tau _{9.7}$/$A_{\rm K}$ ratio. Following our papers on
isolated dense cores \citep{Boogert2011}, and the Lupus cloud
\citep{Boogert2013}, here we study the $\tau _{9.7}$/$A_{\rm K}$ ratio
and ice abundances towards the Perseus and Serpens Molecular
Clouds. These clouds are well studied, low-mass star- forming regions
\citep{Enoch2006, Jorgensen2006, Evans2009}. With the Spitzer Infrared
Array Camera (IRAC), the Legacy Project ``From Molecular Clouds to
Planet Forming Disks” \citep[c2d;][]{Evans2003, Evans2009} mapped 3.86
deg$^2$ of the Perseus molecular cloud and 0.89 deg$^2$ of the Serpens
molecular cloud \citep{Jorgensen2006, Harvey2006}.  We present
\textit{Spitzer} InfraRed Spectrograph (\textit{Spitzer}/IRS) and NASA
InfraRed Telescope Facility SpeX (IRTF/SpeX) and Keck Near InfraRed
SPECtrometer (Keck/NIRSPEC) $K$ and $L$-band spectra of background
stars selected from this survey.

Throughout this paper we will assume {\it Gaia}-derived distances
\citep{Zucker2019} for the relevant clouds: $294\pm 15$ pc for
Perseus, $420\pm 15$ pc for Serpens (Main; see also
\citealt{Ortiz2018}), $189\pm 13$ pc for Lupus, and $141\pm 7$ pc for
Taurus.

In \S \ref{sec:sources} and \S \ref{sec:Observations}, the target
selection and observation data are presented, respectively. In \S
\ref{sec:Methods}, the methods used to fit the spectra are given. \S
\ref{subsec:Si}-\ref{subsec:silakh2o} show the correlations between
the continuum extinction and the 9.7 $\mu$m silicate and 3.0 $\mu$m
ice features. In \S \ref{subsec:CH3OH}, the $\rm{CH_3OH}$ abundances
are analyzed. In \S \ref{subsec:maps}, the results are put into a
spatial context using maps of the extinction and the 3.5 $\mu$m broad
band emission.  In \S\ref{subsec:models} and \S\ref{subsec:variations}
the results are compared to models of grain growth, and implications
of the variations of the observed $\tau_{9.7}/A_{\rm K}$ ratios are
discussed. The results are also combined with those from previous
works. The ice formation thresholds for the sample of clouds are
compared in \S\ref{subsec:ices}, and in \S\ref{subsec:CH3OH_disc}
CH$_3$OH ice abundances are discussed.  Finally, a summary including
mention of future work is given in \S \ref{sec:Summary}.

\begin{figure*}
    \begin{center}
        \includegraphics[width=1\textwidth,height=0.5\textheight]{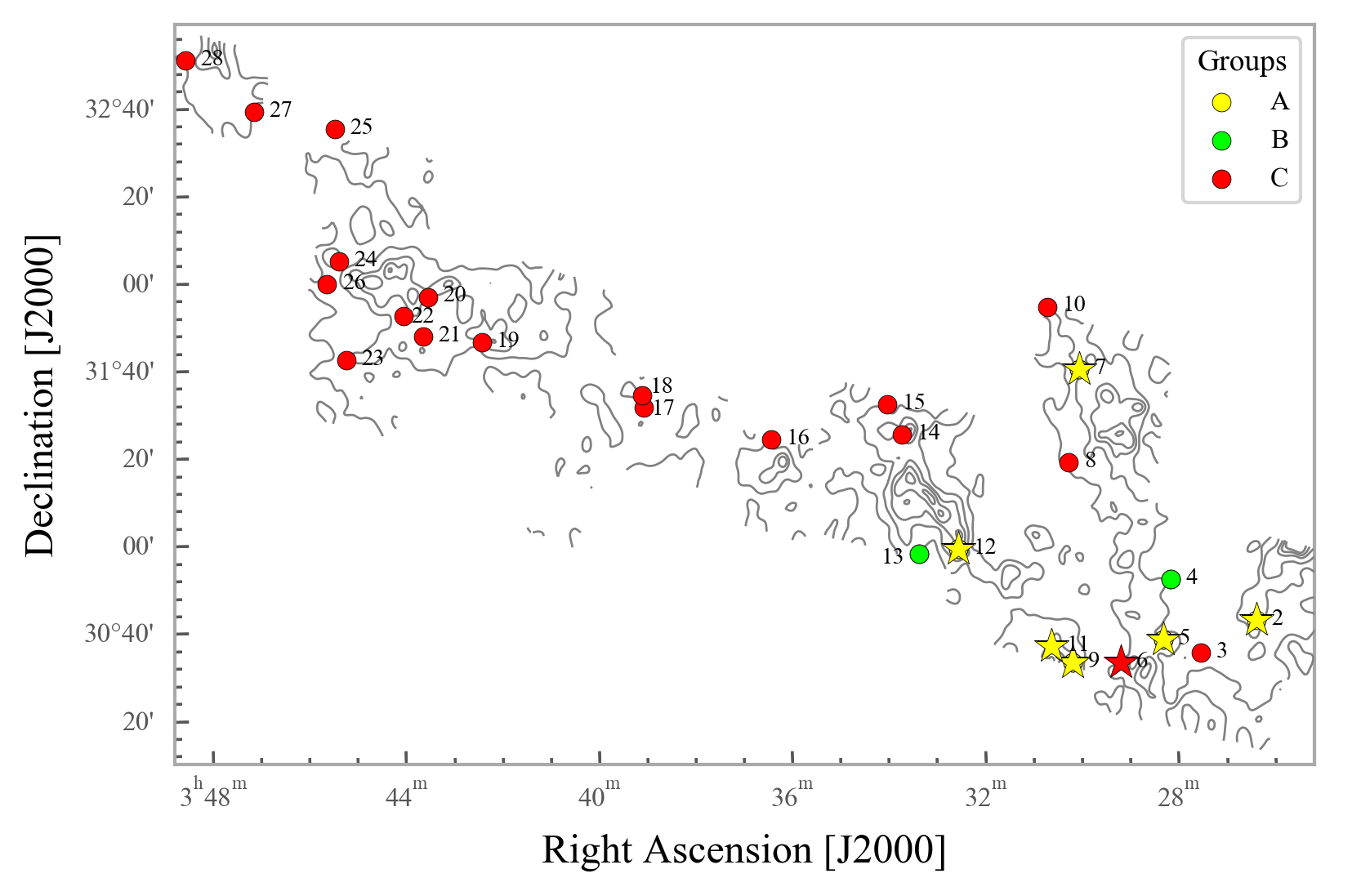}\hfill
        \caption{Perseus target positions labeled by alias and
          overlaid on extinction contours \citep{Evans2009}. The
          contours represent $A_{\rm V}$ extinction levels of 3, 6,
          12, 18, and 24 mag. The target colors refer to the groupings
          determined in \S\ref{subsec:Si}.  Star symbols represent
          lines of sight with a 3.0 $\mu$m ice band optical depth
          ($\tau_{\rm 3.0}$) larger than 0.5 within the
          uncertainties. \label{Fig. per_alias}}
    \end{center}
\end{figure*}

\begin{figure}
    \begin{center}
        \includegraphics[width=0.46\textwidth,height=0.55\textheight]{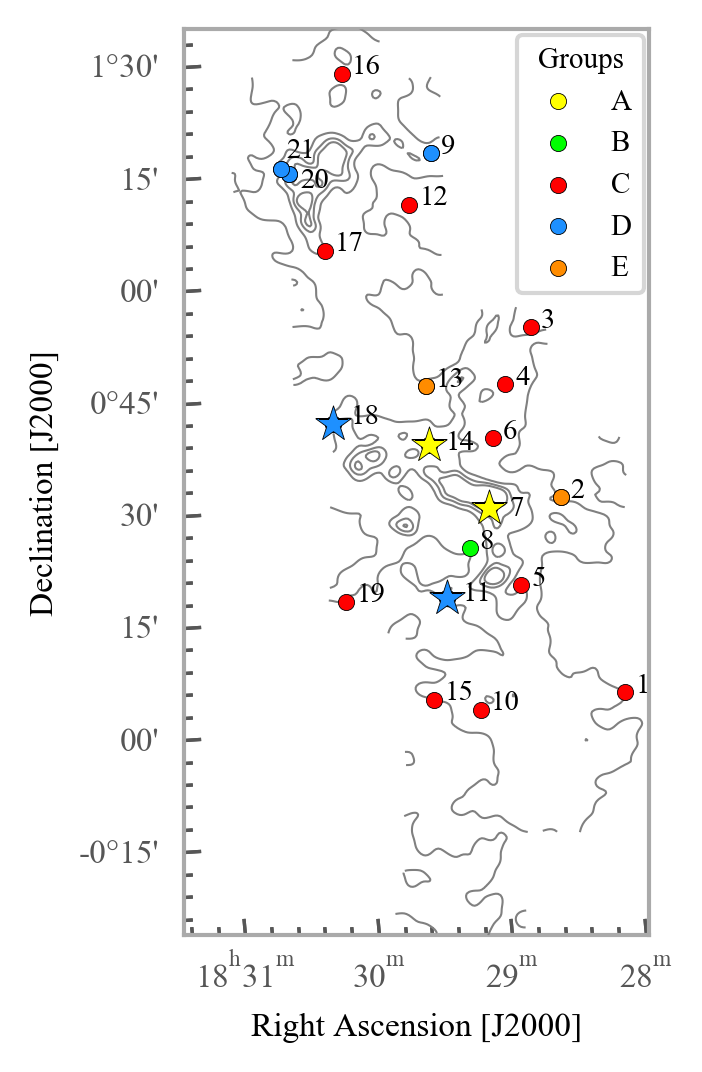}
        \caption{Serpens target positions overlaid on extinction
          contours \citep{Evans2009}. The contours represent $A_{\rm
            V}$ extinction levels of 3, 6, 12, 18, and 24 mag. The
          target colors refer to the groupings determined in
          \S\ref{subsec:Si}.  Star symbols represent lines of sight
          with an ice band optical depth $\tau_{\rm 3.0}>0.5$ within
          the uncertainties.\label{Fig. ser_alias}}
    \end{center}
\end{figure}

\section{Source Selection}\label{sec:sources}

Background stars were selected from the Perseus and Serpens molecular
clouds which were mapped with \textit{Spitzer}/IRAC and MIPS by the
c2d Legacy team \citep{Evans2003, Evans2009}. The maps are complete
down to $A_{\rm V}$ = 6 and $A_{\rm V}$ = 2 for Serpens and Perseus,
respectively \citep{Evans2003}. The selected sources have an overall
spectral energy distribution (SED; 2MASS 1.2–2.2 $\mu$m, IRAC 3–8
$\mu$m, MIPS 24 $\mu$m) of a reddened Rayleigh–Jeans curve. They fall
in the ``star” category in the c2d catalogs and have MIPS 24 $\mu$m to
IRAC 8 $\mu$m flux ratios greater than 4. In addition, fluxes are high
enough ($>$10 mJy at 8.0 $\mu$m) to obtain \textit{Spitzer}/IRS
spectra of high quality (S/N $>$ 50) within $\sim$20 minutes of
observing time per module. This resulted in a list of roughly 100
stars behind Perseus and 400 stars behind Serpens. The list was
reduced by selecting $\sim$10 sources in each interval of $A_{\rm V}$
of 2–5, 5–10, and $>$10 mag for Perseus, and 5–10, 10–15, and $>$15
mag for Serpens (taking $A_{\rm V}$ from the c2d catalogs) and making
sure that the physical extent of the clouds are covered. For Serpens,
there are more background stars to choose from and the overall
extinction is higher, reflecting its lower Galactic latitude.

The final target list contains nearly all high $A_{\rm V}$ lines of
sight. At low extinctions, many more sources were available and the
brightest were selected. The observed samples of 28 targets toward
Perseus, and 21 toward Serpens are listed in Tables \ref{Table 1} and
\ref{Table 2}. The analysis shows that the SEDs of 2 Serpens sources
cannot be fitted with stellar models (\S \ref{sec:Results}), because
they likely have dust shells (silicate band emission). All other
targets were found to be usable (hyper)giants or in a few cases main
sequence stars. One Perseus target (Per-16) was found to be a
foreground, rather than a background star. \citet{Reiners2020} measure
a distance of 13.7 pc to Per-16, which is the closest target in our
sample size by far. Indeed, we find no evidence for dust or ice
extinction in this line of sight (\S \ref{subsec:Si}).

Figures \ref{Fig. per_alias} and \ref{Fig. ser_alias} plot the
location of the observed background stars on extinction map contours
derived from 2MASS and \spitzer photometry \citep{Evans2009}.

%%%%%%%%%% OBSERVATION TABLES %%%%%%%%%%%%%

%commented out 2 rows so table would fit on page for Astroph. They are
%present in the ApJ publication.

\begin{deluxetable*}{cccccc}[h!]
\movetabledown=5in
\tablewidth{0pt}
\centering
\caption{Perseus Target Observations\label{Table 1}}
\tablehead{
  \colhead{Alias\tablenotemark{a}} &  \colhead{Name of Star} & \colhead{Region\tablenotemark{b}} &  \colhead{Date\tablenotemark{c}} & \colhead{AOR key} & \colhead{Modules\tablenotemark{d}}\\
  \colhead{Per-}                   &  \colhead{2MASS}        & \colhead{}                        &  \colhead{Ground-based}          & \multicolumn{2}{c}{Spitzer/IRS}      
}
\startdata
1  & 03245605+3026005 & LDN 1455 IRS 1  & 2008-09-27     & 23083008  & SL, LL2   \\
2  & 03261355+3029223 & IRAS 03222+3034 & 2008-09-30     & 23085824  & SL, LL2   \\
3  & 03272467+3022547 &                 & 2008-09-27     & 23083008  & SL, LL2   \\
4  & 03275729+3040138 & LDN 1455 IRS 1  & 2008-09-30     & 23082752  & SL, LL2   \\
%   &                  &                 & 2010-02-04     &           &           \\
   &                  &                 & 2010-02-05     &           &           \\
5  & 03281034+3026343 & LDN 1455 IRS 1  & 2008-09-29$^*$ & 23086336  & SL, LL2   \\
6  & 03290508+3022080 & LDN 1455 IRS 1  & 2008-09-30     & 23087616  & SL, LL2    \\
7  & 03293654+3129465 & NGC 1333        & 2008-09-25$^*$ & 23088128  & SL, LL2    \\
8  & 03295603+3108454 & NGC 1333        & 2008-09-30     & 23082752  & SL, LL2    \\
9  & 03300474+3023032 & IRAS 03271+3013 & 2008-09-30     & 23086592  & SL        \\
10 & 03301239+3144408 & NGC 1333        & 2008-09-28$^*$ & 23086848  & SL         \\
11 & 03303022+3027087 & IRAS 03271+3013 & 2008-09-25$^*$ & 23087616  & SL, LL2   \\
12 & 03322030+3050485 & IRAS 03292+3039 & 2008-09-25$^*$ & 23088384  & SL, LL2   \\
13 & 03331023+3050177 & IRAS 03292+3039 & 2008-09-26     & 23083776  & SL, LL2    \\
14 & 03332416+3117470 &                 & 2008-09-29$^*$ & 23087360  & SL, LL2   \\
15 & 03334078+3125007 &                 & 2008-09-27     & 23083776  & SL, LL2   \\
   &                  &                 & 2010-02-05     &           &           \\
16 & 03360868+3118398 &                 & 2008-09-25$^*$ & 23084800  & SL, LL2   \\
17 & 03384753+3127345 &                 & 2008-09-26$^*$ & 23084032  & SL, LL    \\
18 & 03384901+3130173 &                 & 2008-07-08$^*$ & 23085312  & SL, LL    \\
19 & 03420993+3144139 &                 & 2008-09-30     & 23086080  & SL        \\
20 & 03431627+3155097 & IC 348          & 2010-02-05$^*$ & 23087872  & SL, LL2   \\
21 & 03432386+3146110 & IC 348          & 2008-09-26$^*$ & 23085568  & SL, LL2   \\
22 & 03434808+3151030 & IC 348          & 2008-09-28$^*$ & 23084544  & SL, LL2   \\
   &                  &                 & 2008-09-29$^*$ &           &           \\
23 & 03450207+3141196 & IC 348          & 2008-09-27     & 23084544 &  SL, LL2   \\
%   &                  &                 & 2010-02-04     &          &            \\
   &                  &                 & 2010-02-05     &          &            \\
24 & 03450796+3204018 & IC 348          & 2008-09-30     & 23084288 &  SL, LL2   \\
25 & 03450839+3234202 &                 & 2008-09-27     & 23083264 &  SL, LL2   \\
   &                  &                 & 2010-02-04     &          &            \\
26 & 03452349+3158573 & IC 348          & 2010-02-04     & 23083520 &  SL, LL2   \\
   &                  &                 & 2010-02-05     &          &            \\
27 & 03465115+3238494 &                 & 2008-09-26     & 23083264 &  SL, LL2   \\
   &                  &                 & 2010-02-04     &          &            \\
28 & 03481723+3250595 &                 & 2008-09-26$^*$ & 23085568 &  SL, LL2   \\
\enddata
\tablenotetext{a}{Alias used throughout this paper.}
\tablenotetext{b}{Named location within the Perseus cloud, if available.}
\tablenotetext{c}{Date of IRTF/SpeX observations. The instrument mode
  used is LongXD1.9 (1.95-4.2 $\mu$m), except for dates labeled with $^*$ for
  which it is LongXD2.1 (2.15-5.0 $\mu$m).}
\tablenotetext{d}{{\it Spitzer}/IRS modules used: SL=Short-Low (5-14~$\mu$m, $R\sim100$), LL2=Long-Low 2 (14-21.3~$\mu$m, $R\sim100$), LL=Long-Low 1 and 2 (14-35~$\mu$m, $R\sim100$)}
\end{deluxetable*}

\begin{deluxetable*}{ccccccc}
\tablewidth{0pt}
\centering
\caption{Serpens Target Observations\label{Table 2}}
\tablehead{
  \colhead{Alias\tablenotemark{a}} &  \colhead{Name of Star} & \colhead{Region\tablenotemark{c}} &  \colhead{Date} & \colhead{Instrument\tablenotemark{d}} & \colhead{AOR key} & \colhead{Modules\tablenotemark{e}}    \\
  \colhead{Ser-}                   &  \colhead{2MASS}        & \colhead{}                        &  \multicolumn{2}{c}{Ground-based}                       & \multicolumn{2}{c}{Spitzer/IRS}\\
  }
\startdata
1  & 18275901+0002337 &                                    & 2009-10-11 & NIRSPEC & 23073536 & SL, LL2  \\
2  & 18282010+0029141 & Ser/G3-G6                          & 2009-10-11 & NIRSPEC & 23073792 & SL, LL2  \\
3  & 18282631+0052133 &                                    & 2009-10-11 & NIRSPEC & 23072768 & SL, LL   \\
4  & 18284038+0044503 & Ser/G3-G6                          & 2009-10-11 & NIRSPEC & 23074816 & SL, LL2  \\
5  & 18284139+0017460 & [EGE2007] Bolo 7 \tablenotemark{b} & 2008-07-09 & SpeX    & 23074048 & SL, LL2  \\
6  & 18284797+0037431 & Ser/G3-G6                          & 2009-10-11 & NIRSPEC & 23074304 & SL, LL2  \\
7  & 18285266+0028242 & Ser/G3-G6                          & 2007-07-05 & NIRSPEC & 13460224 & SL, LL2  \\
   &                  &                                    & 2021-07-27 & SpeX    &          &          \\
8  & 18290316+0023090 & [EGE2007] Bolo 7 \tablenotemark{b} & 2009-10-11 & NIRSPEC & 23076352 & SL, LL2  \\
9  & 18290436+0116207 & Core                               & 2009-10-11 & NIRSPEC & 23074304 & SL, LL2  \\
10 & 18290479-0001301 &                                    & 2008-07-09 & SpeX    & 13210112 & SL, LL \\
11 & 18291546+0016422 & [EGE2007] Bolo 7 \tablenotemark{b} & 2008-07-09 & SpeX    & 23075328 & SL, LL \\
12 & 18291600+0109382 & Core                               & 2008-07-09 & SpeX    & 23073024 & SL, LL \\
13 & 18291619+0045143 & Ser/G3-G6                          & 2009-10-11 & NIRSPEC & 23073024 & SL, LL \\
14 & 18291699+0037191 &                                    & 2009-10-11 & NIRSPEC & 23076096 & SL, LL2  \\
15 & 18292528+0003141 &                                    & 2009-10-11 & NIRSPEC & 23072768 & SL, LL \\
16 & 18294108+0127449 & Core                               & 2009-10-11 & NIRSPEC & 23073536 & SL, LL2  \\
17 & 18295604+0104146 & Core                               & 2008-07-08 & SpeX    & 23075328 & SL, LL \\
18 & 18295940+0041007 &                                    & 2008-07-08 & SpeX    & 23075584 & SL, LL2  \\
   &                  &                                    & 2008-07-09 &         &          &          \\
19 & 18300085+0017069 &                                    & 2009-10-11 & NIRSPEC & 23072768 & SL, LL \\
20 & 18300896+0114441 & Core                               & 2009-10-11 & NIRSPEC & 23074816 & SL, LL2  \\
21 & 18301220+0115341 & Core                               & 2008-07-08 & SpeX    & 23075584 & SL, LL2  \\
\enddata
\tablenotetext{a}{Alias used throughout this paper.}
\tablenotetext{b}{\citet{Enoch2007}}
\tablenotetext{c}{Named location within the Serpens cloud, if available.}
\tablenotetext{d}{Wavelength range covered is 2.83-4.15 $\mu$m for Keck/NIRSPEC and 2.15-5.0 $\mu$m for IRTF/SpeX}
\tablenotetext{e}{{\it Spitzer}/IRS modules used: SL=Short-Low (5-14~$\mu$m, $R\sim100$), LL2=Long-Low 2 (14-21.3~$\mu$m, $R\sim100$), LL=Long-Low 1 and 2 (14-35~$\mu$m, $R\sim100$)}
\end{deluxetable*}

\section{Observations}\label{sec:Observations} 

\textit{Spitzer}/IRS spectra of background stars toward the Perseus
and Serpens clouds were obtained as part of a dedicated Open Time
program (PID 40580). Tables \ref{Table 1} and \ref{Table 2} list all
sources with their astronomical observation request (AOR) keys and the
IRS modules in which they were observed. The SL module, covering the
5–14 $\mu$m range, includes several ice absorption bands, as well as
the 9.7 $\mu$m band of silicates, and has the highest signal-to-noise
values (S/N$>$50). The LL2 module (14–21 $\mu$m) was included in order
to trace the 15 $\mu$m band of solid $\rm{CO_2}$ and for a better
overall continuum determination, although at a lower S/N of $>$30. At
longer wavelengths, the background stars are weaker, and the LL1
module ($\sim$20–35 $\mu$m) was used for only $\sim$30\% of the
sources. The spectra were extracted and calibrated from the
two-dimensional Basic Calibrated Data produced by the standard
\spitzer pipeline (version S16.1.0), using the same method and
routines discussed in \citet{Boogert2011}. Uncertainties (1$\sigma$)
for each spectral point were calculated using the ``func” frames
provided by the \spitzer pipeline.

The \spitzer spectra of Perseus and a subset for Serpens were
complemented by ground-based NASA IRTF/SpeX \citep{Rayner2003} K and
L-band spectra. For the remaining Serpens targets, L-band spectra were
obtained with the NIRSPEC spectrometer \citep{McLean1998} on Keck
II. The SpeX spectra were obtained in the LongXD1.9 or LongXD2.1
modes, offering wavelength ranges of 1.95-4.2 or 2.15-5.0 $\mu$m,
respectively. The M-band portions of these spectra are only presented
if they are of sufficiently high quality. The SpeX observations were
done under observing programs 2008A079, 2008B074, and 2010A107 spread
out over the nights listed in Tables \ref{Table 1} and \ref{Table 2}.
One target, Ser-7, was observed under program 2021A092 in the
LXD\_short mode (1.67-4.2 $\mu$m).  For all SpeX observations, a slit
width of 0.3" was used, yielding a resolving power of R
=$\lambda/\Delta \lambda$ = 2500. The spectra were flat fielded,
wavelength-calibrated, and extracted using the Spextool package
\citep{Cushing2004}. The telluric absorption lines were divided out
and the spectra were flux calibrated using the Xtellcor program
\citep{Vacca2003}. Standard stars of spectral type A0V were used for
this purpose.

The Keck/NIRSPEC spectra of the Serpens targets were observed in the
long-slit mode with the 0.57" wide slit, resulting in a resolving
power of R = 1,500. Two grating settings were observed, providing a
full L-band coverage (2.83-4.2 $\mu$m). The data were reduced from the
raw frames in a way standard for ground-based long-slit spectra with
the same IDL routines described in \citet{Boogert2008}. Sky emission
lines were used for the wavelength calibration and bright, nearby main
sequence stars were used as telluric and photometric standards. For
the division over the standard star, the S/N was optimized by
carefully matching the wavelength scale to that of the science
targets.

In the end, all SpeX, NIRSPEC, and \spitzer spectra were combined with
2MASS \textit{$J$}, \textit{$H$}, and \textit{$K_s$} broadband
photometry \citep{Skrutskie2006}, 3.55, 4.49, 5.73, and 7.87 $\mu$m
photometry from c2d \textit{Spitzer}/IRAC, 24 $\mu$m photometry from
\textit{Spitzer}/MIPS \citep{Evans2003}, and WISE photometry
\citep{Wright2010}.  The 1-5 $\mu$m spectra were matched to the
photometry by convolving them with the $K_s$-band filter profile and
then multiplying them along the flux scale. Similarly, the
\textit{Spitzer}/IRS spectra were matched to the IRAC 7.87 $\mu$m
photometric flux. The same photometry was used in the continuum
determination discussed in \S \ref{sec:Methods}. Catalog flags were
taken into account, such that the photometry of sources listed as
being confused within a 2" radius or being located within 2" of a
mosaic edge were treated as upper limits. The c2d catalogs do not
include flags for saturation. Therefore, photometry exceeding the IRAC
saturation limit (at the appropriate integration times) was flagged as
a lower limit. In those cases, the nearby WISE photometric points were
used instead. Finally, as the relative photometric calibration is
important for this work, the uncertainties in the \spitzer c2d and
2MASS photometry were increased with the zero-point magnitude
uncertainties listed in Table 21 of \citet{Evans2009} and further
discussed in Section 3.5.3 of that paper.

\section{Methods}\label{sec:Methods}

To determine the interstellar $A_{\rm K}$, $\tau_{3.0}$, and
$\tau_{9.7}$ values, contributions from the stellar photosphere to the
observed spectra need to be removed. We did so, following the methods
described in \citet{Boogert2011, Boogert2013}, and
\citet{Chu2020}. The fits to the observed spectra were significantly
more constrained by combining the spectra with broad-band photometric
data (\S\ref{sec:Observations}).

The fitting process was done in two steps. First, all data available
in the 1-5 $\mu$m range were fitted using a large (224) database of
template spectra \citep{Rayner2009} to derive accurate spectral types,
as well as $A_{\rm K}$ and $\tau_{3.0}$ values. Second, all data in
the full 1-20 $\mu$m range were fitted using a small (13) sample of
model spectra \citep{Decin2004, Boogert2011} to provide
$\tau_{9.7}$. Throughout this paper, the $A_{\rm K}$ and $\tau_{3.0}$
values derived from the 1-5 $\mu$m template fitting process were
reported, as they are most accurate considering the much larger
database of template spectra available.

All fit parameters are determined simultaneously, and any dependencies
are taken into account in the uncertainty estimates. For the 1-5
$\mu$m template fits, this is described in detail in
\citet{Chu2020}. In short, the key fit parameters are:

1. Spectral Type. The CO overtone lines between 2.25-2.60 $\mu$m
provide a sensitive tracer of spectral type. The near-infrared
\textit{JHK} photometry and absorption features in the 3.8-4.1 $\mu$m
spectral range also prove to be important for this. The \textit{JHK}
photometry depends on the extinction as well, which is discussed in
point 2 below.

2. Continuum Extinction ($A_{\rm K}$). We adopt the commonly used
extinction curve from \citet{Indebetouw2005}. Due to the steepness of
the extinction curve in the 1.0-2.5 $\mu$m region, $A_{\rm K}$ and its
uncertainty are primarily determined by the \textit{JHK} photometry
and the shape of the un-reddened spectral template spectrum. The
dependency on flux values at longer wavelengths is weak.

3. $\rm{H_2O}$ Absorption Feature ($\tau_{3.0}$). $\rm{H_2O}$ ice has
a prominent broad feature at 3.0 $\mu$m. $\tau_{3.0}$ and its
uncertainty are determined by the flux values and the observational
noise in the 2.9-3.2 $\mu$m range, including dependencies on the
accuracy of the baseline surrounding the ice feature. For the fitting,
an $\rm{H_2O}$ ice absorption spectrum for spherical, pure ice grains
using optical constants of amorphous ice at a temperature of 10 K
\citep{Hudgins1993} is assumed. A grain size of 0.4 $\mu$m is
chosen. This particular ice temperature and grain size have no
significance other than that they match the shape of the 3.0 $\mu$m
band well, providing a tool to derive $\tau_{3.0}$.

Besides the reduced $\chi^2$ values derived for the individual
wavelength regions discussed above, a total reduced $\chi^2$
($\chi_{\nu}^2$) is calculated across the wavelength range of 1-4
$\mu$m, using the IRTF template database. Where this $\chi_{\nu}^2$ is
lowest, the model template is chosen as the best fit to the star. The
final errors for $A_{\rm K}$ and $\tau_{3.0}$ are increased by
including all of the model templates that have $\chi_{\nu}^2$ (1-4
$\mu$m) within a factor of 2 of the best template. This represents a
confidence level of at least 3$\sigma$. In some cases, the
uncertainties on $A_{\rm K}$ and $\tau_{3.0}$ are much smaller than
expected from the $\chi_{\nu}^2$ across the full 1-4 $\mu$m wavelength
range, because $A_{\rm K}$ and $\tau _{3.0}$ depend strongly on only a
sub-set of this wavelength range.

For the full 1-21 $\mu$m fits, using a much smaller set of stellar
models, $\tau_{3.0}$ is kept fixed. The strength and shape of the
longer wavelength $\rm{H_2O}$ ice bands is set by assuming the
$\rm{H_2O}$ ice model discussed above. The 9.7 $\mu$m silicate band is
fitted for grains small compared to the wavelength, having a pyroxene
to olivine optical depth ratio of 0.62 at the 9.7 $\mu$m peak
\citep{Boogert2011}. A key factor in the fitting process is the
photospheric SiO band at $\sim$8 $\mu$m, as it overlaps with the 9.7
$\mu$m band of silicate dust. For some targets, the model spectral
types were optimized to fit that photospheric band best.  Also, in
some cases the models were normalized to the data at a wavelength of
8.0 $\mu$m, instead of the default of 5.5 $\mu$m. This provides a
better local baseline for the silicate feature, while reducing the fit
quality at other wavelengths.  Accurate $\tau_{9.7}$ values are more
important for this work than a good fit over the larger wavelength
range.

\section{Results}\label{sec:Results}

Using the models described in \S \ref{sec:Methods}, the values for
$A_{\rm K}$, $\tau_{3.0}$, and $\tau_{9.7}$ are derived. The best
fitting models are shown in Fig. \ref{Fig. appendix} and the derived
values in Tables \ref{Table 3} and \ref{Table 4}.

%%%%% RESULTS DATA TABLES %%%%%%

\begin{deluxetable*}{ccccccccp{3cm}}[p]
\tablewidth{0pt}
\centering
\caption{Perseus Target Results\label{Table 3}}
\tablehead{
  \colhead{Alias}  & \colhead{IRTF Template} & \colhead{Model}       &  \colhead{$A_{\rm K}$}&  \colhead{$\tau_{3.0}$} & \colhead{$\tau_{9.7}$} & \colhead{$\chi_{\nu}^2$} & \colhead{N($\rm{H_2O_{ice}}$)\tablenotemark{a}} & \colhead{Notes} \\
  \colhead{(Per-)} & \colhead{1-5 $\mu$m}    & \colhead{1-30 $\mu$m} &  \colhead{mag}       &  \colhead{}             & \colhead{}           & \colhead{}            & \colhead{$10^{17} \rm{cm}^{-2}$}                & \colhead{}      \\
}
\startdata
1 & HD16068 (K3.5II) & K4III & 0.34 $\pm$ 0.07 & 0.0 $\pm$ 0.12 & 0.26 $\pm$ 0.09 & 2.92 & $<$1.94 & L-band = 1.15 \\
2 & HD170820 (G9II) & K0III & 0.92 $\pm$ 0.08 & 0.49 $\pm$ 0.17 & 0.23 $\pm$ 0.05 & 0.76 & 7.87 $\pm$ 2.73 & L-band = 1.15 \\
3 & HD44391 (K0I) & K0III\tablenotemark{b} & 0.30 $\pm$ 0.08 & 0.01 $\pm$ 0.05 & 0.24 $\pm$ 0.07 & 0.57 & $<$0.83 & ... \\
4 & HD222093 (G9III) & G8III & 0.22 $\pm$ 0.06 & 0.0 $\pm$ 0.05 & 0.22 $\pm$ 0.05 & 2.02 & $<$0.81 & L-band = 1.02 \\
5 & HD202314 (G6I) & G8III & 1.27 $\pm$ 0.09 & 0.49 $\pm$ 0.06 & 0.33 $\pm$ 0.05 & 0.51 & 7.91 $\pm$ 0.97 & L-band = 1.13 \\
6 & HD44391 (K0I) & G8III\tablenotemark{b} & 0.75 $\pm$ 0.07 & 0.39 $\pm$ 0.08 & 0.38 $\pm$ 0.05 & 1.70 & 6.29 $\pm$ 1.36 & L-band = 0.93 \\
7 & HD35620 (K3.5III) & K4III & 1.63 $\pm$ 0.05 & 0.55 $\pm$ 0.05 & 0.35 $\pm$ 0.05 & 0.24 & 8.97 $\pm$ 0.82 & L-band = 0.97 \\
8 & HD44391 (K0I) & K0III\tablenotemark{b} & 0.43 $\pm$ 0.15 & 0.01 $\pm$ 0.11 & 0.26 $\pm$ 0.05 & 0.5 & $<$1.78 & L-band = 1.07 \\
9 & HD192713 (G3I) & G8III & 1.41 $\pm$ 0.05 & 0.52 $\pm$ 0.05 & 0.41 $\pm$ 0.05 & 0.67 & 8.47 $\pm$ 0.86 & L-band = 1.06 \\
10 & HD9852 (K0.5III) & K4III & 1.32 $\pm$ 0.14 & 0.24 $\pm$ 0.08 & 0.57 $\pm$ 0.07 & 1.11 & 3.81 $\pm$ 1.27 & ... \\
11 & HD132935 (K2III) & K5III & 1.69 $\pm$ 0.07 & 0.84 $\pm$ 0.05 & 0.43 $\pm$ 0.05 & 1.24 & 13.66 $\pm$ 0.82 & L-band = 1.07 \\
12 & HD182694 (G7III) & G8III & 1.87 $\pm$ 0.12 & 0.92 $\pm$ 0.11 & 0.38 $\pm$ 0.07 & 1.11 & 14.9 $\pm$ 1.79 & L-band = 0.92 \\
13 & HD9852 (K0.5III) & G8III\tablenotemark{b} & 0.21 $\pm$ 0.17 & 0.0 $\pm$ 0.05 & 0.19 $\pm$ 0.05 & 1.08 & $<$0.81 & ... \\
14 & HD91810 (K1III) & K3III & 1.43 $\pm$ 0.05 & 0.29 $\pm$ 0.11 & 0.57 $\pm$ 0.09 & 0.81 & 4.76 $\pm$ 1.80 & ... \\
15 & HD120477 (K5.5III) & K7III & 0.42 $\pm$ 0.11 & 0.0 $\pm$ 0.10 & 0.24 $\pm$ 0.08 & 3.24 & $<$1.62 & L-band = 1.03 \\
16 & Gl581 (M2.5V) & G8III & 0.0 $\pm$ 0.05 & 0.09 $\pm$ 0.05 & 0.0 $\pm$ 0.05 & 3.65 & 1.41 $\pm$ 0.78 & L-band = 0.93 \\
17 & HD35620 (K3.5III) & K7III\tablenotemark{b} & 0.67 $\pm$ 0.12 & 0.08 $\pm$ 0.05 & 0.29 $\pm$ 0.07 & 0.59 & 1.32 $\pm$ 0.82 & L-band = 0.95 \\
18 & HD213893 (M0III) & K4III & 0.58 $\pm$ 0.09 & 0.04 $\pm$ 0.05 & 0.24 $\pm$ 0.06 & 0.71 & $<$0.84 & L-band = 0.98 \\
19 & HD108519 (F0V) & G8III & 0.98 $\pm$ 0.05 & 0.18 $\pm$ 0.06 & 0.35 $\pm$ 0.07 & 1.17 & 2.91 $\pm$ 0.97 & local 9.7 $\mu$m baseline \\
20 & HD160365 (F6III) & G8III & 1.23 $\pm$ 0.13 & 0.1 $\pm$ 0.06 & 0.54 $\pm$ 0.06 & 4.09 & 1.63 $\pm$ 0.98 & L-band = 0.95 \\
21 & HD222093 (G9III) & G8III\tablenotemark{b} & 0.95 $\pm$ 0.08 & 0.06 $\pm$ 0.05 & 0.34 $\pm$ 0.05 & 0.68 & 0.97 $\pm$ 0.81 & L-band = 0.98 \\
22 & HD4408 (M4III) & M6III & 0.95 $\pm$ 0.06 & 0.19 $\pm$ 0.05 & 0.33 $\pm$ 0.08 & 2.99 & 3.01 $\pm$ 0.79 & L-band=0.83, local 9.7 $\mu$m baseline \\
23 & HD132935 (K2III) & K4III & 0.71 $\pm$ 0.1 & 0.13 $\pm$ 0.08 & 0.37 $\pm$ 0.05 & 0.80 & 2.09 $\pm$ 1.29 & L-band = 0.90 \\
24 & HD10697 (G3V) & G8III & 0.61 $\pm$ 0.11 & 0.09 $\pm$ 0.09 & 0.27 $\pm$ 0.05 & 1.18 & 1.49 $\pm$ 1.49 & L-band = 0.95 \\
25 & HD100006 (K0III) & K0III & 0.35 $\pm$ 0.09 & 0.0 $\pm$ 0.06 & 0.12 $\pm$ 0.05 & 1.77 & $<$0.97 & ... \\
26 & HD16139 (G8.5III) & G8III & 0.49 $\pm$ 0.09 & 0.04 $\pm$ 0.05 & 0.20 $\pm$ 0.05 & 0.89 & $<$0.83 & ... \\
27 & HD132935 (K2III) & K3III & 0.37 $\pm$ 0.05 & 0.05 $\pm$ 0.05 & 0.22 $\pm$ 0.05 & 0.49 & 0.83 $\pm$ 0.83 & ... \\
28 & HD35620 (K3.5III) & K4III\tablenotemark{b} & 0.76 $\pm$ 0.12 & 0.23 $\pm$ 0.11 & 0.26 $\pm$ 0.05 & 0.51 & 3.72 $\pm$ 1.78 & L-band = 0.95 \\
\enddata

\tablenotetext{a}{Column density upper limits are 3$\sigma$.}
\tablenotetext{b}{Poor fit 8 $\mu$m photospheric SiO band.}
\tablecomments{Perseus targets observed by alias along with the fitted
  IRTF templates, full spectra templates assuming spectral types,
  extinctions, water ice optical depths, silicate optical depths,
  $\chi_{\nu}^2$ values for the IRTF template database fits, water ice
  column densities, and extra notes.}
\end{deluxetable*}

\begin{deluxetable*}{ccccccccp{3cm}}[p]
\tablewidth{0pt}
\centering
\caption{Serpens Target Results\label{Table 4}}
\tablehead{
  \colhead{Alias}  & \colhead{IRTF Template} & \colhead{Model}       &  \colhead{$A_{\rm K}$}&  \colhead{$\tau_{3.0}$} & \colhead{$\tau_{9.7}$} & \colhead{$\chi_{\nu}^2$} & \colhead{N($\rm{H_2O_{ice}}$)\tablenotemark{a}} & \colhead{Notes} \\
  \colhead{(Ser-)} & \colhead{1-5 $\mu$m}    & \colhead{1-30 $\mu$m} &  \colhead{mag}       &  \colhead{}             & \colhead{}           & \colhead{}            & \colhead{$10^{17} \rm{cm}^{-2}$}                & \colhead{}      \\
}
%
%\tablehead{\colhead{\makecell{Alias \\ (Ser-)}} & \colhead{IRTF
%    Template} & \colhead{\makecell{2-21 $\mu$m \\ Model}} &
%  \colhead{\makecell{$A_{\rm K} \pm$ error \\ (mag)}} &
%  \colhead{$\tau_{3.0} \pm$ error} & \colhead{$\tau_{9.7} \pm$ error}
%  & \colhead{$\chi_{\nu}^2$} & \colhead{\makecell{N($\rm{H_2O_{ice}}$)
%      \\ $\pm$ error \tablenotemark{a} \\ $\left(10^{17}
%      \rm{cm}^{-2}\right)$}} & Notes}
%
\startdata
1 & HD204724 (M1III) & M1III\tablenotemark{c} & 0.73 $\pm$ 0.23 & 0.10 $\pm$ 0.10 & 0.24 $\pm$ 0.05 & 0.86 & 1.64 $\pm$ 1.64 & ... \\
2 & HD196610 (M6III) & M7III & 0.87 $\pm$ 0.16 & 0.10 $\pm$ 0.18 & --- \tablenotemark{b} & 1.25 & $<$2.93 & ... \\
3 & HD179870 (K0II) & K3III\tablenotemark{c} & 0.75 $\pm$ 0.19 & 0.03 $\pm$ 0.06 & 0.31 $\pm$ 0.05 & 1.91 & $<$0.97 & L-band = 0.95 \\
4 & HD28487 (M3.5III) & M3III & 1.12 $\pm$ 0.17 & 0.24 $\pm$ 0.12 & 0.38 $\pm$ 0.08 & 1.18 & 3.81 $\pm$ 1.91 & ... \\ 
5 & HD64332 (S4.5) & M1III & 1.09 $\pm$ 0.07 & 0.22 $\pm$ 0.08 & 0.36 $\pm$ 0.05 & 0.24 & 3.62 $\pm$ 1.32 & L-band = 1.10 \\
6 & HD4408 (M4III) & M1III & 1.01 $\pm$ 0.05 & 0.21 $\pm$ 0.09 & 0.44 $\pm$ 0.07 & 0.64 & 3.32 $\pm$ 1.42 &  local 9.7 $\mu$m baseline \\
7 & HD18191 (M6III) & M6III & 4.75 $\pm$ 0.44 & 1.63 $\pm$ 0.06 & 1.18 $\pm$ 0.06 & 0.87 & 26.33 $\pm$ 0.97 &  L-band = 1.10, local 9.7 $\mu$m baseline \\
8 & HD4408 (M4III) & M6III & 0.73 $\pm$ 0.05 & 0.09 $\pm$ 0.05 & 0.42 $\pm$ 0.05 & 0.64 & 1.52 $\pm$ 0.85 & ... \\
9 & HD201065 (K4I) & M0III & 0.92 $\pm$ 0.40 & 0.08 $\pm$ 0.07 & 0.37 $\pm$ 0.05 & 1.01 & 1.38 $\pm$ 1.20 & ... \\
10 & HD39045 (M3III) & M1III & 0.73 $\pm$ 0.08 & 0.13 $\pm$ 0.13 & 0.31 $\pm$ 0.05 & 0.26 & 2.18 $\pm$ 2.18 & L-band = 0.95 \\
11 & HD204724 (M1III) & M6III & 1.44 $\pm$ 0.12 & 0.47 $\pm$ 0.13 & 0.54 $\pm$ 0.05 & 0.48 & 7.66 $\pm$ 2.12 & ... \\
12 & HD204724 (M1III) & M0III & 0.71 $\pm$ 0.13 & 0.03 $\pm$ 0.11 & 0.19 $\pm$ 0.06 & 3.03 & $<$1.81 & L-band = 1.10 \\
13 & HD194193 (K7III) & M1III & 0.78 $\pm$ 0.15 & 0.05 $\pm$ 0.12 & --- \tablenotemark{b} & 0.94 & $<$2.02 & L-band = 1.15 \\
14 & HD4408 (M4III) & M6III & 2.03 $\pm$ 0.09 & 0.63 $\pm$ 0.20 & 0.61 $\pm$ 0.09 & 0.89 & 10.16 $\pm$ 3.22 & local 9.7 $\mu$m baseline  \\
15 & HD201065 (K4I) & K7III\tablenotemark{c} & 0.69 $\pm$ 0.33 & 0.03 $\pm$ 0.05 & 0.29 $\pm$ 0.08 & 0.81 & $<$0.82 & ... \\
16 & HD201065 (K4I) & M1III & 0.83 $\pm$ 0.24 & 0.06 $\pm$ 0.07 & 0.28 $\pm$ 0.08 & 1.13 & $<$1.15 & ... \\
17 & HD94705 (M5.5III) & M6III & 1.40 $\pm$ 0.05 & 0.08 $\pm$ 0.18 & 0.54 $\pm$ 0.13 & 0.39 & $<$2.88 & ... \\
18 & HD108477 (G4III) & G8III & 1.91 $\pm$ 0.20 & 0.50 $\pm$ 0.05 & 0.74 $\pm$ 0.06 & 0.54 & 8.13 $\pm$ 0.81 & L-band = 1.08 \\
19 & HD11443 (F6IV) & G8III & 0.98 $\pm$ 0.24 & 0.0 $\pm$ 0.05 & 0.43 $\pm$ 0.06 & 3.50 & $<$0.81 & ... \\
20 & HD28487 (M3.5III) & M6III & 1.19 $\pm$ 0.15 & 0.24 $\pm$ 0.12 & 0.43 $\pm$ 0.05 & 0.93 & 3.87 $\pm$ 1.94 & ... \\
21 & HD44391 (K0I) & G8III & 1.61 $\pm$ 0.10 & 0.29 $\pm$ 0.05 & 0.57 $\pm$ 0.05 & 0.48 & 4.63 $\pm$ 0.80 & L-band = 1.07 \\
\enddata

\tablenotetext{a}{Column density upper limits are 3$\sigma$.}

\tablenotetext{b}{Poor full spectrum fit, resulting in discarded
  $\tau_{9.7}$ value.}

\tablenotetext{c}{Poor fit 8 $\mu$m photospheric SiO band.}

\tablecomments{Serpens targets observed by alias along with the fitted
  IRTF templates, full spectra templates assuming spectral types,
  extinctions, water ice optical depths, silicate optical depths, the
  total reduced $\chi_{\nu}^2$ values for the IRTF template database
  fits, water ice column densities, and extra notes.}
\end{deluxetable*}

The fits to the 1-5 $\mu$m wavelength region, using the IRTF template
database, are generally good. In a fair number of cases, however, we
did notice deficiencies in fitting the continuum region near 4.0
$\mu$m, even though the data at shorter wavelengths were fitted
well. This could be due to calibration errors in the slope of the 2-4
$\mu$m SpeX spectra or in the \spitzer or WISE photometry used to
calibrate the NIRSPEC spectra (for which no $K-$band portion is
available). It could also reflect an uncertainty in the extinction
curve or incompleteness of the spectral database. Regardless of the
origin, we reduced the effect on the derived $\tau_{3.0}$ by scaling
the \textit{L-}band spectrum to match the best-fitting model. The
scaling factors are provided in Tables \ref{Table 3} and \ref{Table
  4}. In five cases, this correction factor is less than 5\%, and in
twenty-three cases 5-17\%. We are confident in the derived values of
$\tau_{3.0}$, as the absorption band can be distinguished by the shape
to the $L-$band spectrum, and the adjustments are essentially local
baseline corrections.

Of all 49 targets, 7 have deviating slopes in the 8-12 $\mu$m region,
even though the IRTF template fits in the 1-5 $\mu$m range are
good. This may reflect the small number of 1-20 $\mu$m photospheric
models available. In order to improve the fit of the 9.7 $\mu$m
silicate feature, a local baseline correction was applied by changing
the wavelength range for which the model is normalized to the
observations from the default of 5.34-5.50 $\mu$m to 7.3-7.5
$\mu$m. However, for two of these targets (Ser-2 and Ser-13), this
approach was insufficient. The deviations from the model are very
large, and there is a hint of silicate emission affecting the shape of
the 9.7 $\mu$m absorption feature. For these, probably more evolved,
mass losing stars, we discarded the derived $\tau_{9.7}$ values.  For
another 10 targets, while the 8-12 $\mu$m slopes are in good agreement
with the spectral types derived from the 1-5 $\mu$m range, the 8
$\mu$m SiO photospheric absorption band suggests later spectral
types. Such inconsistency was also noted for several Lupus background
stars \citep{Boogert2013}. The affected targets are indicated in
Tables \ref{Table 3} and \ref{Table 4}, and the effect on the reported
$\tau_{9.7}$ values was taken into account.

\subsection{$\tau_{9.7}$ and $A_{\rm K}$ Correlation}\label{subsec:Si}

\begin{figure}[h]
  \center
  %\begin{flushright}
  %\includegraphics[width=0.445\textwidth,right]{PerseusSi.png}\vspace{-5pt}
  %\includegraphics[width=0.445\textwidth]{PerseusSi.png}\vspace{-5pt}
    
  %\includegraphics[width=0.47\textwidth,left]{SerpensSi(zoomed).png}\vspace{-6pt}
  %  \includegraphics[width=0.47\textwidth]{SerpensSi(zoomed).png}\vspace{-6pt}  
    
    %\includegraphics[width=0.445\textwidth,right]{SerpensSi.png}
    %  \includegraphics[width=0.445\textwidth]{SerpensSi.png}
  \includegraphics[width=0.445\textwidth]{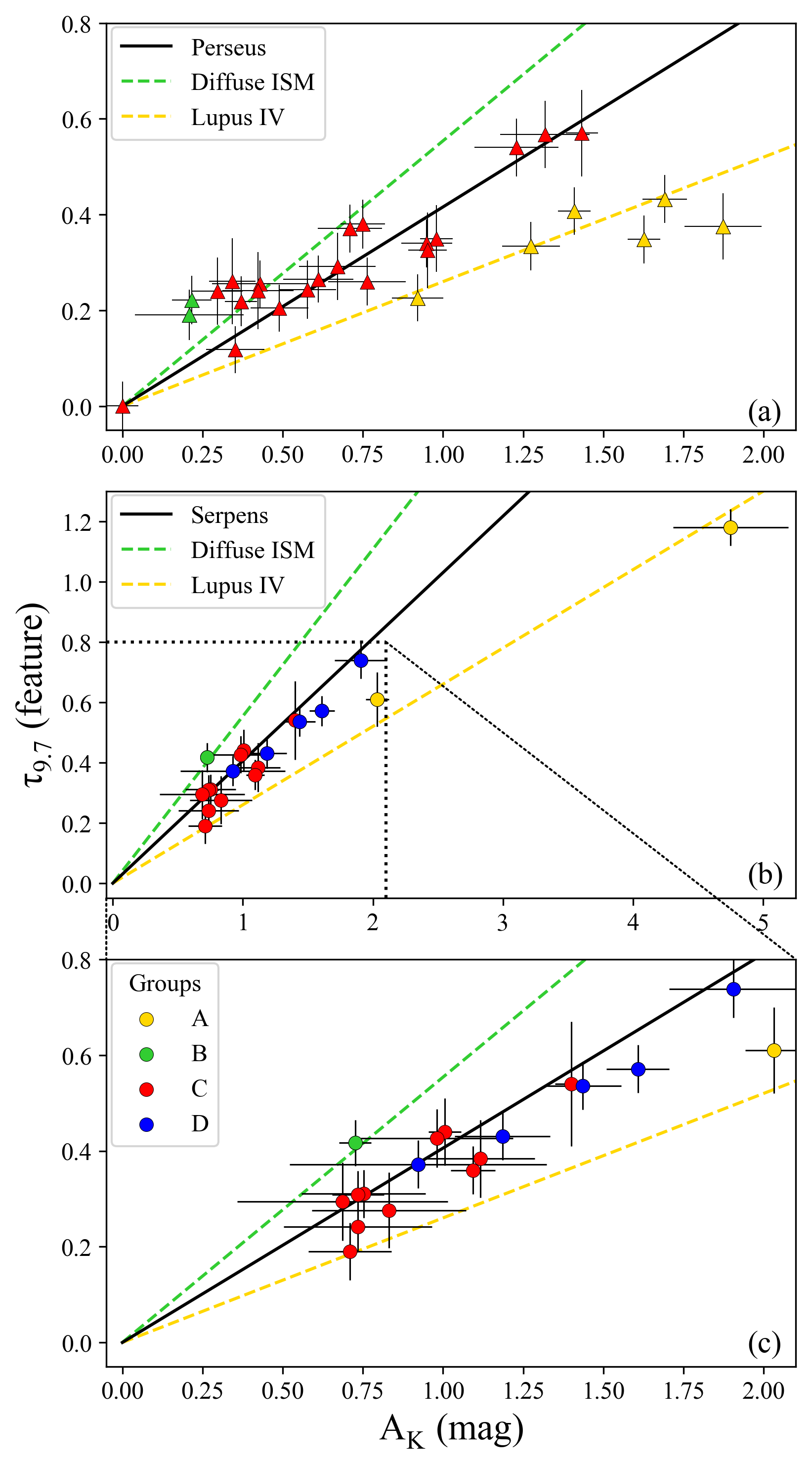}    
  %\end{flushright}
    \caption{Silicate band optical depth and extinction correlation
      for Perseus (a) and Serpens (b and c) targets. The error bars
      are of 3$\sigma$ significance. The solid black line is a least
      square fit to all targets in the cloud. The dashed yellow line
      is the Lupus IV correlation from \citet{Boogert2013}. The dashed
      green line is the Diffuse ISM correlation from
      \citet{Whittet2003}. {\bf Panel c} zooms in on the lower
      extinction Serpens targets. The colors of the bullets represent
      the groups defined in \S \ref{subsec:Si}, and are indicated in
      panel c. They apply to panels a and b as well as to the other
      Figures in this paper.\label{Fig. Si_correlation}}
\end{figure}

Analogous to earlier work \citep{Chiar2007, Boogert2013}, we plot the
derived values for $\tau_{9.7}$ against $A_{\rm K}$
(Fig. \ref{Fig. Si_correlation}). The rising trend is fitted with:

\begin{equation}
    \tau_{9.7} = m \times A_{\rm K}
\end{equation}

We find that for Perseus, $m = 0.416 \pm 0.026$, and for Serpens, $m =
0.406 \pm 0.041$. A significant amount of scatter is visible in these
plots, especially for Perseus. In order to investigate the origin of
these variations, we separate the target into different
groups. Throughout this work, we will refer to these as Groups A to E.

The targets that fall at least $6\sigma$ below the linear fit are
referred to as ``Group A." Those that are at least $6\sigma$ above
this relation will be referred to as ``Group B." As can be seen in
Fig. \ref{Fig. Si_correlation}, Group A targets follow the Lupus IV
dense cloud relation \citep[$m = 0.26$;][]{Boogert2013}, and Group B
targets follow the diffuse ISM relation \citep[$m =
  0.554$;][]{Whittet2003}. Based on this, targets that fall within
$6\sigma$ of more than one of these three relations are assigned to
``Group C," and targets that exclusively follow the fits over all
targets are in ``Group D." Only Serpens exhibits this latter Group.
For Perseus, several Group C targets follow the overall fit, but the
uncertainties are too large to assign them to Group D.  Finally,
targets for which no reliable $\tau_{9.7}$ values could be derived,
but that do have $A_{\rm K}$ and $\tau_{3.0}$ measurements, are in
``Group E." Also, note that one target (Per-16) has $A_{\rm K} = 0.0
\pm 0.05$ (Table \ref{Table 3}), which is consistent with its distance
of 13.7 pc \citep{Reiners2020}, indicating this target is actually a
Perseus foreground, rather than background, star.

\begin{figure}[h]
  \center
  %\begin{flushright}
  %%\includegraphics[width=0.445\textwidth,right]{PerseusH2O.png}\vspace{-6pt}
  %\includegraphics[width=0.445\textwidth]{PerseusH2O.png}\vspace{-6pt}
  %  
  %%\includegraphics[width=0.47\textwidth,left]{SerpensH2O(zoomed).png}\vspace{-6pt}
  %\includegraphics[width=0.47\textwidth]{SerpensH2O(zoomed).png}\vspace{-6pt}
  %  
  %%\includegraphics[width=0.445\textwidth,right]{SerpensH2O.png}
  %\includegraphics[width=0.445\textwidth]{SerpensH2O.png}
    \includegraphics[width=0.445\textwidth]{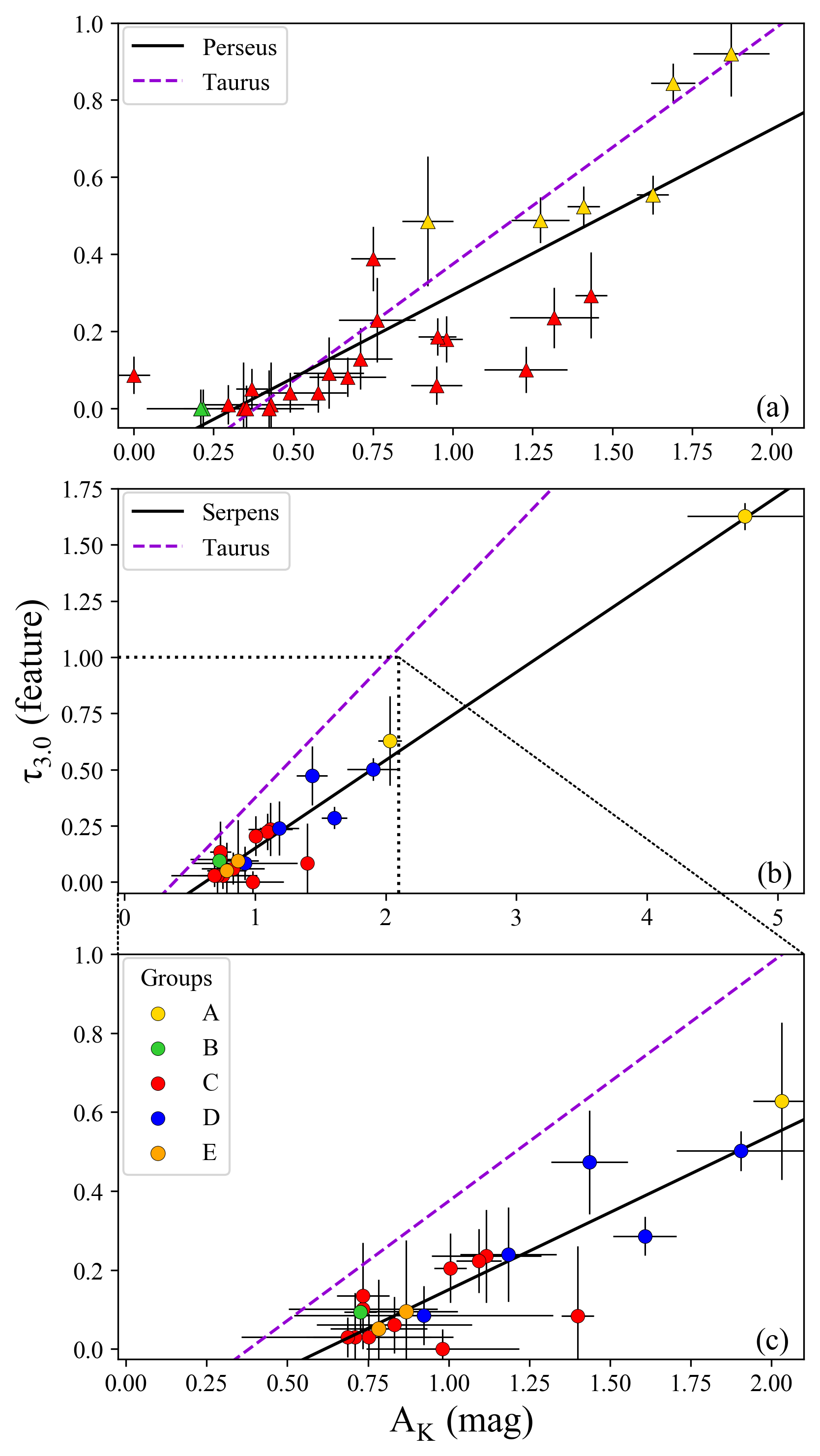}    
  %\end{flushright}  
    \caption{Water ice band optical depth and extinction correlations
      for Perseus ({\bf a}) and Serpens ({\bf b and c}) targets. The
      error bars are of 3$\sigma$ significance. The solid black line
      is a least square fit to all targets in the cloud. The dashed
      purple line is the Taurus molecular cloud correlation from
      \citet{Whittet2001}. The colors of the data points correspond to
      the groupings defined in \S \ref{subsec:Si} and
      Fig. \ref{Fig. Si_correlation}. {\bf Panel c} zooms in on the
      lower extinction Serpens targets.\label{Fig. H2O_correlation}}
\end{figure}

\begin{figure}[h]
  \center
  %\begin{flushright}
  %\includegraphics[width=0.445\textwidth]{Perseus_97v30.png}\vspace{-6pt}
  %  
  %\includegraphics[width=0.47\textwidth]{Serpens_97v30_zoomed.png}\vspace{-6pt}
  %  
  %\includegraphics[width=0.445\textwidth]{Serpens_97v30.png}
  %\end{flushright}
    \includegraphics[width=0.445\textwidth]{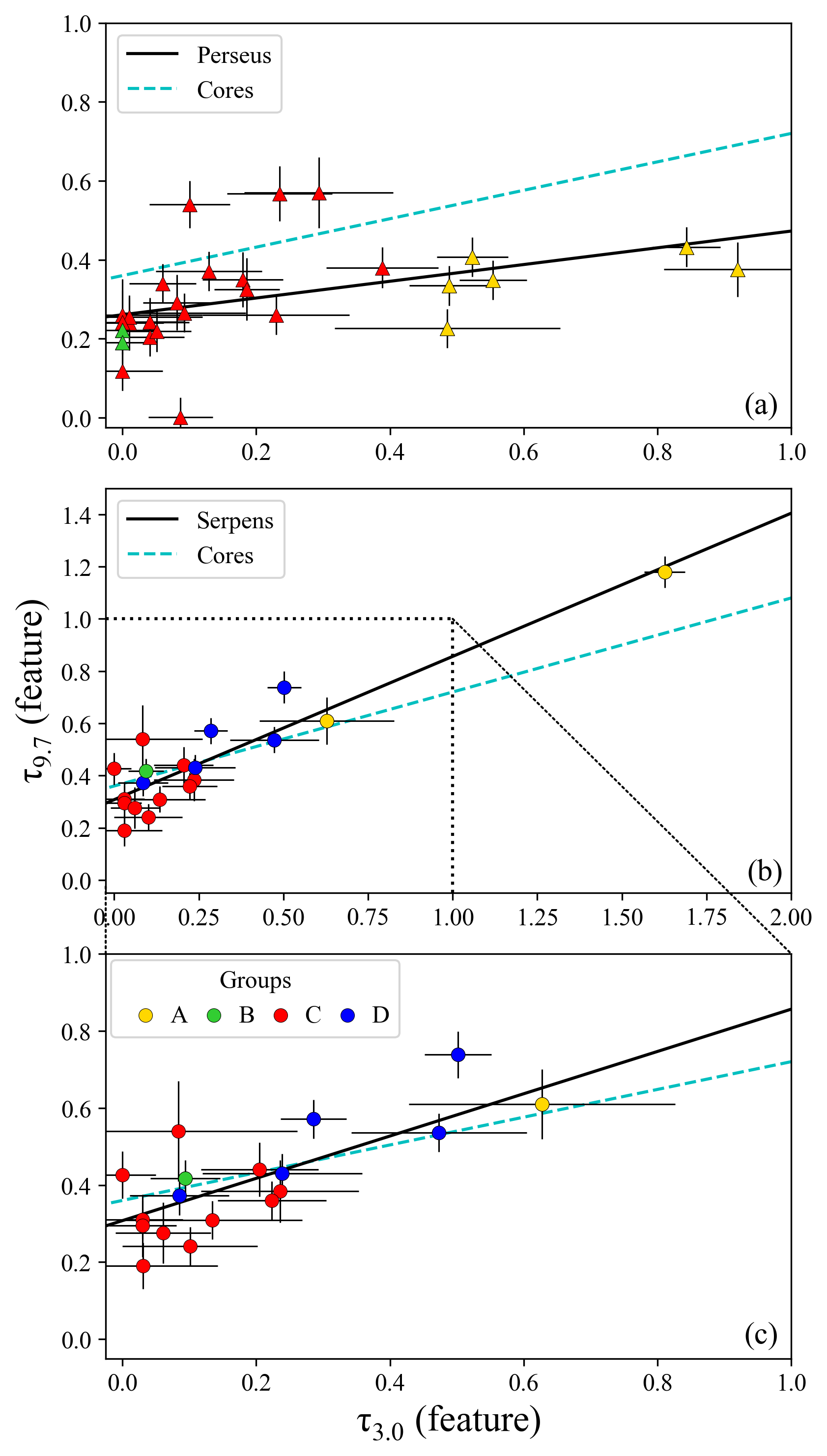}      
    \caption{Water ice and silicate band optical depth correlations
      for Perseus ({\bf a}) and Serpens ({\bf b and c}) targets. The
      error bars are of 3$\sigma$ significance. The solid black line
      is a least square fit to all targets in the cloud. The dashed
      blue line is the correlation derived from dense cores in
      \citet{Boogert2011}. The colors of the data points correspond to
      the groupings defined in \S \ref{subsec:Si} and
      Fig. \ref{Fig. Si_correlation}. {\bf Panel c} zooms in on the
      lower extinction Serpens
      targets. \label{Fig. H2Osil_correlation}}
\end{figure}

\begin{figure}[h]
  \centering
  \includegraphics[width=0.445\textwidth]{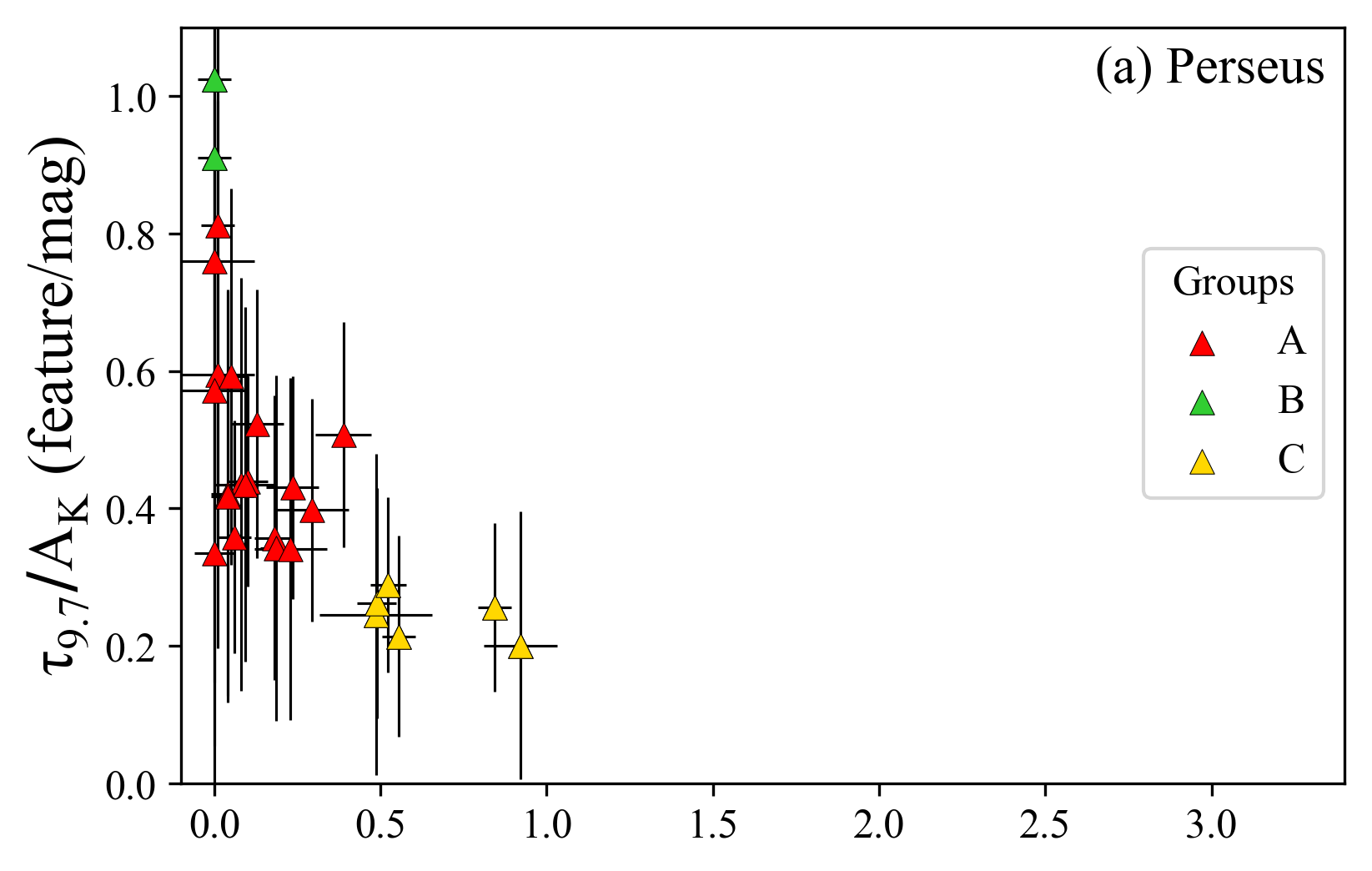}\vspace{-6pt}
    
  \includegraphics[width=0.445\textwidth]{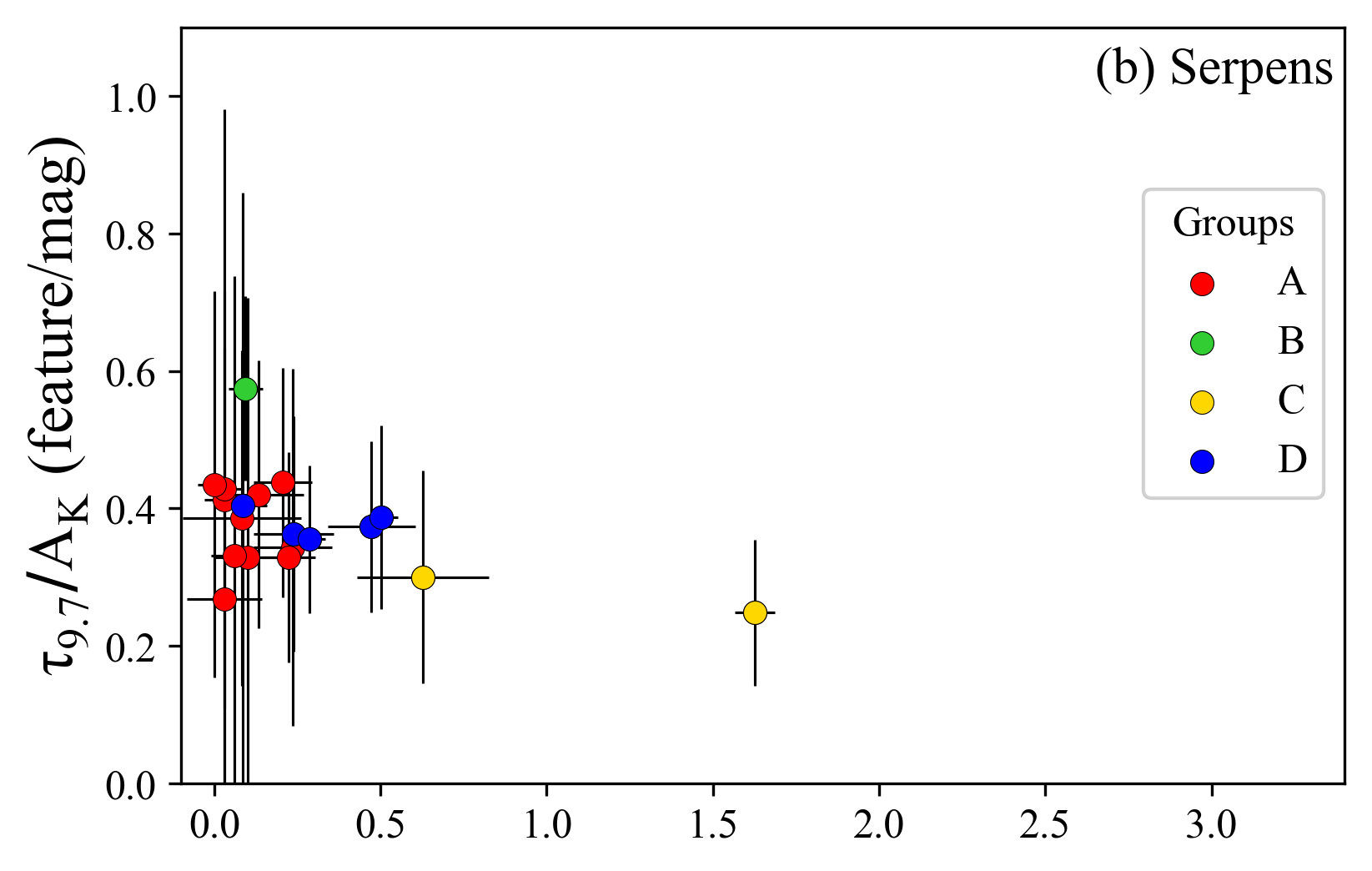}\vspace{-6pt}
    
  \includegraphics[width=0.445\textwidth]{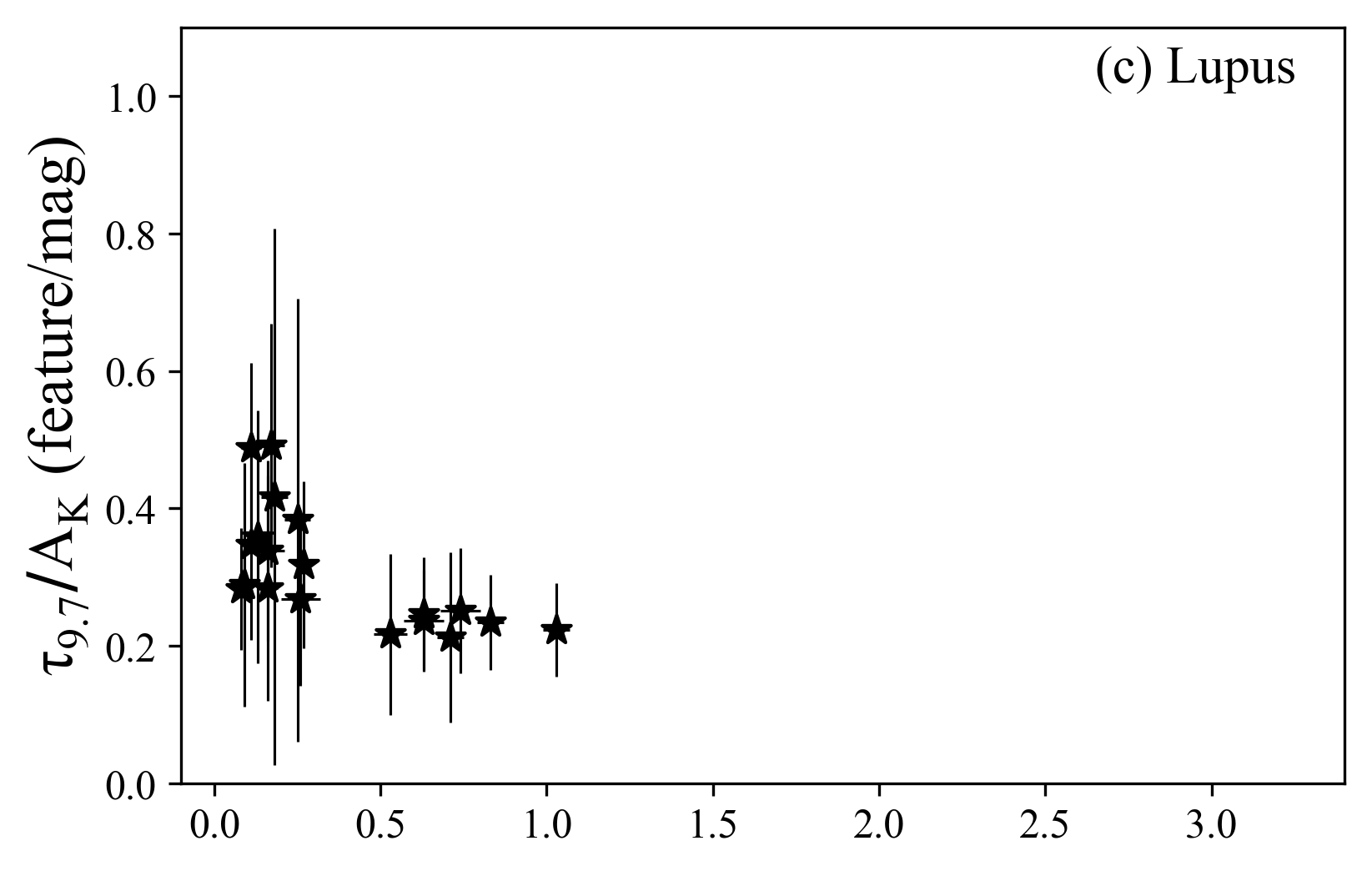}\vspace{-6pt}

  \includegraphics[width=0.445\textwidth]{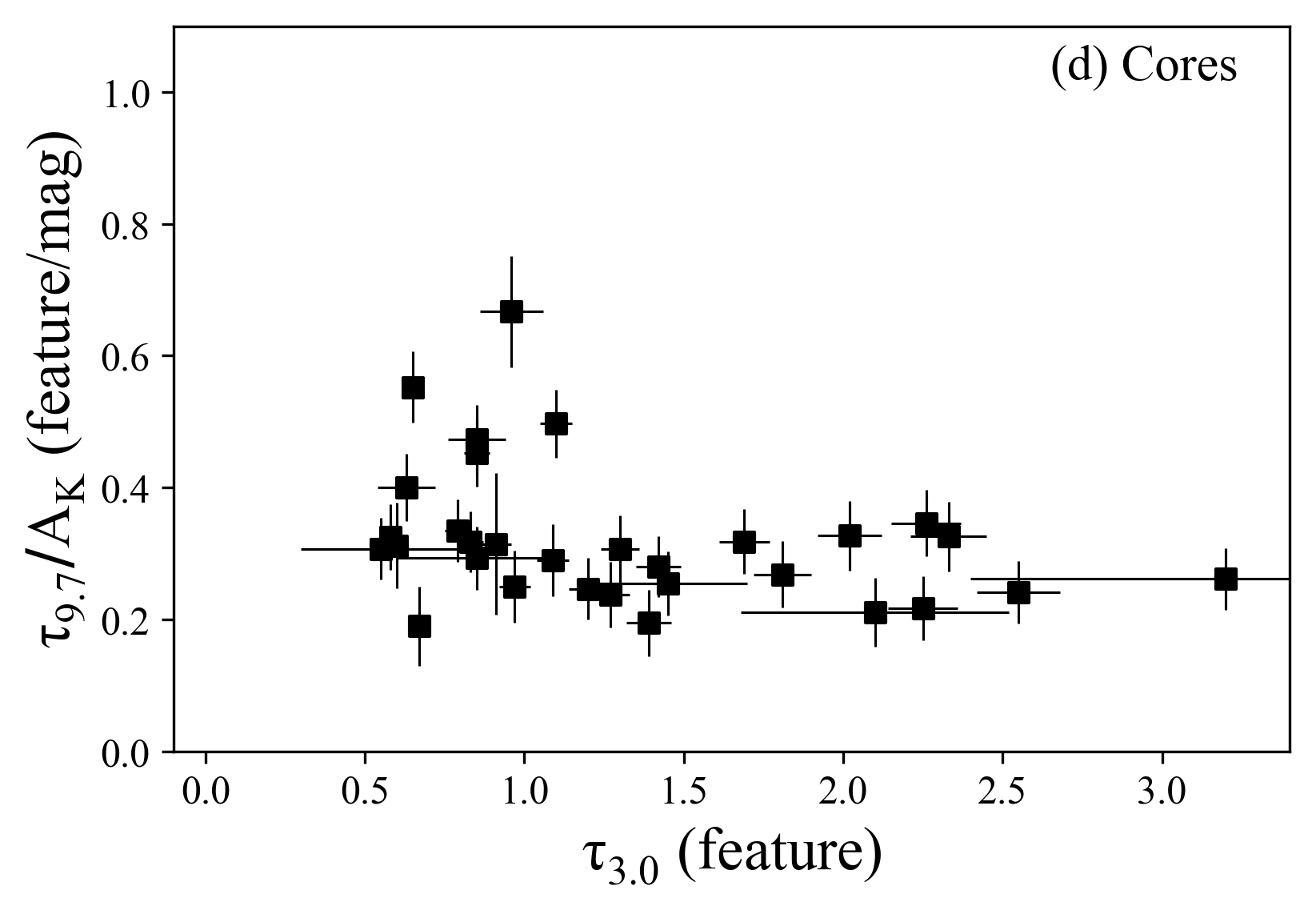}  
    \caption{Relation between the $\tau_{9.7}$/$A_{\rm K}$ ratio and
      $\tau_{3.0}$ for Perseus ({\bf a}), Serpens ({\bf b}), and Lupus
      ({\bf c}) clouds, as well as isolated dense cores ({\bf d}). The
      error bars are of 3$\sigma$ significance. The colors of the data
      points in panels {\bf a} and {\bf b} correspond to the groupings
      in
      Fig. \ref{Fig. Si_correlation}.\label{Fig. silakh2o_correlation}}
\end{figure}

\subsection{$\tau_{3.0}$ and $A_{\rm K}$ Correlation}\label{subsec:H2O}

The relation between $\tau_{3.0}$ and $A_{\rm K}$ can be used to
determine the ice formation threshold \citep[e.g.,][]{Whittet2001} and
to determine the abundance of ices relative to dust. This relation is
plotted for Perseus and Serpens in
Fig. \ref{Fig. H2O_correlation}. Generally, $\tau_{3.0}$ increases as
a function of $A_{\rm K}$, and the relation does not go through the
origin. The data points are therefore fitted with the function:

\begin{equation}
    \tau_{3.0} = a \times A_{\rm K} + b
\end{equation}

We find that for Perseus, $a = 0.430 \pm 0.054$ and $b = -0.136 \pm
0.052$, and for Serpens, $a = 0.391 \pm 0.021$ and $b = -0.241 \pm
0.032$. The abscissa of this relation gives the ice formation
threshold. For Perseus this is $A_{\rm K} = 0.315 \pm 0.127$, and for
Serpens $A_{\rm K} = 0.616 \pm 0.087$.  The Perseus values are
comparable to those for Lupus IV: $a = 0.44 \pm 0.03$ and $b = -0.11
\pm 0.03$, and an ice formation threshold of $A_{\rm K} = 0.25 \pm
0.07$ \citep{Boogert2013}.  The Serpens ice formation threshold is
almost twice that of the other clouds, which may relate to unrelated
foreground extinction at the larger distance of Serpens
(\S\ref{subsec:ices}; \citealt{Zucker2019}).

For Perseus, $\tau_{3.0}$ shows a significant scatter as a function of
$A_{\rm K}$ (Fig. \ref{Fig. H2O_correlation}a). By indicating the
different groups identified in the $\tau_{9.7}$ versus $A_{\rm K}$
correlation (\S\ref{subsec:Si}), it is evident that the Group A
targets, i.e., those with the most suppressed, ``dense core-like"
silicate bands, have deeper water ice bands at a given $A_{\rm K}$
than the Group C targets. These Group A targets closely follow the
Taurus ice correlation, within $3\sigma$ ($a=0.60 \pm 0.02$, $b=-0.23
\pm 0.01$; \citealt{Whittet2001}).

For Serpens (Fig. \ref{Fig. H2O_correlation}b and c), all targets have
$\tau_{3.0}$ values that fall significantly below the Taurus
correlation. This is not only reflected in the higher extinction
threshold derived above, but also in the shallower slope of the
correlation. The latter might indicate lower $\rm{H_2O}$ ice
abundances in Serpens compared to Taurus.  It might also be a
reflection of smaller grain sizes in Serpens, enhancing $A_{\rm K}$
relative to $\tau_{3.0}$ (\S\ref{subsec:models}). Here again, the
different groups, distinguished in the $\tau_{9.7}$ versus $A_{\rm K}$
correlation, are indicated. All groups follow the same correlation,
but Group A targets are higher on the correlation than Group C and D
targets, respectively. It is also worth noting that all targets, with
the exception of Ser-19 (Group C), have $\rm{H_2O}$ ice detections,
including the Group B ``diffuse ISM" target Ser-8.

\subsection{$\tau_{3.0}$ and $\tau_{9.7}$ Correlation}\label{subsec:H2Osil}

The relations between $\tau_{3.0}$ and $\tau_{9.7}$
(Fig.~\ref{Fig. H2Osil_correlation}) are linear, although they are
distincly different for Serpens and Perseus.  Least square fits yield
for Perseus

\begin{equation}
\tau_{9.7} = (0.21 \pm 0.09) \times \tau_{3.0} + (0.26 \pm 0.03)
\end{equation}

and for Serpens

\begin{equation}
\tau_{9.7} = (0.55 \pm 0.06) \times \tau_{3.0} + (0.31 \pm 0.03)
\end{equation}

For comparison, the relation in a sample of isolated dense cores
\citep{Boogert2011} is

\begin{equation}
\tau_{9.7} = (0.36 \pm 0.06) \times \tau_{3.0} + (0.36 \pm 0.09)
\end{equation}

The relation is significantly steeper in Serpens compared to Perseus.
In fact, for $\tau_{3.0}>0.15$ the Perseus relation is almost flat,
i.e., while $\tau_{3.0}$ increases with a factor of 5, $\tau_{9.7}$
increases by at most a factor of 1.5.  Also, compared to the dense
cores, the Perseus $\tau_{9.7}$ values are systematically a factor of
2 lower.  The Serpens relation is in better agreement with that of
dense cores \citep{Boogert2011}, although the slope in Serpens is
steeper.

\subsection{$\tau_{9.7}$/$A_{\rm K}$ and $\tau_{3.0}$ Correlation}\label{subsec:silakh2o}

The Group A targets are well separated from the other groups along the
$\tau_{3.0}$ axis in Fig.~\ref{Fig. H2Osil_correlation}. This is
confirmed in the relations between $\tau_{9.7}$/$A_{\rm K}$ and
$\tau_{3.0}$ (Fig.~\ref{Fig. silakh2o_correlation}), where the Group A
targets have the lowest $\tau_{9.7}$/$A_{\rm K}$ ratios and the
largest $\tau_{3.0}$ values. Such trend is also visible for Lupus. The
isolated dense cores are biased to the highest $A_{\rm K}$ and
$\tau_{3.0}$ values.  The group of points with the higher
$\tau_{9.7}$/$A_{\rm K}$ ratios are from the core L328, which has a
known diffuse dust foreground component.

\subsection{$\rm{CH_3OH}$ Ice}\label{subsec:CH3OH}

\begin{figure}
    \centering
    \includegraphics[width=0.47\textwidth,height=0.25\textheight]{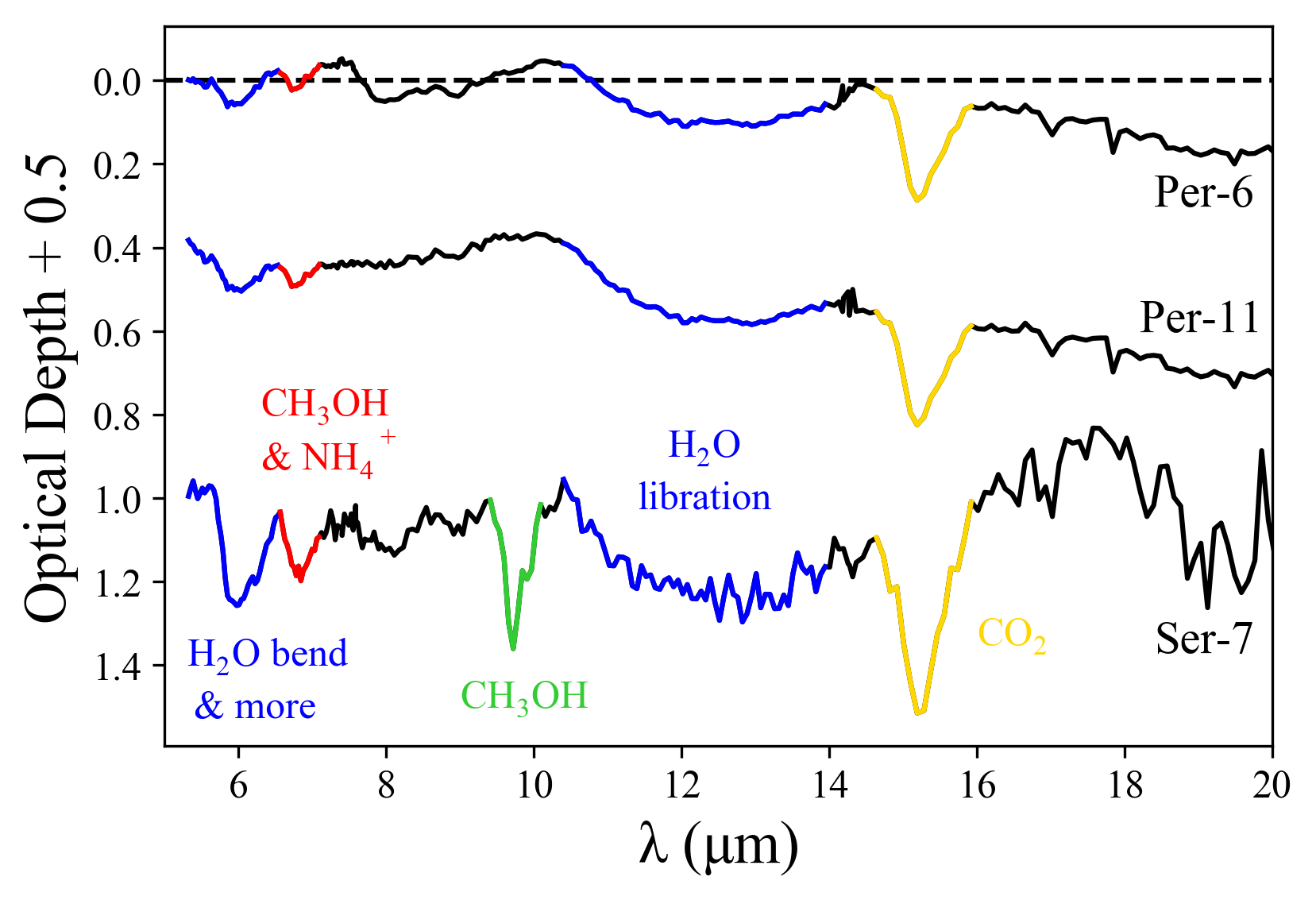}
    \caption{5-20 $\mu$m optical depth spectra of the targets with the
      strongest ice absorption features: Per-6, Per-11, and Ser-7,
      from top to bottom, respectively. A model for the 9.7 $\mu$m and
      18 $\mu$m silicate bands has been subtracted. Ice absorption
      feature identifications are indicated. The 6 $\mu$m and 6.85
      $\mu$m absorption features consist of overlapping ice bands, and
      not all components have been securely identified. The structure
      between 8-9 $\mu$m toward Per-6 is likely due to insufficiently
      corrected photospheric absorption. Offsets of 0.5 along the
      y-axis were applied for clarity.\label{Fig. ices}}
\end{figure}

\begin{figure}
    \centering
    \includegraphics[width=0.47\textwidth,height=0.24\textheight]{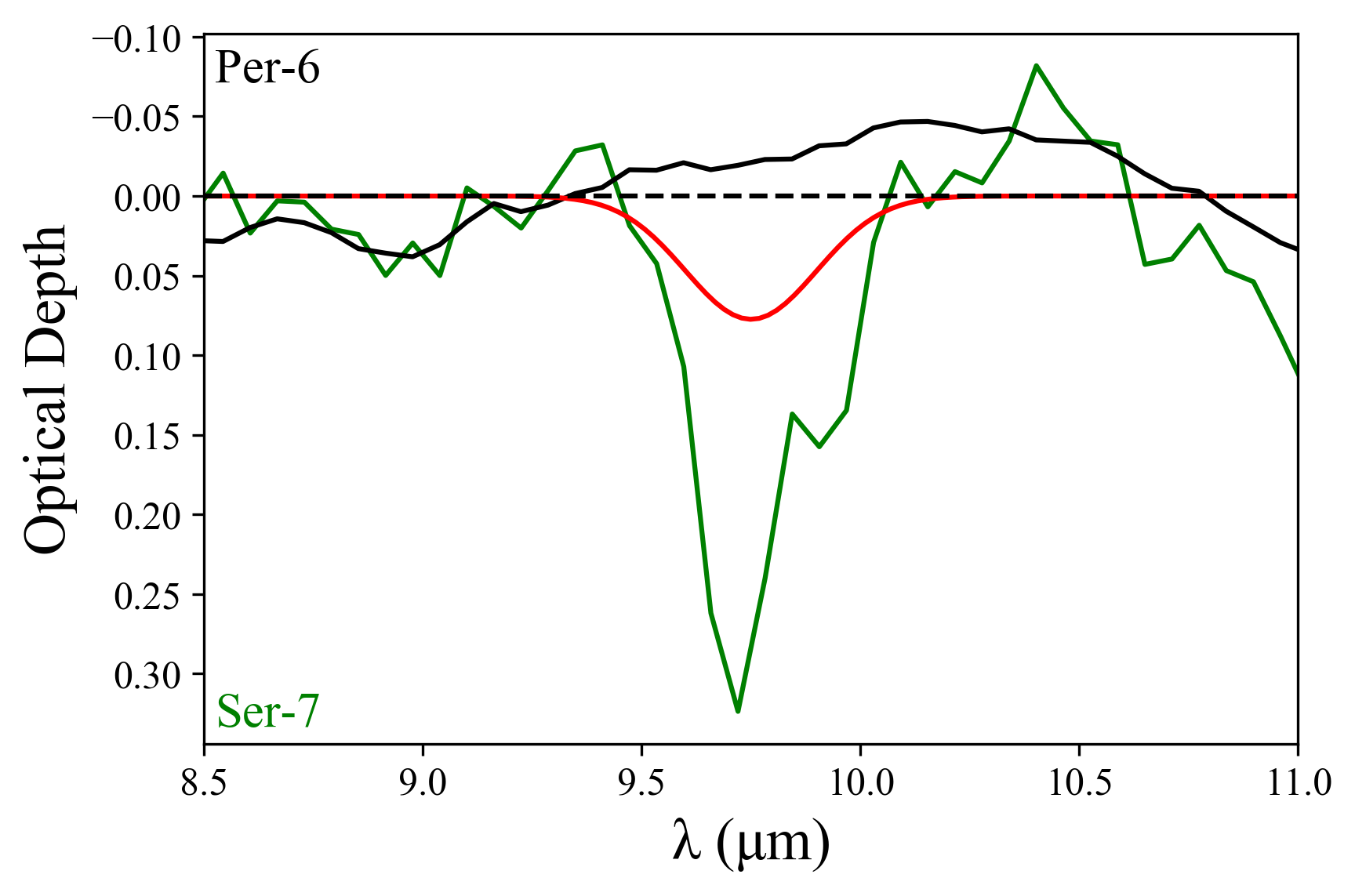}
    \includegraphics[width=0.47\textwidth,height=0.24\textheight]{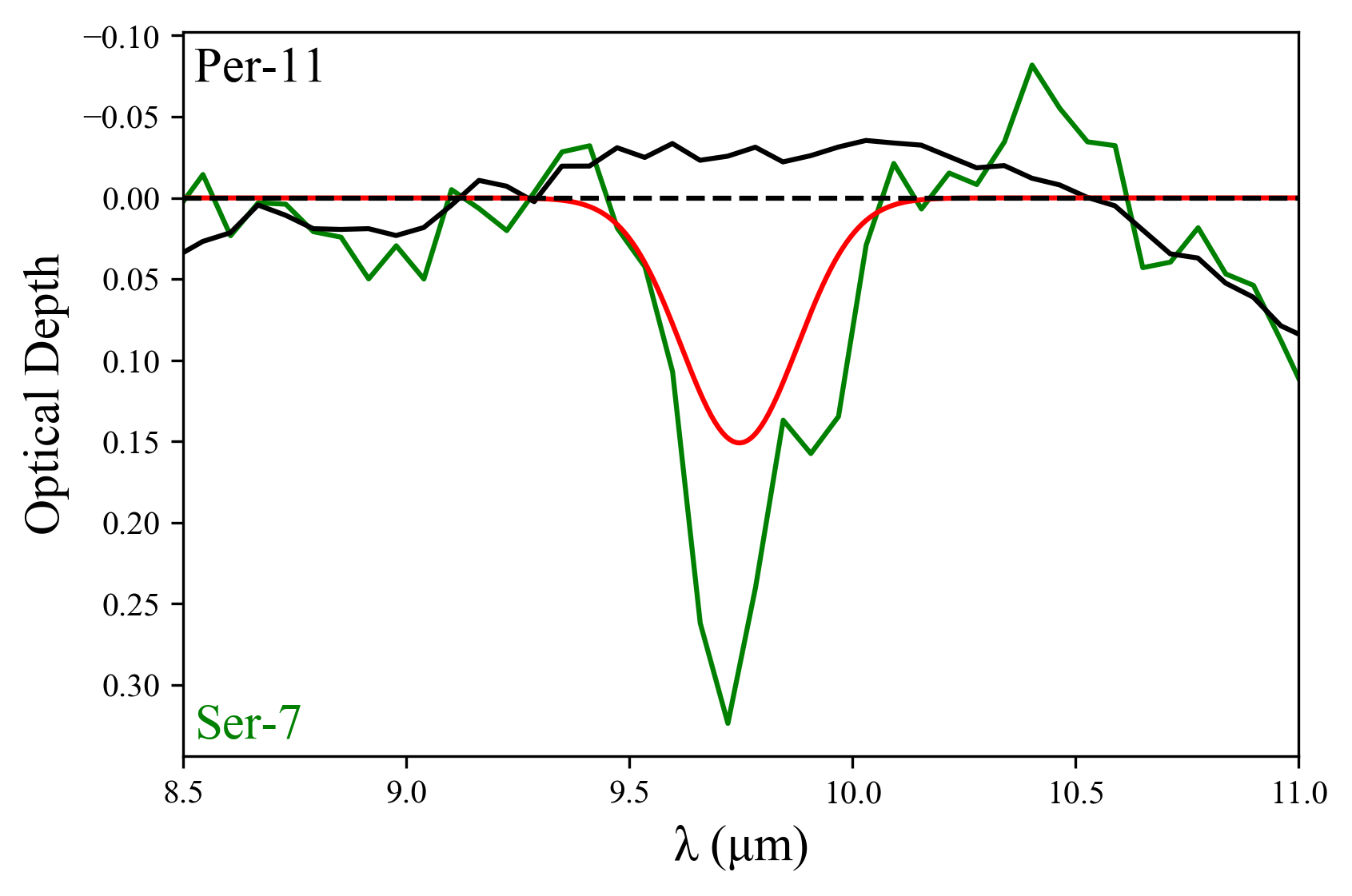}
    \caption{Comparison of the deep $\rm{CH_3OH}$ feature in Ser-7
      (green) to a selection of other targets with deep 3.0 $\mu$m
      $\rm{H_2O}$ ice bands, Per-6 {\bf (left)} and Per-11 {\bf
        (right)} in black. A model for the 9.7 $\mu$m silicate feature
      has been subtracted.  Ser-7 has an extinction of $A_{\rm K} =
      4.75 \pm 0.44$, Per-6 of $A_{\rm K} = 0.75 \pm 0.07$, and Per-11
      of $A_{\rm K} = 1.69 \pm 0.07$. Gaussians (red) demonstrate the
      expected peak depths if the CH$_3$OH abundance relative to
      $\rm{H_2O}$ were the same in these targets as in Ser-7. CH$_3$OH
      is clearly absent, and upper limits are listed in Table
      \ref{Table CH3OH}.\label{Fig. CH3OH}}
\end{figure}

A number of targets show ice absorption features in the 5-20 $\mu$m
wavelength range, most commonly the 6 $\mu$m O-H bending mode of
$\rm{H_2O}$. Three of them, two behind Perseus and one behind Serpens,
show particularly deep ice absorption features
(Fig. \ref{Fig. ices}). The 9.7 $\mu$m CO stretch mode arising from
$\rm{CH_3OH}$ is only detected in Ser-7. Despite the good S/N of the
spectra for the other two ice targets, Per-6 and Per-11, there is no
significant detection of $\rm{CH_3OH}$ (Fig. \ref{Fig. CH3OH}).  This
also shows that the 6.85 $\mu$m absorption feature, which is detected
in all three targets, is for at most a small fraction due to the C-H
deformation mode of CH$_3$OH. The most promising candidate is solid
NH$_4^+$ \citep{Schutte2003, Raunier2003, Gibb2004, Boogert2008}. For
other potential contributors, we refer to \citet{Keane2001}.

The $\rm{CH_3OH}$ ice column density is calculated using an integrated
band strength of $A = 1.6 \times 10^{-17}$ cm $\cdot$ molecule$^{-1}$
\citep{Kerkhof1999}. This $A$-value is an average across different
mixtures with a variation of $\sim$20\%. The abundance relative to
$\rm{H_2O}$ is 21$\pm $2\% for Ser-7 (Table \ref{Table CH3OH}). The
uncertainty reflects baseline fluctuations, and does not take into
account the uncertainty in the $A$-value.  Ser-7 has the highest
$\rm{CH_3OH}$ abundance measured toward a background star of a dense
cloud or core to date \citep{Boogert2011, Chu2020}. It is also
significantly higher (factors of 2-6) than the upper limits derived
for the other two ice targets (Table \ref{Table CH3OH}).

\begin{deluxetable}{ccccc}[b]
\centering
\caption{CH$_3$OH Ice Abundances\label{Table CH3OH}}
\tablehead{
  \colhead{Alias} & \colhead{$A_{\rm K}$} & \colhead{N($\rm{H_2O}$)}     & \colhead{N($\rm{CH_3OH}$)}  & \colhead{$\frac{N(CH_3OH)}{N(H_2O)}$}\\
  \colhead{}      & \colhead{mag}        & \colhead{$(10^{17} cm^{-2})$} & \colhead{$(10^{17} cm^{-2})$} & \colhead{}                           \\
}
\startdata
Ser-7  & 4.75$\pm$0.44 & 26.33$\pm$0.97 & 5.4$\pm$0.5 & 0.21$\pm$0.02 \\
Per-6  & 0.75$\pm$0.07 & 6.29$\pm$1.36  & $<$0.64       & $<$0.10 \\
Per-11 & 1.69$\pm$0.07 & 13.66$\pm$0.82 & $<$0.39       & $<$0.03 \\
\enddata

\tablecomments{The three targets with deep ice bands, as in
  Fig. \ref{Fig. ices}, including their aliases, extinctions, water
  ice column densities, methanol column densities, and their methanol
  to water ice ratios.}

\end{deluxetable}

\subsection{Spatial Context}\label{subsec:maps}

In order to put the variations in the dust and ice properties observed
toward the Perseus and Serpens background stars in a spatial context,
we compare them to infrared images, extinction maps, and the YSO
population.

\begin{figure*}
\begin{center}
    \includegraphics[width=1\textwidth,height=0.42\textheight]{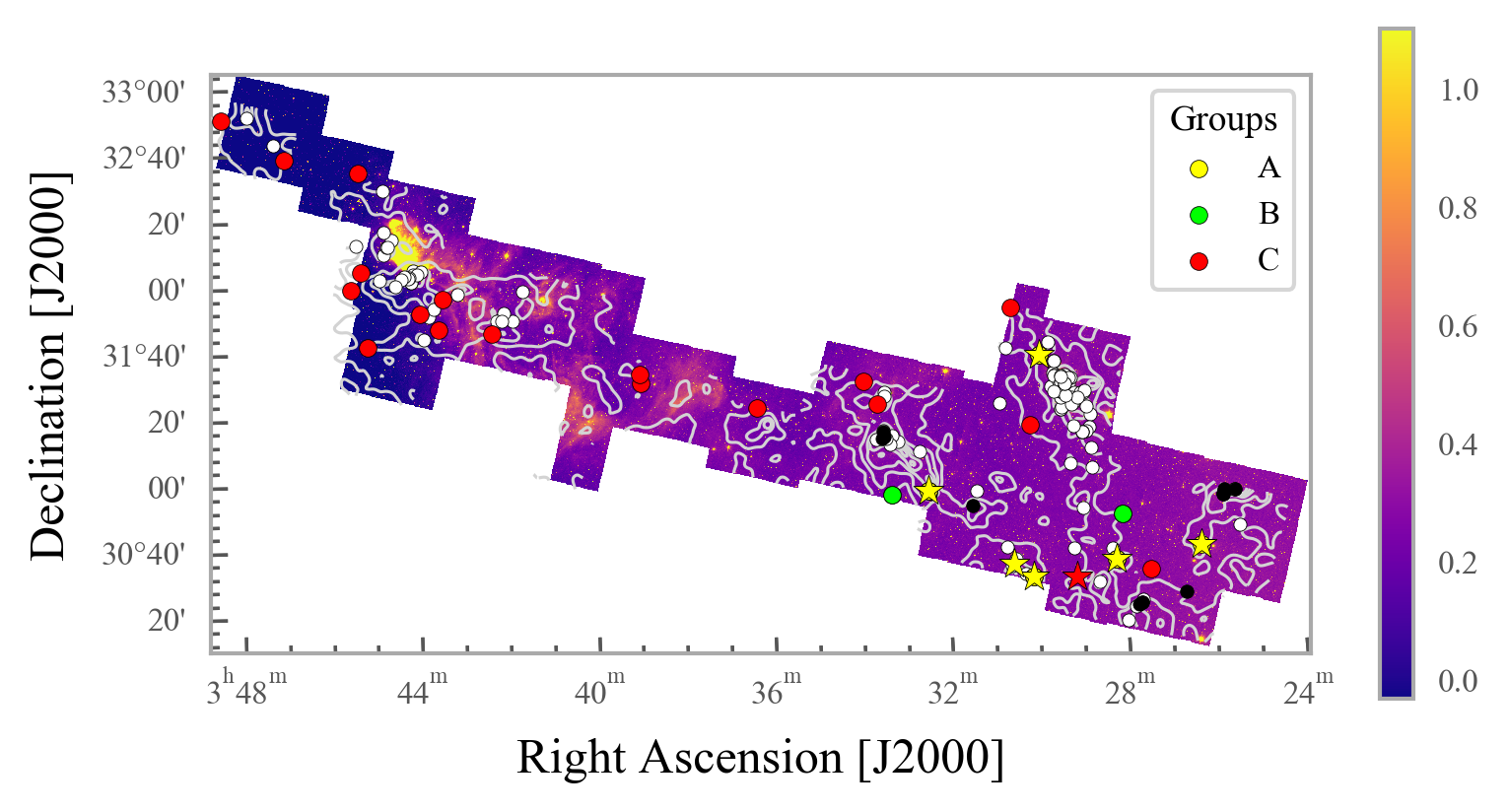}\hfill
    \includegraphics[width=0.46\textwidth,height=0.20\textheight]{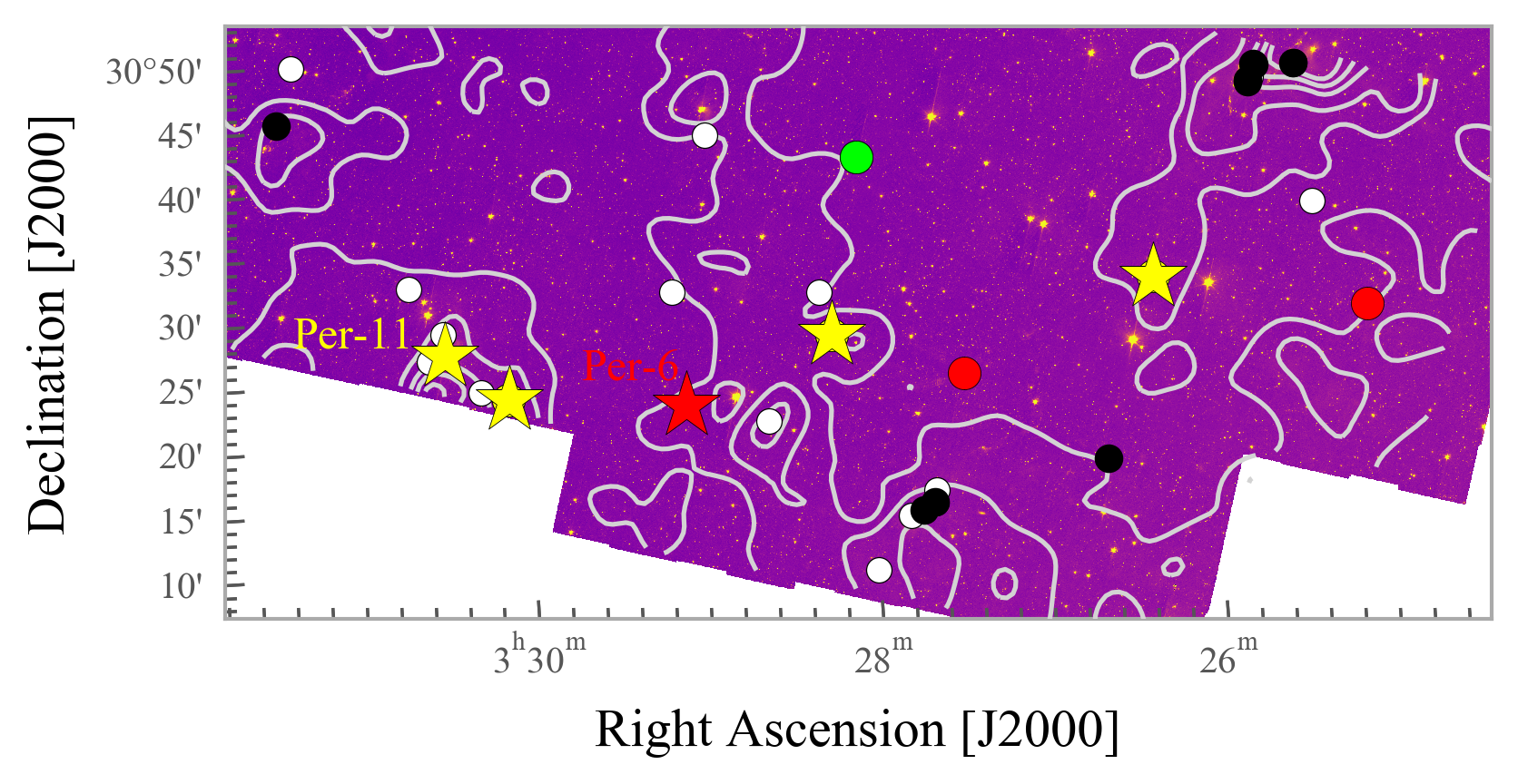}\hfill
    \includegraphics[width=0.53\textwidth,height=0.205\textheight]{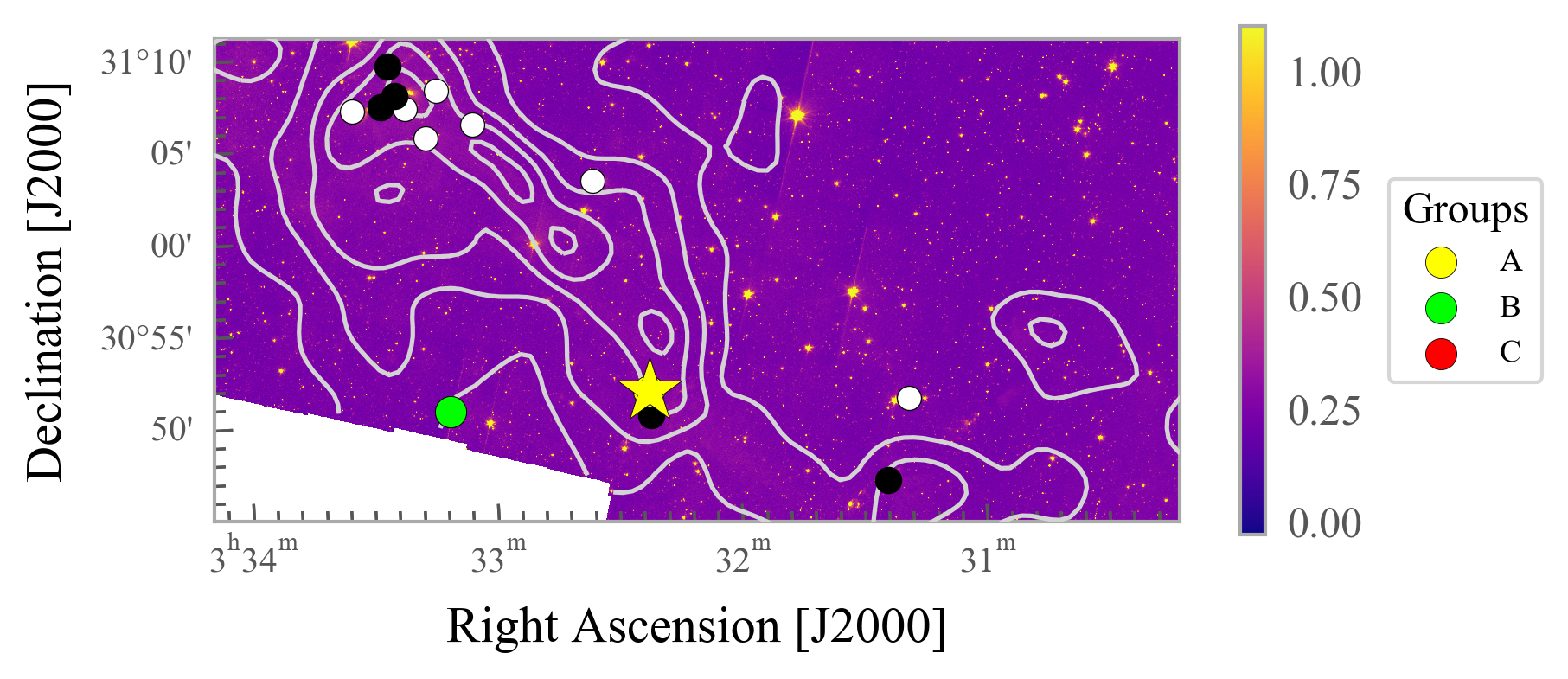}
    \caption{{\bf Top panel}: Location of the background stars
      (yellow, green, and red circles and {\bf star symbols}, as
      defined in Fig. \ref{Fig. Si_correlation}) compared to Class 0
      (black) and Class I and ``flat spectrum" (white) YSOs observed
      toward Perseus overlaid the IRAC1 map (3.6 $\mu$m). The colorbar
      is in units of MJy/sr. The contours represent the extinction map
      of $A_{\rm V}$ levels of 3, 6, 12, 18, and 24 mag.  The star
      symbols represent the targets that exhibit strong ice features,
      at $\tau_{3.0}>0.5$ within the uncertainties. {\bf Bottom
        panels}: Zoomed in regions of top panel, The red star symbol
      in the left panel represents Per-6, and the yellow star symbol
      to the far left Per-11, both of which have particularly deep ice
      absorption bands.\label{Fig. per_map}}
\end{center}
\end{figure*}

\begin{figure}
\begin{center}
  \includegraphics[width=0.56\textwidth,height=0.51\textheight]{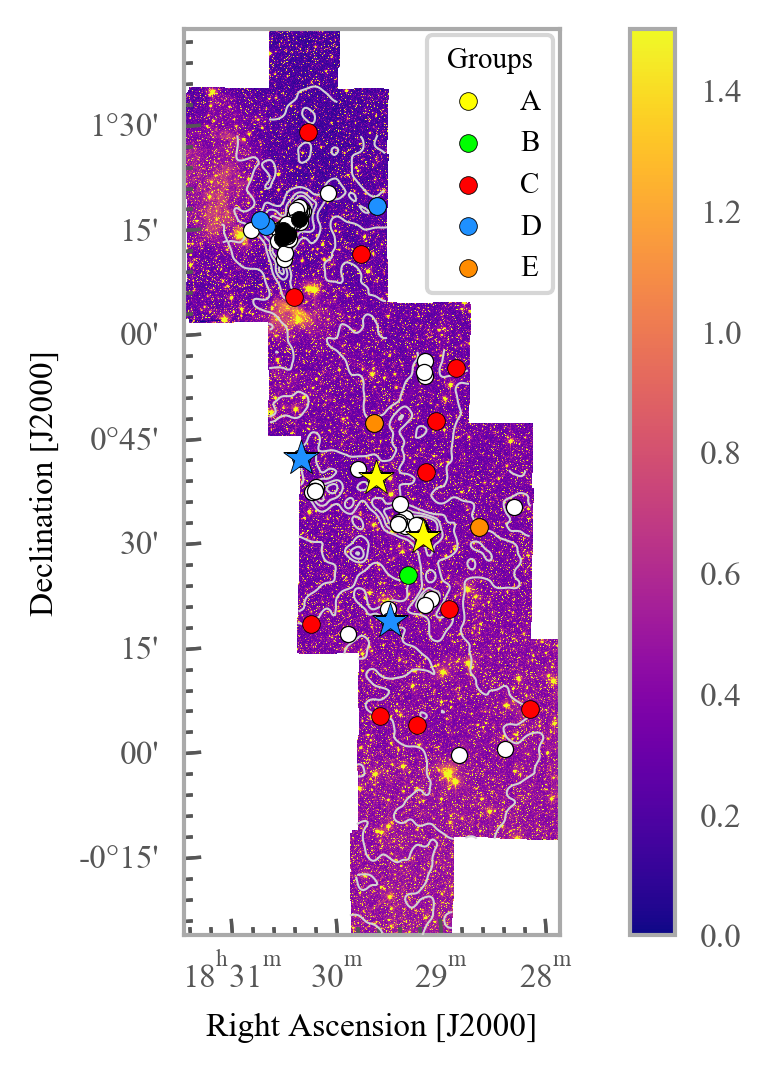}
    \includegraphics[width=0.45\textwidth,height=0.29\textheight]{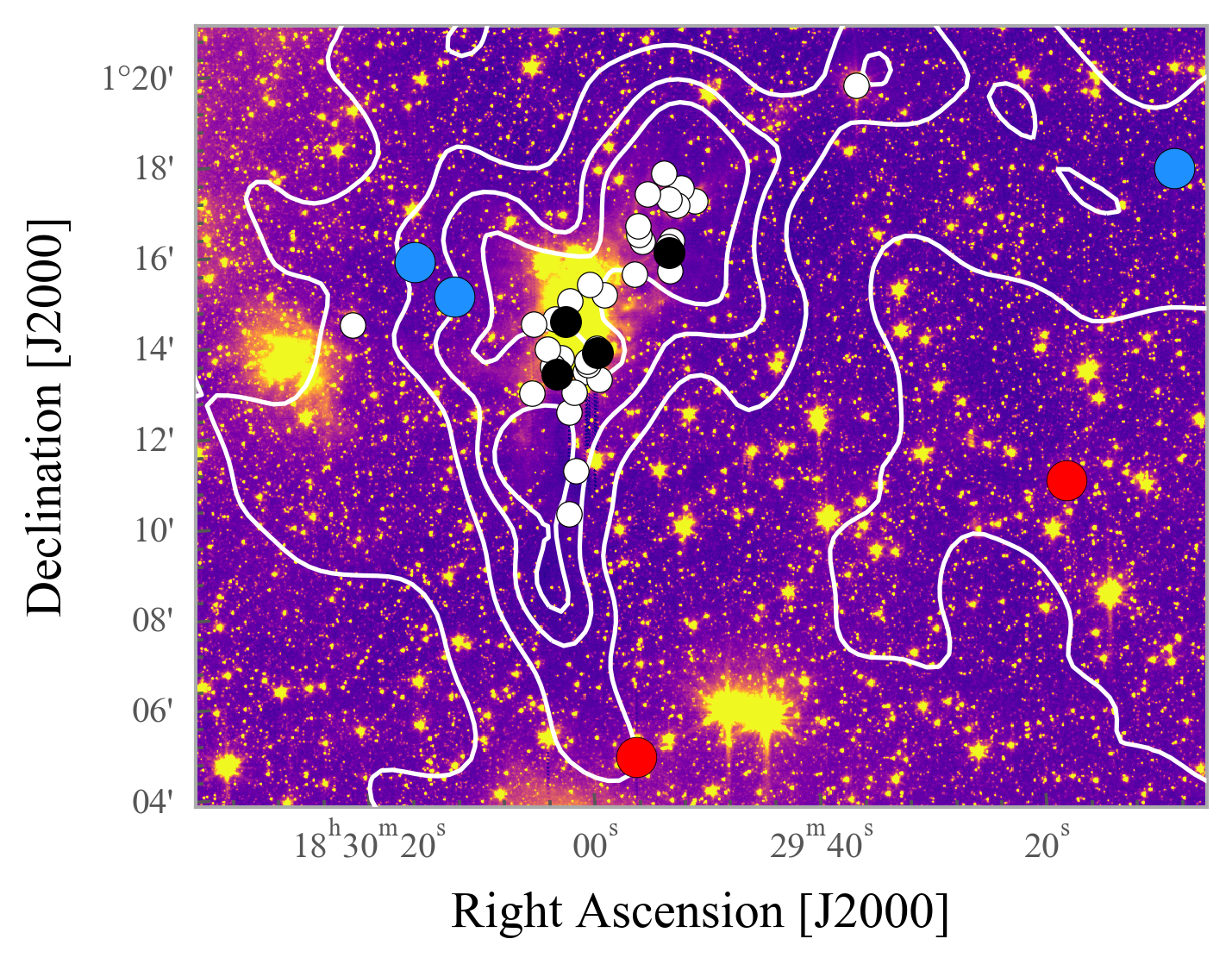}\hfill
    \includegraphics[width=0.52\textwidth,height=0.295\textheight]{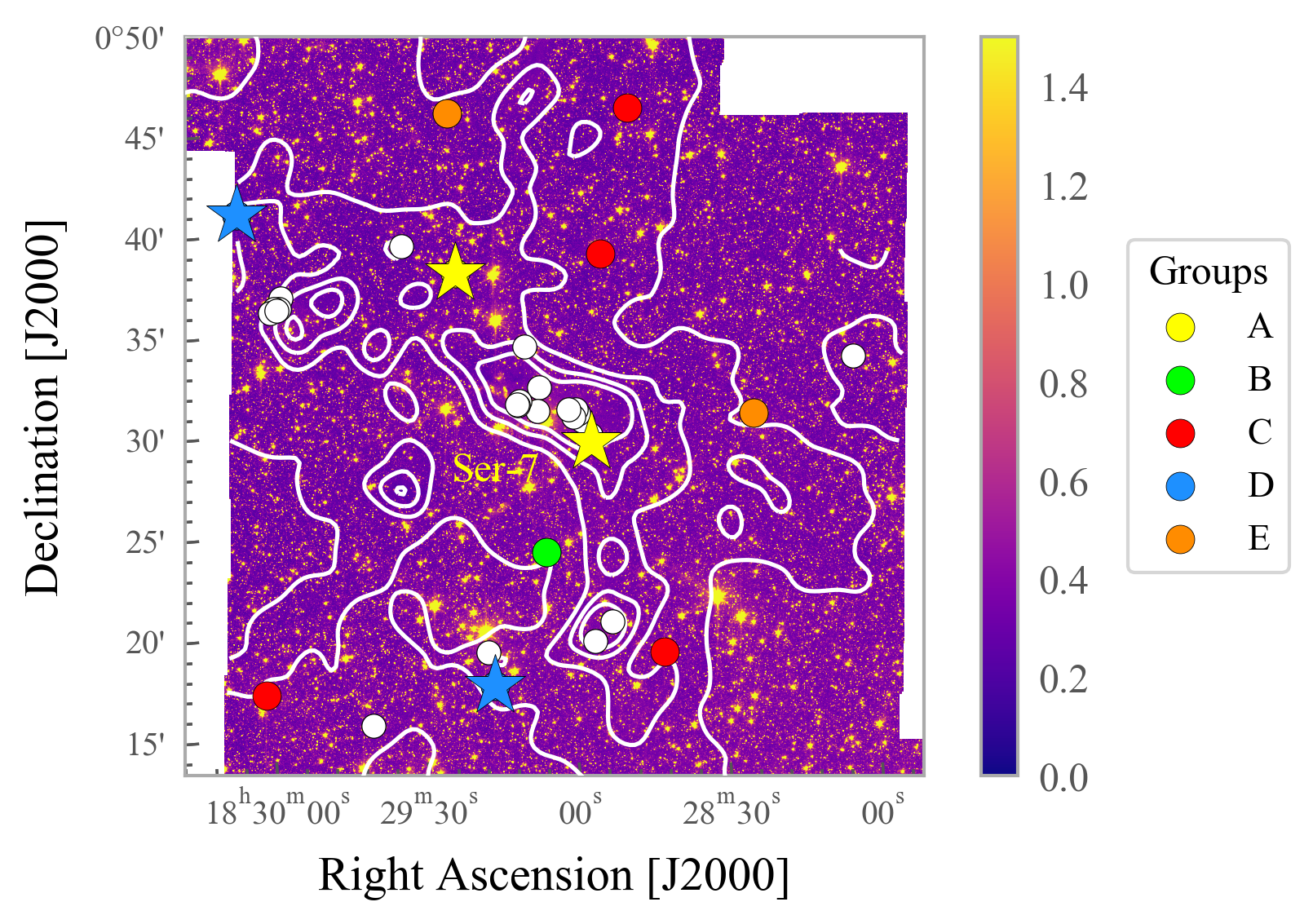}  
    \caption{{\bf Top panel}: Location of the background stars
      (colored bullets and {\bf star symbols}, as defined in
      Fig. \ref{Fig. Si_correlation}) compared to Class 0 (black) and
      Class I and ``flat spectrum" (white) YSOs observed toward
      Serpens overlaid the IRAC1 map (3.6 $\mu$m). The colorbar is in
      units of MJy/sr. The contours represent the extinction map of
      $A_{\rm V}$ levels of 3, 6, 12, 18, and 24 mag.  The star
      symbols represent the targets that exhibit strong ice features,
      at $\tau_{3.0}>0.5$ within the uncertainties {\bf Bottom
        panels}: Zoomed in regions for known clusters in Serpens, the
      ``core" (left) and ``Ser/G3-G6" (right).  The yellow asterisk
      near the center represents Ser-7, which exhibits particularly
      strong ice features.\label{Fig. ser_map}}
\end{center}
\end{figure}

We use \spitzer extinction maps and infrared images that were produced
by the c2d Legacy project \citep{Evans2003, Evans2009}. The Perseus
extinction map, derived from near-infrared colors, has a resolution of
180" and has a minimum of $A_{\rm V} = 2$ mag. For Serpens, the
resolution is 90" and starts at the $A_{\rm V} = 5$ mag cloud
boundary. For the infrared images, we chose the IRAC1 filter at 3.6
$\mu$m, as this is more sensitive to dust scattering than the longer
wavelengths. To select the protostellar population, we used the
\spitzer c2d catalogs \citep{Evans2009} available through the InfraRed
Sky Archive (IRSA) interface. All Perseus and Serpens targets with
infrared spectral indices $\alpha > -0.3$ were selected, where
$\alpha$ follows the definition of \citet{Lada1987}. These represent
the most embedded population of stars. We then designated targets with
$-0.3 \leq \alpha < 0.3$ as ``flat spectrum" YSOs, and targets with
$\alpha \geq 0.3$ as Class I targets. We also added the more embedded
Class 0 sources, that are harder to identify using \spitzer photometry
alone due to their weakness. Thirteen Class 0 targets in Perseus were
obtained from \citet{Jorgensen2006}, and four in Serpens from
\citet{Hogerheijde1999}. The YSO populations and the background star
groups defined by the $\tau_{9.7}$ versus $A_{\rm K}$ correlation
(\S\ref{subsec:Si}) are indicated on the IRAC images and extinction
maps in Figures \ref{Fig. per_map} and \ref{Fig. ser_map} for Perseus
and Serpens, respectively.

For Perseus, the majority of the Group A (``dense" cloud-type) targets
are located in the Western portion of the cloud
(Fig. \ref{Fig. per_map}). This is also where the five Class 0 YSOs
are all located. Class 0 envelopes extend to typically 1 arcmin in
Perseus \citep{Enoch2006}. Group A target Per-9 is located 143"
(35,631 AU) from the Class I YSO IRAS 03271+3013. Another Group A
target, Per-12, is located 67" (16,819 AU) from the Class 0 YSO IRAS
03292+3039. The smallest distance is 53" between the background star
Per-11 and the Class 0 target [DCE2008] 081, and, therefore, it
appears that none of the background stars significantly trace Class 0
envelopes.

It is noteworthy that the two Group B (``diffuse ISM-like") targets
are located in the Western part of Perseus as well. They are located
at the edges of extinction enhancements, and somewhat further away
from YSOs compared to several Group A targets.  The number of targets
is small, however, and these trends are not statistically significant.
The median distance between the six Group A targets and nearest YSOs
is 176" with a standard deviation of 240"; the distances of the two
Group B targets and their nearest YSOs are 657" and 671".

The Perseus targets with the deepest ice bands, Per-6 (Group C) and
Per-11 (Group A), are also located in the Western region of
Perseus. Per-11 is located among a small cluster of Class I and ``flat
spectrum" YSOs, but Per-6 is also located near such YSOs. Overall,
these results indicate a significant variety of dust properties within
Perseus. Dense-cloud like dust is found within the vicinity
($\sim$15,000 AU) of YSOs but also in regions without YSOs. Diffuse
ISM-like dust may have strong ice absorption, and is also found in the
denser regions.

In Serpens, most of the Class I and ``flat spectrum" YSOs are
concentrated in two clusters, one located in the upper half of
Serpens, and the other, ``Ser/G3-G6," located around the middle of the
molecular cloud \citep{Zhang1988}. The four Class 0 YSOs are toward
the center of the upper cluster (Fig. \ref{Fig. ser_map}). The Group A
target Ser-7 is located at the edge of Ser/G3-G6, but within an
arcminute of the YSO cluster. Ser-7 has very deep ice bands, and a
very large $\rm{CH_3OH}$ ice abundance. This likely reflects a high
density associated with the star formation in this cluster. The single
Group B target is located in a local extinction minimum. Group D
targets (intermediate between ``diffuse ISM-like" and ``dense-like"
dust) are spread out along the molecular cloud, in some cases located
near the edges of local cores or near YSOs, but certainly not in all
cases.

Overall, as for Perseus, no clear correlations stand out for Serpens,
and the dust properties are governed by local physical conditions not
evident in these tracers. The distances between the two Group A
targets and their nearest YSOs are 180" (standard deviation 23"); for
the one Group B target, there is a distance of 285"; and for the six
Group D targets, there is a median distance of 118" with a standard
deviation of 105".

\section{Discussion}\label{sec:Discussion}

\subsection{The $A_{\rm K}$/$\tau _{9.7}$  Ratio in Grain Growth Models}\label{subsec:models}

Grain growth, starting when most grains are much smaller than the
near-infrared wavelength, i.e., in the Rayleigh limit, will increase
$A_{\rm K}$ initially. But when the grain sizes become comparable to
or larger than $\sim0.4$ $\mu $m ($\lambda/2\pi$), gray extinction
becomes important, and $A_{\rm K}$ (per dust mass), derived from the
near-IR color excess, will decrease. This implies that the observed
$A_{\rm K}$/$\tau _{9.7}$ reduction in dense clouds
(Fig.~\ref{Fig. Si_correlation}) could trace a moderate amount of
grain growth. This is also concluded by \citet{Ormel2011} based on
grain growth models. These models further infer that the grain growth
is in the form of aggregates of ice coated silicate and ice coated
carbonaceous grains.  The increased stickiness of ice coated grains is
key in the coagulation process.  The observed relation between $A_{\rm
  K}$/$\tau _{9.7}$ and the ices ($\tau_{3.0}$) will be discussed in
\S\ref{subsec:variations}.

\begin{figure}
    \centering
    \includegraphics[width=0.47\textwidth,height=0.25\textheight]{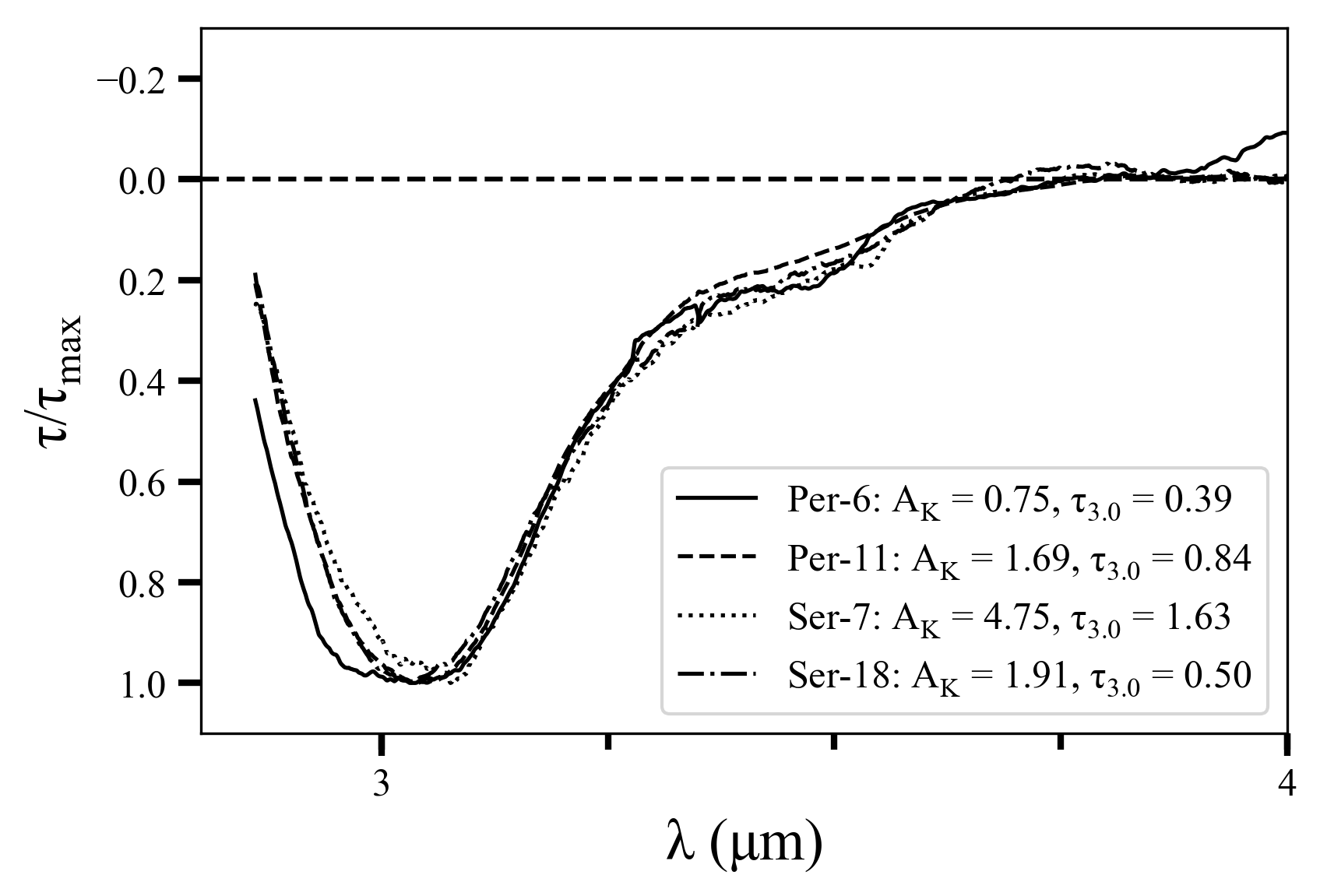}
    \caption{Profile of the 3 $\mu$m water ice bands observed toward
      four targets tracing a range of extinctions ($A_K =
      0.75-4.75$). Per-6 is Group C, Per-11 is Group A, Ser-7 is Group
      A, and Ser-18 is Group D. The spectra have been normalized and
      smoothed to a resolving power of 100. Little variation is
      evident in the long-wavelength wing, indicating insignificant
      growth of the relative population of grains larger than
      $\sim$0.5.\label{Fig. coagulation}}
\end{figure}

Such moderate grain growth is consistent with the invariant profiles
of the 9.7 $\mu$m absorption bands. This was also the primary
conclusion of the extensive study by \citet{vanBreemen2011}.  The
coagulated grains are too small to affect the profile of the 3.0
$\mu$m ice band as well, as little variation is observed as a function
of extinction (Fig. \ref{Fig. coagulation}). The long-wavelength wing,
which is thought to be affected by large grain scattering, limits the
grains to sizes of $\frac{3}{2\pi} \approx$ 0.5 $\mu$m and less. Some
variation is observed on the short wavelength wing, which might be a
result of NH$_3$ abundance variations due to the N-H stretching mode
at 2.9 $\mu$m. Indeed, the target with the strongest short-wavelength
wing also shows a more pronounced 3.47 $\mu$m feature, which results
from ammonia hydrates \citep{Dartois2001}.

A different scenario was discussed in \citet{Chiar2007} and
\citet{Ormel2011}. Considering that the $K-$band extinction is
primarily caused by carbonaceous grains, the reduction of the $\tau
_{9.7}$/$A_{\rm K}$ ratio (Fig.~\ref{Fig. Si_correlation}) is possible
if growth is limited to silicate grains in the inner cloud
regions. These grains would need to be much larger than 1 $\mu$m so
that they do not contribute to the $\tau_{9.7}$ absorption feature
(and also not to $A_{\rm K}$), and not change its profile. In the
models of \citet{Ormel2011}, the silicate absorption feature
disappears for time scales longer than 1 Myr at a density of 10$^5$
cm$^{-3}$, or at shorter time scales at higher densities. Thus, in
this scenario, large ice-rich silicate grains reside in the inner,
dense regions of the clouds, while the small silicate grains in the
outer regions are responsible for the 9.7 $\mu$m absorption feature.
Carbonaceous grains would not form such large grains, perhaps due to
not acquiring ice mantles.  It is unclear why this would be the case.
In fact, the increase of $\tau_{3.0}$ as a function of $A_{\rm K}$, as
observed in both Perseus and Serpens
(Fig.~\ref{Fig. H2O_correlation}), is difficult to explain with
ice-less carbonaceous grains. If only silicate grains, tracing the
outer cloud layers, are covered with ice mantles, the $\tau_{3.0}$ as
a function of $A_{\rm K}$ would flatten, which is clearly not the
case. Thus, overall, this scenario seems less likely.

\subsection{$A_{\rm K}$/$\tau _{9.7}$ Variations and Relations with Ices and Dense Core Formation}\label{subsec:variations}

Striking variations in the relation between $A_{\rm K}$ and $\tau
_{9.7}$ are observed (Fig. \ref{Fig. Si_correlation}). At the highest
extinctions, the targets are trending to lie systematically below the
linear fit, i.e., $\tau_{9.7}$ is suppressed relative to $A_{\rm
  K}$. For Perseus, this inflection occurs at $A_{\rm K} \sim 1.2$
($A_{\rm V} \sim 10$, assuming $A_{\rm V}$/$A_{\rm K}$ = 8.4), while
for Serpens, it is near $A_{\rm K} \sim 2$ ($A_{\rm V} \sim
17$). Below this inflection point, the data points are located between
the diffuse and dense cloud relations. Above this inflection point,
the data points tend to follow the dense cloud relation. The same is
observed for the third cloud in our survey, Lupus, which shows an
inflection point similar to Perseus \citep{Boogert2013}.

\begin{figure}
    \centering
    \includegraphics[width=0.47\textwidth]{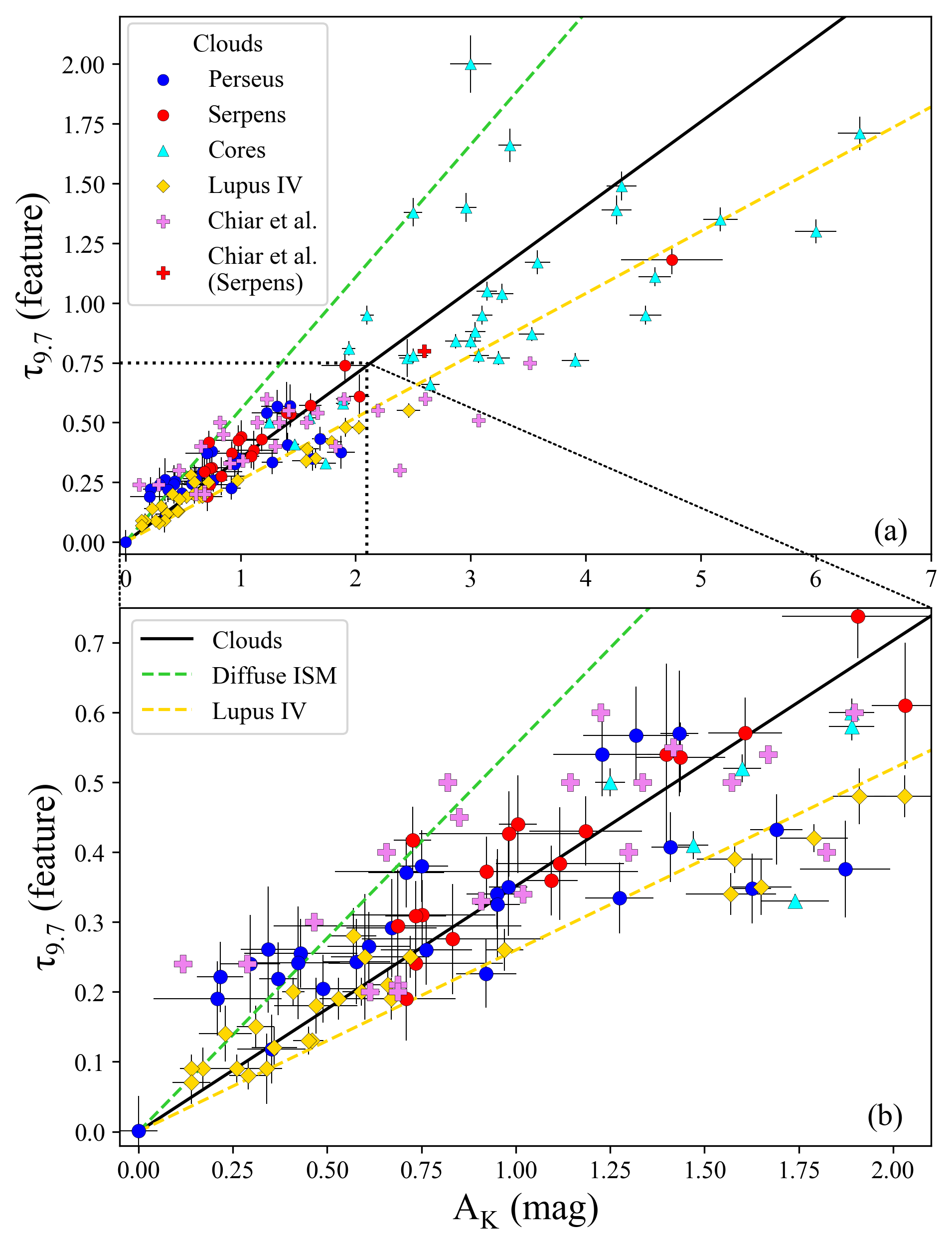}\vspace{-5pt}    
    \caption{Silicate ice and extinction correlation for the Perseus,
      Serpens, Lupus, Taurus, and Chameleon I molecular clouds, as
      well as several dense cores, taken from \citet{Chiar2007}, and
      \citet{Boogert2011, Boogert2013}.  The bottom panel zooms in on
      the lower extinctions. The data points are color-coordinated by
      cloud, as indicated. The black line is the linear fit to all
      targets. The green dotted line is the diffuse ISM correlation
      \citep{Whittet2003}, and the yellow dotted line is the Lupus IV
      correlation \citep{Boogert2011}.  The error bars, when
      available, are 3$\sigma$. Four Lupus IV sources have upper
      limits that are consistent with the general correlation but are
      not included in this graph.  The isolated dense core targets
      near the diffuse ISM relation trace L328, which is thought to be
      strongly contaminated by foreground dust
      absorption.}\label{Fig. summary}
\end{figure}

In Fig. \ref{Fig. summary}, we accumulate all $\tau_{9.7}$ and $A_{\rm
  K}$ data of other clouds and cores known to us \citep{Chiar2007,
  Boogert2011, Boogert2013}. In addition to Perseus, Serpens, and
Lupus (I and IV), this includes targets tracing Taurus, IC 5146,
Barnard 68, Chameleon I, and a range of isolated dense cores. The
isolated dense cores extend the $A_{\rm K}$ range to much higher
values. An overall correlation is visible, but the scatter is
large. The fitted line to all the data lies between the diffuse ISM
and dense core correlations, just as for our individual cloud fits.
An overall inflection point, after which the targets center around the
dense cloud correlation, is also visible, near $A_{\rm K} \sim 1.5$
($A_{\rm V} \sim 13$).  Thus, overall, assuming that the
$\tau_{9.7}/A_{\rm K}$ ratio is a measure of grain sizes
({\S\ref{subsec:models}}), Fig. \ref{Fig. summary} shows that the dust
coagulation process follows a smilar pattern across different clouds.

It does appear that these $\tau _{9.7}$/$A_{\rm K}$ variations relate
to the ice band optical depths. For Perseus, the $\tau _{3.0}$ values
for Group A targets (lowest $\tau _{9.7}$/$A_{\rm K}$) are
significantly higher compared to all other targets, even those at
similar $A_{\rm K}$ (Fig. \ref{Fig. H2O_correlation}). Indeed, the
Group A targets are well separated in the $\tau _{9.7}$/$A_{\rm K}$
versus $\tau _{3.0}$ correlation plots
(Fig.~\ref{Fig. silakh2o_correlation}). The Lupus cloud shows a
similar behavior. For Serpens, targets with the largest $\tau
_{9.7}$/$A_{\rm K}$ ratios, have the largest $\tau _{3.0}$ values,
although very few lines of sight with deep 3.0 $\mu$m ice bands are
available.

The threshold for ice formation is similarly low for all clouds
($A_{\rm V}\sim 2.6-3.4$; \S\ref{subsec:ices}), and thus it appears
that not only the mere presence of ice on the grains, but also the ice
column density (traced by $\tau _{3.0}$) is an important factor in the
grain growth process. This, in turn, relates to the cloud density, as
lines of sight with deeper ice bands, likely trace higher densities
deeper into the cloud.
  
Indeed, for both scenarios discussed in \S\ref{subsec:models}, the
$\tau _{9.7}$ versus $A_{\rm K}$ relation is a measure of the density
structure of the clouds. The inflection point in this relation
reflects a transition to higher density inner cloud regions where
grain growth accelerates. This implies that the density structure for
Serpens is shallower than for Perseus, Lupus, and other dense cores,
although the density at the cloud edge is similar for all these
clouds, as evidenced by their similar ice formation thresholds
(\S\ref{subsec:ices}). Within Perseus, however, coagulation is
strongest for the targets with the largest ice column densities.

\begin{figure}
  \centering
  %\subfloat{\includegraphics[width=0.44\textwidth,height=0.23\textheight]{compare_2MASS03420993+3144139.png}}
  \includegraphics[width=0.44\textwidth,height=0.23\textheight]{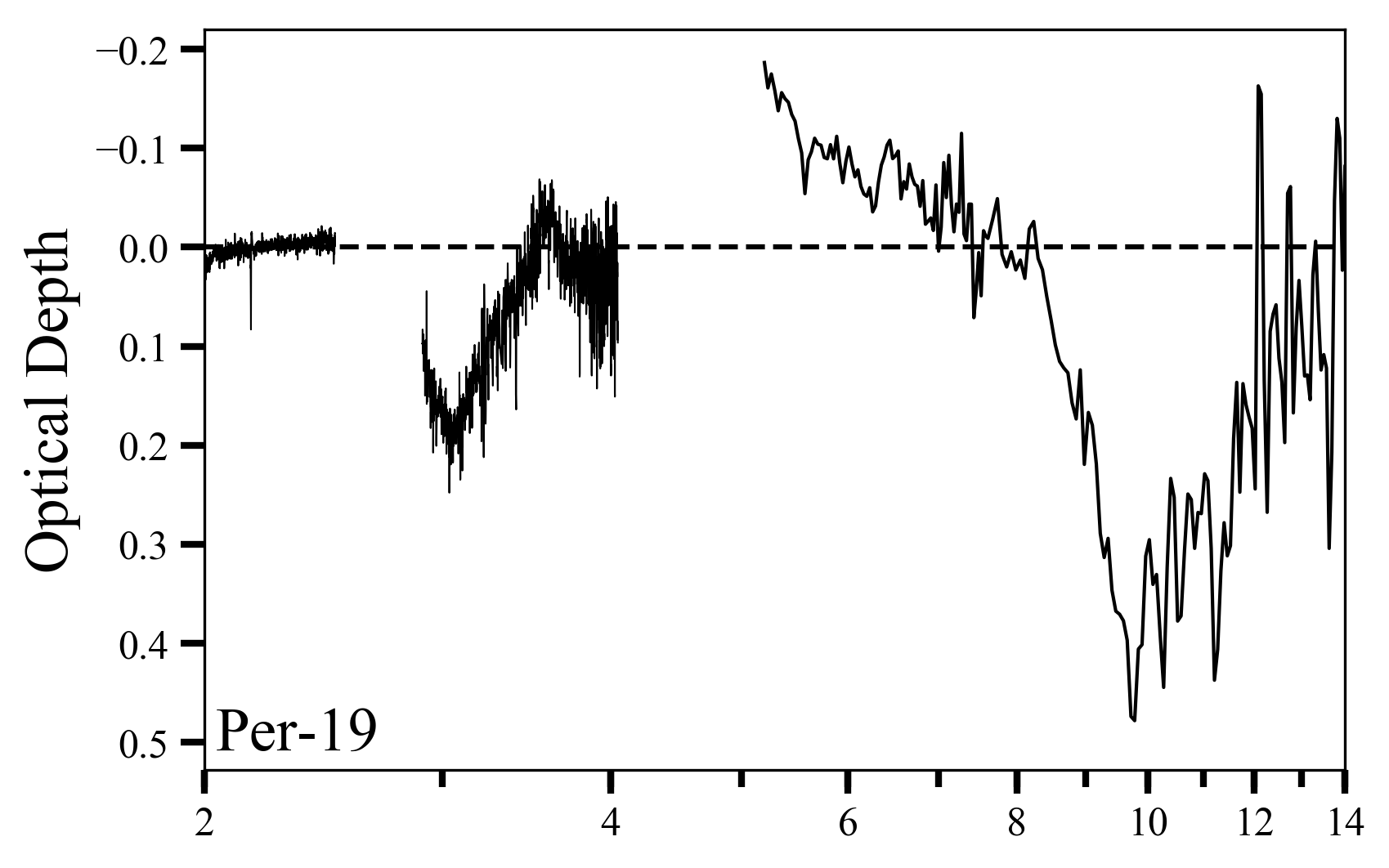}    

  %\subfloat{\includegraphics[width=0.44\textwidth,height=0.25\textheight]{compare_2MASS03261355+3029223.png}}\hfill
  \includegraphics[width=0.44\textwidth,height=0.25\textheight]{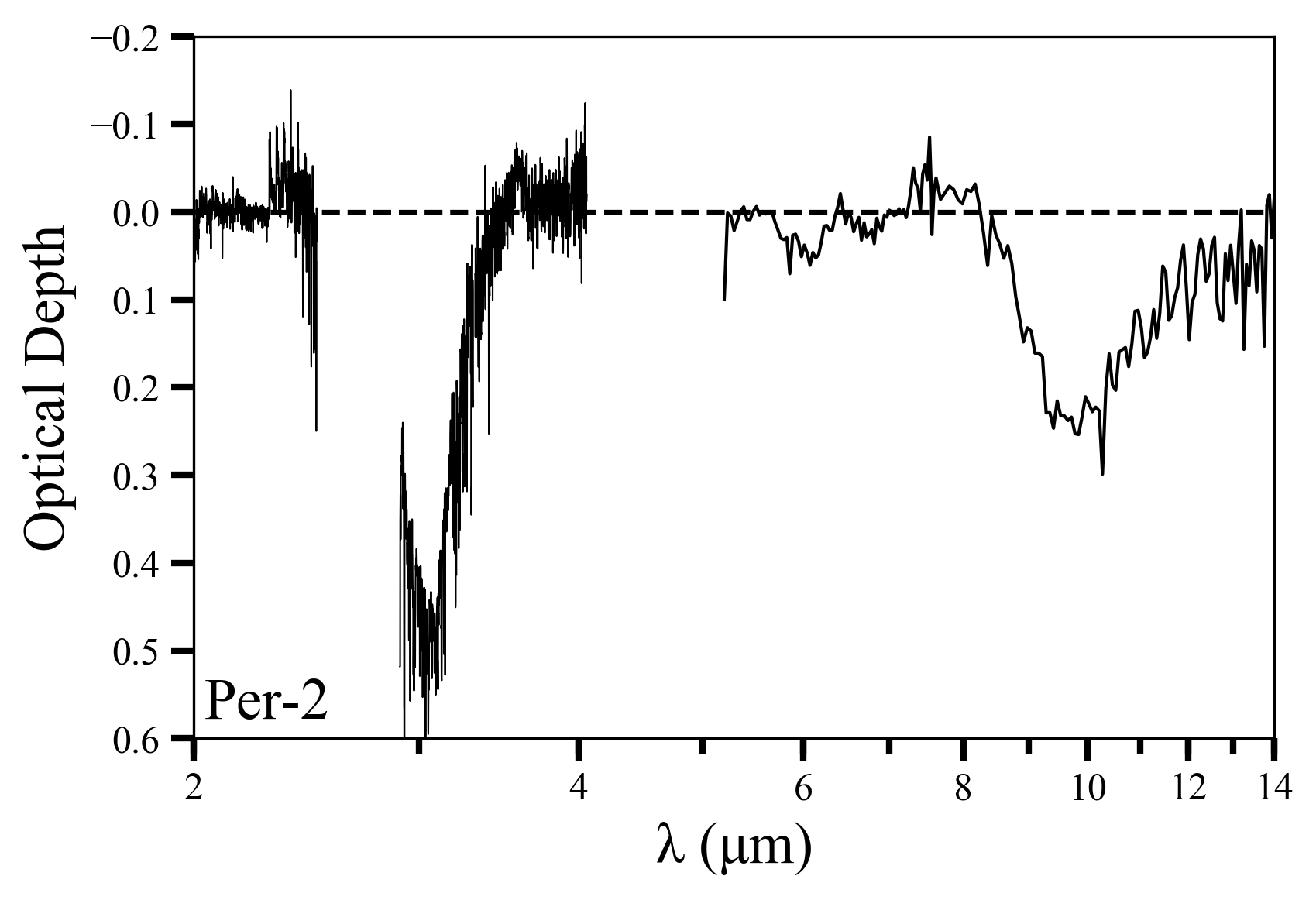}\hfill
  
    \caption{Comparison of the 3.0 and 9.7 $\mu$m ice and silicate
      bands of the targets Per-19 and Per-2. Both targets have similar
      $A_{\rm K}$ values, but Per-19 has a much deeper 9.7 $\mu$m
      silicate band (and thus belongs to ``diffuse ISM" Group
      B). Conforming with its diffuse ISM nature, Per-19 has a factor
      of $\sim$2.5 less water ice than Per-2, as seen by the shallower
      3.0 $\mu$m $\rm{H_2O}$ feature.  Such variations might point to
      a relation between ice formation and grain coagulation. The
      negative optical depth values between 5-8 $\mu$m for Per-19 are
      due to the choise of a local baseline for the 9.7 $\mu$m
      absorption feature, as noted in Table~\ref{Table
        3}.\label{Fig. per19_per2}}
\end{figure}

\begin{figure}[t]
    \centering
    \includegraphics[width=0.44\textwidth,height=0.23\textheight]{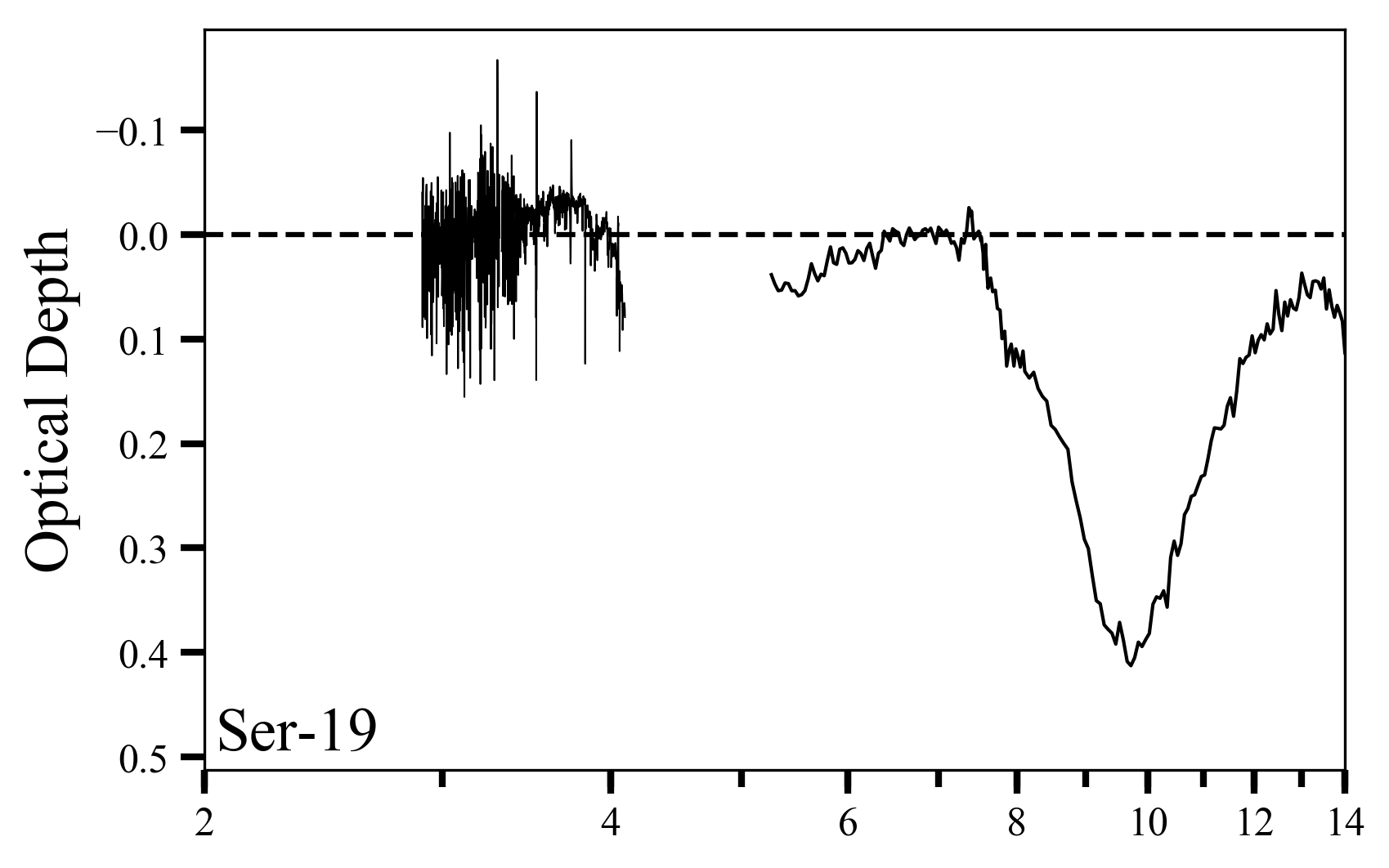}    

    \includegraphics[width=0.44\textwidth,height=0.25\textheight]{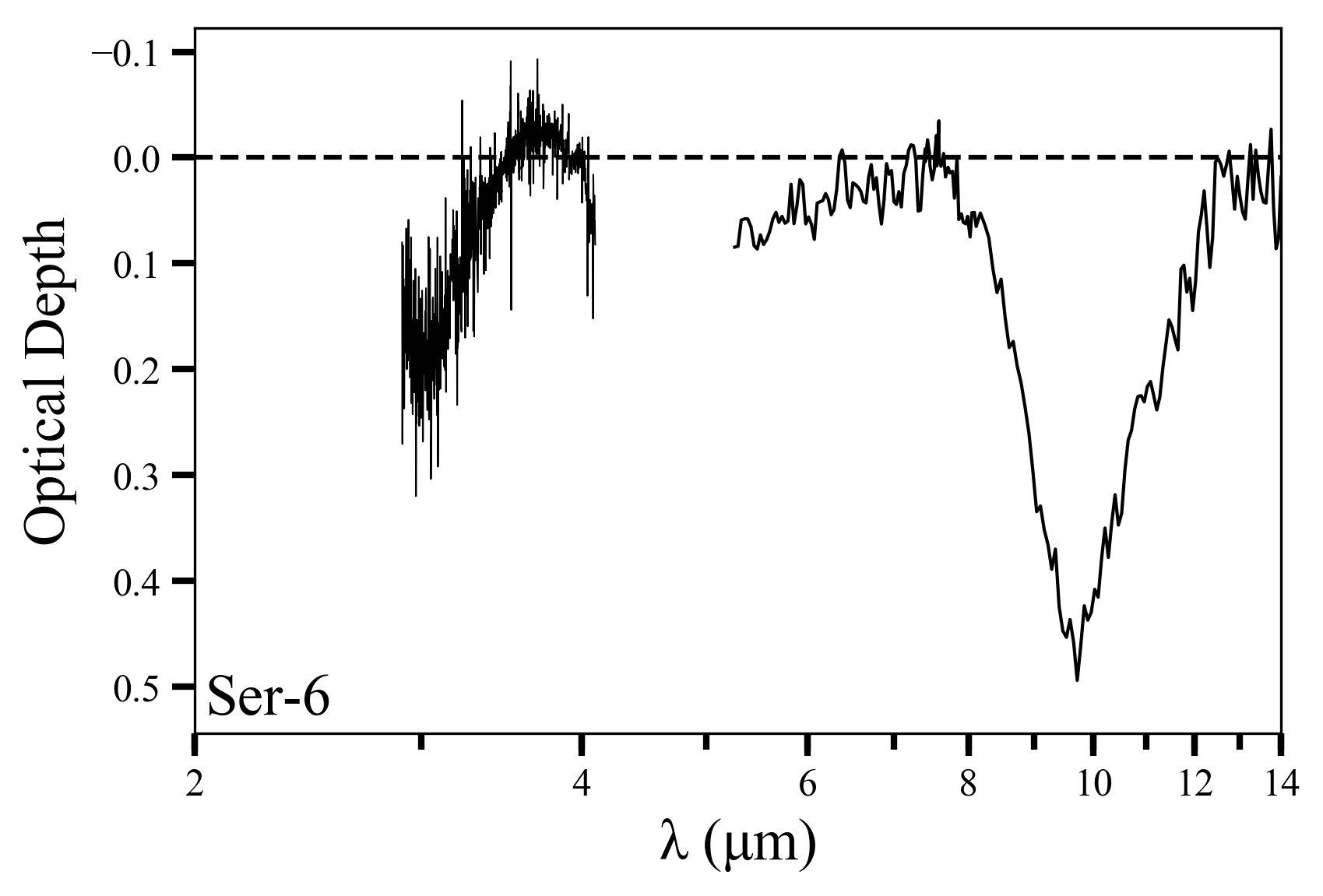}
    \caption{The targets Ser-19 and Ser-6 have similar $A_{\rm K}$
      values and 9.7 $\mu$m silicate band depths, but only Ser-6
      displays the 3.0 $\mu$m $\rm{H_2O}$ feature. The
      \textit{$L$-}band portions have been smoothed to a resolving
      power of 25. This comparison indicates that ice formation is not
      the only factor that affects coagulation as traced by the
      $A_{\rm K}$/$\tau _{9.7}$ ratio.\label{Fig. ser19_ser6}}
\end{figure}

In addition to the overall trends described above, there are
deviations, indicating that local conditions, such as density
fluctuations, across the cloud also matter. Such local scatter in the
$\tau_{9.7}$ versus $A_{\rm K}$ plots was also noted for Lupus
\citep{Boogert2013}. Fig. \ref{Fig. per19_per2} compares the optical
depth spectra of Per-19 (Group C) and Per-2 (Group A), which have
similar $A_{\rm K}$ values, but Per-19 has a 50\% deeper 9.7 $\mu$m
silicate band. Conforming with its diffuse ISM nature, Per-19 has a
factor of $\sim$2.5 less water ice than Per-2.  Thus, Per-19 appears
to trace lower density cloud material.  In contrast, Ser-19 and Ser-6
have similar $\tau_{9.7}$ and $A_{\rm K}$ at very different
$\tau_{3.0}$ (Fig. \ref{Fig. ser19_ser6}).  Ser-19 is in fact the only
Serpens target without ice. It is relatively isolated on the edge of
the cloud, but it is unclear if this plays a role.  Unfortunately,
Ser-19 and Ser-6 have uncertainn $\tau_{9.7}/A_{\rm K}$ ratios (Group
C), precluding a distinction between dense-like (Group A) or
diffuse-like (Group B) dust.

It is worthwhile to note that the inflection points in the $A_{\rm K}$
versus $\tau_{9.7}$ relation of $A_{\rm V} \sim 10$ and $A_{\rm V}
\sim 17$ for Perseus and Serpens, respectively, are comparable to the
dense core formation thresholds in these clouds. \citet{Enoch2007}
derive dense core formation thresholds of $A_{\rm V} \sim 8$ and
$A_{\rm V} \sim 15$ from comprehensive infrared and sub-millimeter
surveys. This reinforces the idea that the dust coagulation process is
enhanced at higher densities.

\subsection{Ice Formation Threshold}\label{subsec:ices}

The relation between $A_{\rm K}$ and $\tau_{3.0}$ (\S\ref{subsec:H2O},
Fig.~\ref{Fig. H2O_correlation}) for Serpens shows a cut-off along the
$A_{\rm K}$ axis of $0.616 \pm 0.087$ ($A_{V}=5.17\pm0.06$), which is
approximately twice that of Perseus ($A_{V}= 2.65\pm0.25$), Lupus IV
($A_{V}=2.10\pm0.59$; \citealt{Boogert2011}), and Taurus
($A_{V}=3.2\pm0.1$; \citealt{Whittet2001}). This might be attributed
to extinction by unrelated foreground dust. Serpens is located at a
larger distance ($420\pm 15$ pc) than all these other clouds: $294\pm
15$ pc for Perseus, $141\pm 7$ pc for Taurus, and $189\pm 13$ pc for
Lupus \citep{Ortiz2018, Zucker2019}. To the first order, the
foreground extinction due to diffuse dust in the Galactic Plane may be
estimated by scaling to the extinction of $A_{\rm V}\sim 30$ towards
the Galactic Center (\citealt{Rieke1989} and references therein) at
$\sim 8$ kpc. This results in foreground contributions of $A_{\rm
  V}\sim 1.6$, 1.1, 0.7, and 0.5 mag for Serpens, Perseus, Lupus IV,
and Taurus, respectively. Models using {\it Gaia}-derived distances
\citep{Zucker2019}, however, point to a larger foreground contribution
($A_{\rm V}= 3.0$) for Serpens than for all the other clouds ($A_{\rm
  V}= 1.0)$. This could be related to its location closer to the
Galactic Plane ($b\sim +4^{\circ}$), more directed to the Galactic
Center ($l\sim 30^{\circ}$) than the other clouds.

Overall, when correcting for the larger foreground extinction towards
Serpens, it thus seems that H$_2$O ice is formed at similar depths for
all studied clouds. This would be at $A_{\rm V}\sim 2.6-3.4$, or $\sim
1.6-2.4$ when subtracting the 1.0 mag foreground for all
\citep{Zucker2019}.  If we further correct for the fact that the
observations trace both the front and back of the clouds, the cloud
depth at which H$_2$O ice is abundantly formed is $A_{Vf}\sim 0.8-1.2$
mag. Following \citet{Hollenbach2009}, this could indicate that the
cloud edges have similar densities ($n$), provided that they
experience similar interstellar radiation fields $F_0 G_0$

\begin{equation}
        A_{Vf} = 0.56 \times
        ln\left(\frac{G_0F_0Y}{n(O)\nu_0}\right)\label{Eq_AV}
\end{equation}

where Y is the total yield of photodesorbing $\rm{H_2O}$ ice, and
$\nu_O$ is the vibrational frequency of O atoms bound to a grain
surface related to dust temperature. For typical conditions
\citep{Hollenbach2009}, this corresponds to $n\sim 3\times 10^4$ at
$A_{Vf}\sim 0.8-1.2$ mag

A higher threshold was observed toward the Ophiuchus cloud
\citep{Tanaka1990} and is usually ascribed to the presence of a nearby
O and B-stars, increasing $F_0$, photodesorbing the ices. No such
sources for a high UV radiation field near the other clouds are
known. Note that if the larger cut-off in the relation between $A_{\rm
  K}$ and $\tau_{3.0}$ is not caused by foreground dust,
Eq.~\ref{Eq_AV} would imply a lower Serpens cloud edge density
compared to the other clouds by a factor of $\sim$7.

\subsection{$\rm{CH_3OH}$}\label{subsec:CH3OH_disc}

One of our targets, Ser-7, shows a very high $\rm{CH_3OH}$ abundance
of $\sim$21\% relative to $\rm{H_2O}$, surpassing previous records for
background stars of L694 \citep[$\sim$14\%;][]{Chu2020} and L429-C
\citep[$\sim$12\%;][]{Boogert2011}. This target was also noted as
containing high abundances of other organic molecules, such as methane
($\rm{CH_4}$), and, tentatively, solid acetylene ($\rm{C_2H_2}$) at
13.5 $\mu$m \citep{Knez2008,Knez2012}.

The $\rm{CH_3OH}$ abundance toward Ser-7 is significantly larger than
the upper limits derived for other Serpens and Perseus background
stars (Fig. \ref{Fig. CH3OH}).  It approaches the largest CH$_3$OH ice
abundances observed toward YSOs in Serpens (28\%;
\citealt{Pontoppidan2004, Perotti2020}).  Of all background stars in
our sample, Ser-7 is closest to a YSO (2MASSJ18285277+0028463), which
is a member of the Ser/G3-G6 cluster of Class I and ``flat spectrum"
YSOs. At an angular distance of 23” (5,725 AU), the large CH$_3$OH
abundance might trace the very outer edges of the envelope of this
flat spectrum YSO, or the high density core material within which the
cluster formed.  While $\rm{CH_3OH}$ is also expected to form at low
densities (e.g., \citealt{Qasim2018}), its formation is strongly
enhanced at high densities ($10^5$ cm$^{-3}$) when CO rapidly freezes
out \citep{Cuppen2009}.  Indeed, so far all CH$_3$OH ice detections
toward background stars trace sightlines with very high extinction,
above a threshold of $A_{\rm V}\sim 18$ ($A_{\rm K} \sim 2$;
\citealt{Boogert2015, Chu2020}). Ser-7 is the only target in our
Serpens and Perseus sample with an extinction ($A_{\rm K}=4.75\pm
0.44$) above this threshold.

\section{Summary and Future Work}\label{sec:Summary}

We present 2-4 and 5-20 $\mu$m spectra of a sample of 28 stars behind
the Perseus and 21 stars behind the Serpens molecular clouds. We
fitted the target spectra using a combination of template spectra from
the IRTF spectral database and photospheric model spectra, combined
with extinction curves, laboratory $\rm{H_2O}$ ice spectra, and
silicate absorption spectra to derive $A_{\rm K}$, $\tau_{3.0}$, and
$\tau_{9.7}$ values. Correlation plots of $\tau_{9.7}$ versus$A_{\rm
  K}$ show a variation of a factor of $\sim$2 for both
clouds. Combining our $\tau_{9.7}$ and $A_{\rm K}$ values with those
available in the literature indicates that such scatter is common.

In general, the $\tau_{9.7}$/$A_{\rm K}$ ratios are reduced relative
to the diffuse ISM.  The largest reductions, up to a factor of 2, are
visible above $A_K\sim$1.2 for Perseus and Lupus, and above
$A_K\sim2.0$ for Serpens.  A picture emerges that grains, after
acquiring ice mantles (at $A_K\sim$0.2-0.4), grow to moderate sizes
due to higher densities deeper in the cloud, especially above
$A_K\sim$1.2-2.  A significant population of grains larger than $\sim
0.5$ $\mu$m is unlikely, however, as this would increase the
$\tau_{9.7}$/$A_{\rm K}$ ratio, and would also change the profiles of
the 3.0 and 9.7 $\mu$m absorption profiles, which is not observed.

The regions with the lowest $\tau_{9.7}$/$A_{\rm K}$ ratios are also
the regions where dense core (and thus star) formation will take
place, considering similar dense core formation extinction
thresholds.  Indeed, Serpens stands out by having a factor of $\sim$2
higher inflection in the $\tau_{9.7}$ versus $A_{\rm K}$ relation, and
also a factor of 2 larger dense core formation threshold.  These
aspects may indicate that Serpens has an overall shallower density
profile than the other clouds.

We derived H$_2$O ice formation thresholds of $A_{\rm V}\sim 2.6-3.4$
for all studied clouds, after correction for a 2 mag larger foreground
extinction towards Serpens. This threshold is well below the
extinctions where the lowest $\tau_{9.7}$/$A_{\rm K}$ ratios are
observed. Thus we conclude that, in agreement with the grain growth
models by \citet{Ormel2011}, grain coagulation is facilitated by ice
mantles, and enhanced at higher densities. Targets tracing the highest
ice column densities (proportional to $\tau_{3.0}$), and thus likely
densities, have the lowest $\tau_{9.7}/A_{\rm K}$ ratios.
  
Besides the overall trends, we also found a large scatter in the
$\tau_{9.7}$/$A_{\rm K}$ ratio across small $A_{\rm K}$
intervals. Using extinction maps, infrared images, YSO and molecular
outflow locations, we did not find strong correlations of these
variations with cloud location. Finally, we found three targets
(Per-6, Per-11, Ser-7) with particularly deep ice bands, of which
Ser-7 has an especially high $\rm{CH_3OH}$ ice abundance of 21\%
relative to $\rm{H_2O}$. This is significantly higher than the upper
limits in the other sources, which is attributed to high densities in
a local star formation region.

A larger sample of sight-lines, fully covering a wide range of $A_{\rm
  K}$ values and molecular cloud conditions is needed to further
constrain the relation between ice formation, the silicate band, and
continuum extinction. A confirmation of the inflection in the $A_{\rm
  K}$ versus $\tau _{9.7}$ relation is needed, as well as studies of
the origin of the scatter in the $\tau _{9.7}/A_K$ ratio, in
particular the relation with local density. Future work will rely
heavily on observations with the James Webb Space Telescope (JWST),
enabling the construction of detailed maps of $A_{\rm K}$, $\tau
_{3.0}$, and $\tau _{9.7}$, facilitating an assessment of the relation
between grain coagulation and other cloud properties.

\section{Acknowledgments}\label{sec:Acknowledgements}

We thank Lee Mundy and Ewine van Dishoeck for their help with the
early stages of this project.  We thank Megan Kiminki who worked on
this project while an REU student at the SETI Institute, supported by
NSF grant AST-1359346. MCLM thanks the Institute for Astronomy at UH
Manoa for their hospitalty, and Columbia University in the City of New
York and the I.I. Rabi Scholars Program for their generous financial
support.  We thank the referee for insightful comments, in particular
on the cloud distances and foreground extinction correction, that
helped improve this paper. This work is based on observations made at
the IRTF and Keck telescopes. We wish to extend our special thanks to
those of Hawaiian ancestry on whose sacred mountain of Maunakea we are
privileged to be guests. The observations presented herein would not
have been possible without their generosity. The authors recognize
that the summit of Maunakea has always held a very significant
cultural role for the indigenous Hawaiian community. We are thankful
to have the opportunity to use observations from this mountain.

%%%%%%%%%%%% BIBLIOGRAPHY %%%%%%%%%%%%%
%\newpage
\bibliography{real.bib}{}

%%%%%%%%%%%% APPENDIX %%%%%%%%%%%
\clearpage
\newpage

\section{Appendix}\label{sec:Appendix}

%SPECTRA PLOTS

%Comment to AAS Data Editor: we use the standard figure environment to
%      show all figures for the referee. In the published version
%      those should be replaced by the Figure Set given at the end

\begin{figure*}[h!]
%%    \captionsetup[subfloat]{farskip=2pt,captionskip=1pt}
%    {\includegraphics[width=6cm,height=3.9cm]{IRTF_2MASS03245605+3026005.png}}\hfil
%    {\includegraphics[width=6cm]{Full_2MASS03245605+3026005.png}}\hfil 
%    {\includegraphics[width=6cm]{OHCH3OH_2MASS03245605+3026005.png}}
%    
%    {\includegraphics[width=6cm,height=3.9cm]{OH_2MASS03245605+3026005.png}}\hfil   
%    {\includegraphics[width=6cm]{Si_2MASS03245605+3026005.png}}\hfil
%    {\includegraphics[width=6cm,height=3.9cm]{Tau_2MASS03245605+3026005.png}}
%
%    {\includegraphics[width=6cm,height=3.9cm]{IRTF_2MASS03261355+3029223.png}}\hfil
%    {\includegraphics[width=6cm]{Full_2MASS03261355+3029223.png}}\hfil 
%    {\includegraphics[width=6cm]{OHCH3OH_2MASS03261355+3029223.png}}
%    
%    {\includegraphics[width=6cm,height=4.2cm]{OH_2MASS03261355+3029223.png}}\hfil 
%    {\includegraphics[width=6cm]{Si_2MASS03261355+3029223.png}}\hfil
%    {\includegraphics[width=6cm,height=4.2cm]{Tau_2MASS03261355+3029223.png}}

\includegraphics[width=18cm]{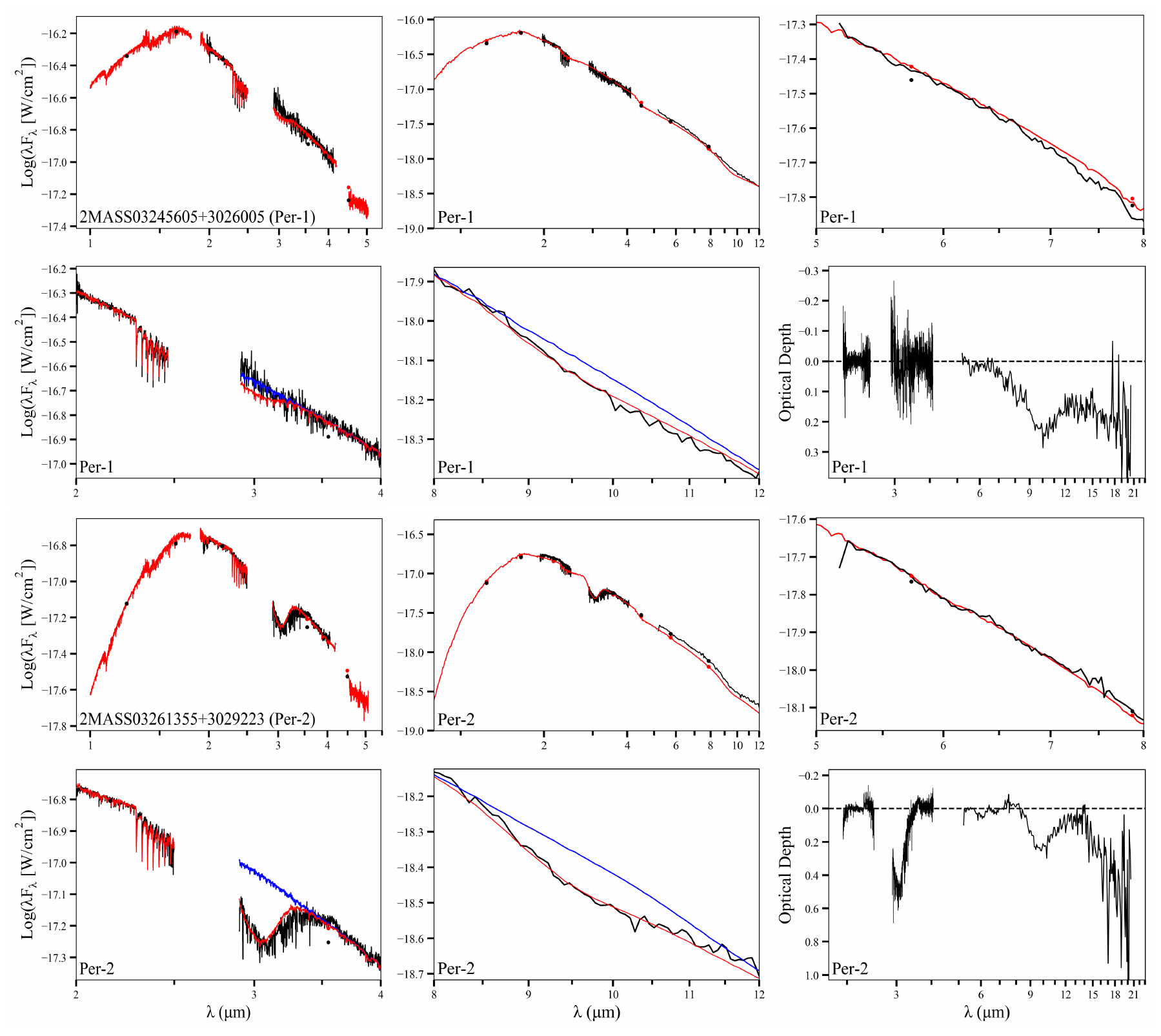}    
    \caption{The model fits and optical depth plots of the 28 Perseus
      and 21 Serpens targets. Each target has a 2x6 format of its
      plots: short wavelength IRTF model ({\bf top left}), full
      wavelength IRTF and \spitzer model ({\bf top middle}), 5-8
      $\mu$m ({\bf top right}), 2-4 $\mu$m ({\bf bottom left}), 8-12
      $\mu$m ({\bf bottom middle}), and optical depth ({\bf bottom
        right}). Black indicates the observed data, and red the model
      data. The 2-4 $\mu$m plot indicates any presence of the 3 $\mu$m
      OH stretch mode, which is made more prominent by the blue model
      flux which has the $\rm{H_2O}$ ice contribution omitted; the
      models do not include parameters to fit the wings around 3.4
      $\mu$m. The 8-12 $\mu$m plot indicates any presence of the 9.7
      $\mu$m silicates, which is made more prominent by the blue model
      flux which has the silicates contribution omitted. The 5-8
      $\mu$m plot indicates any presence of the 6 $\mu$m OH stretch
      mode and the 6.85 $\mu$m absorption feature due to, tentatively,
      NH$_4^+$.  \label{Fig. appendix}}
\end{figure*}

\setcounter{figure}{14}

\begin{figure*}
%    {\includegraphics[width=6cm,height=3.9cm]{IRTF_2MASS03272467+3022547.png}}\hfil
%    {\includegraphics[width=6cm]{Full_2MASS03272467+3022547.png}}\hfil 
%    {\includegraphics[width=6cm]{OHCH3OH_2MASS03272467+3022547.png}}
%    
%    {\includegraphics[width=6cm,height=3.9cm]{OH_2MASS03272467+3022547.png}}\hfil 
%    {\includegraphics[width=6cm]{Si_2MASS03272467+3022547.png}}\hfil
%    {\includegraphics[width=6cm,height=3.9cm]{Tau_2MASS03272467+3022547.png}}
%
%    {\includegraphics[width=6cm,height=3.9cm]{IRTF_2MASS03275729+3040138.png}}\hfil
%    {\includegraphics[width=6cm]{Full_2MASS03275729+3040138.png}}\hfil 
%    {\includegraphics[width=6cm]{OHCH3OH_2MASS03275729+3040138.png}}
%    
%    {\includegraphics[width=6cm,height=3.9cm]{OH_2MASS03275729+3040138.png}}\hfil   
%    {\includegraphics[width=6cm]{Si_2MASS03275729+3040138.png}}\hfil
%    {\includegraphics[width=6cm,height=3.9cm]{Tau_2MASS03275729+3040138.png}}
%
%    {\includegraphics[width=6cm,height=3.9cm]{IRTF_2MASS03281034+3026343.png}}\hfil
%    {\includegraphics[width=6cm]{Full_2MASS03281034+3026343.png}}\hfil 
%    {\includegraphics[width=6cm]{OHCH3OH_2MASS03281034+3026343.png}}
%    
%    {\includegraphics[width=6cm,height=4.2cm]{OH_2MASS03281034+3026343.png}}\hfil   
%    {\includegraphics[width=6cm]{Si_2MASS03281034+3026343.png}}\hfil 
%    {\includegraphics[width=6cm,height=4.2cm]{Tau_2MASS03281034+3026343.png}}

  \includegraphics[width=18cm]{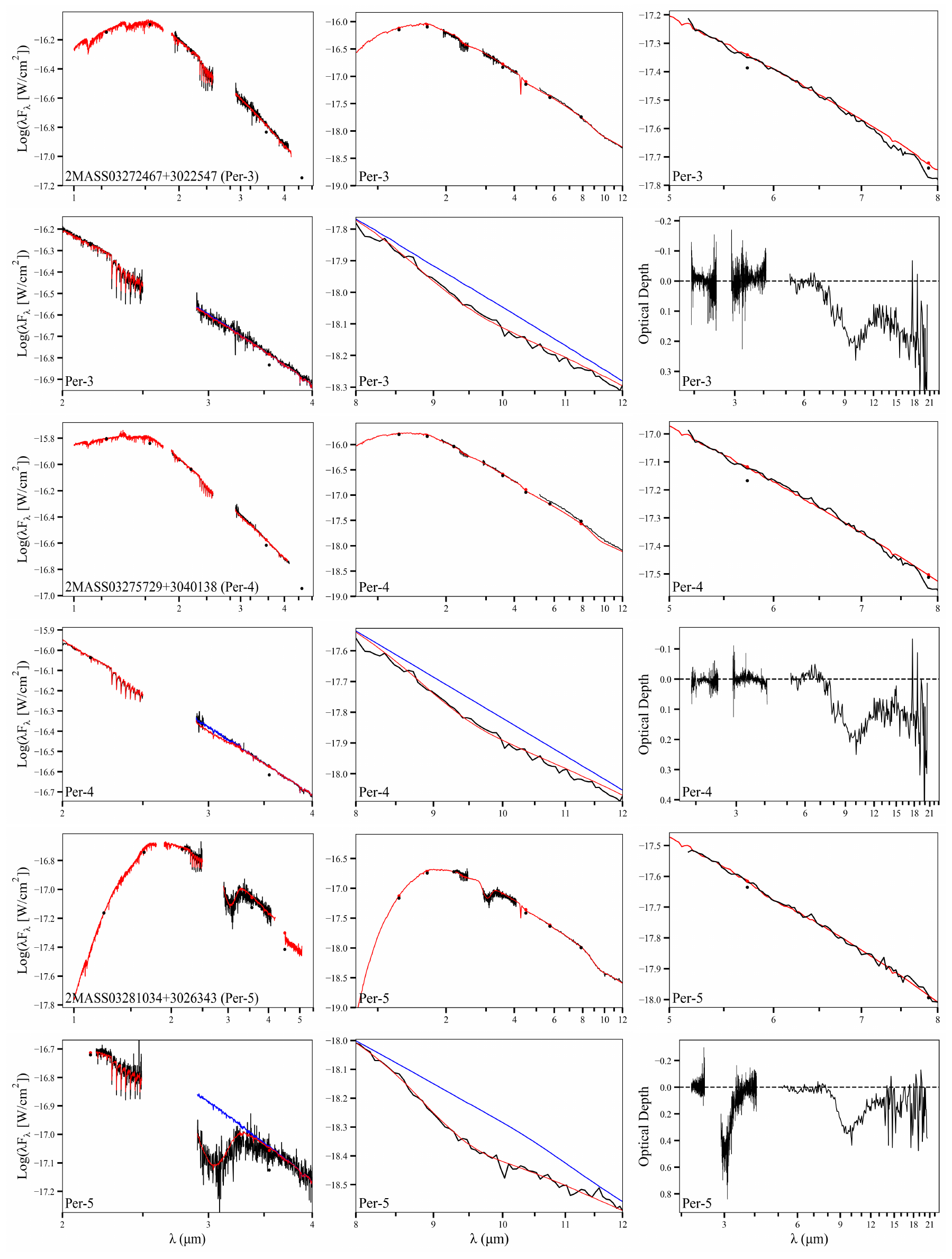}    
  \caption{(Continuation)}
\end{figure*}

\setcounter{figure}{14}

\begin{figure*}
%%%    \captionsetup[subfloat]{farskip=2pt,captionskip=1pt}
%    {\includegraphics[width=6cm,height=3.9cm]{IRTF_2MASS03290508+3022080.png}}\hfil
%    {\includegraphics[width=6cm]{Full_2MASS03290508+3022080.png}}\hfil 
%    {\includegraphics[width=6cm]{OHCH3OH_2MASS03290508+3022080.png}}
%    
%    {\includegraphics[width=6cm,height=3.9cm]{OH_2MASS03290508+3022080.png}}\hfil 
%    {\includegraphics[width=6cm]{Si_2MASS03290508+3022080.png}}\hfil
%    {\includegraphics[width=6cm,height=3.9cm]{Tau_2MASS03290508+3022080.png}}
%
%    {\includegraphics[width=6cm,height=3.9cm]{IRTF_2MASS03293654+3129465.png}}\hfil
%    {\includegraphics[width=6cm]{Full_2MASS03293654+3129465.png}}\hfil 
%    {\includegraphics[width=6cm]{OHCH3OH_2MASS03293654+3129465.png}}
%    
%    {\includegraphics[width=6cm,height=3.9cm]{OH_2MASS03293654+3129465.png}}\hfil 
%    {\includegraphics[width=6cm]{Si_2MASS03293654+3129465.png}}\hfil
%    {\includegraphics[width=6cm,height=3.9cm]{Tau_2MASS03293654+3129465.png}}
%
%    {\includegraphics[width=6cm,height=3.9cm]{IRTF_2MASS03295603+3108454.png}}\hfil
%    {\includegraphics[width=6cm]{Full_2MASS03295603+3108454.png}}\hfil 
%    {\includegraphics[width=6cm]{OHCH3OH_2MASS03295603+3108454.png}}
%    
%    {\includegraphics[width=6cm]{OH_2MASS03295603+3108454.png}}\hfil   
%    {\includegraphics[width=6cm,height=4.2cm]{Si_2MASS03295603+3108454.png}}\hfil 
%    {\includegraphics[width=6cm,height=4.2cm]{Tau_2MASS03295603+3108454.png}}

  \includegraphics[width=18cm]{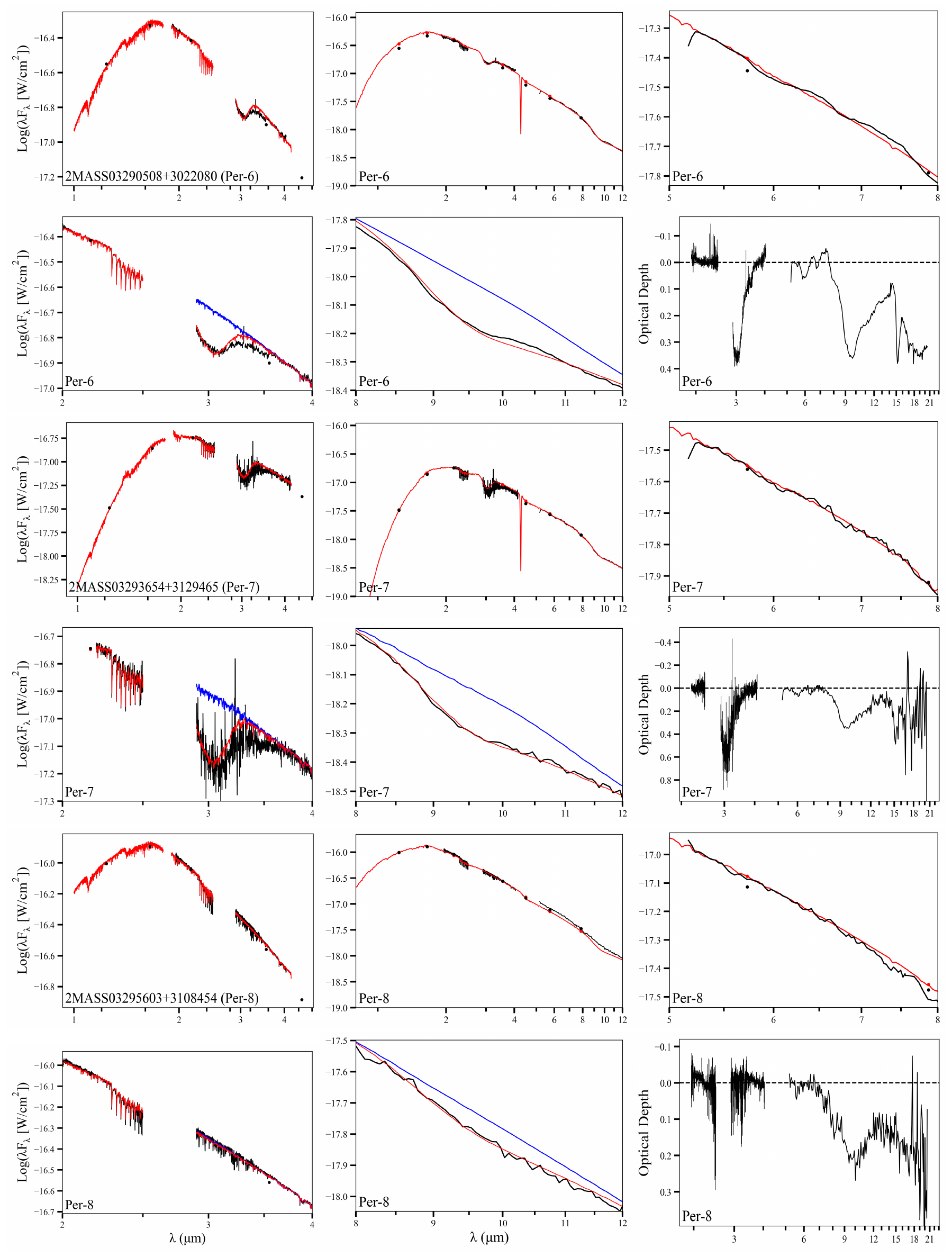}    
  \caption{(Continuation)}
\end{figure*}

\setcounter{figure}{14}

\begin{figure*}
%%    \captionsetup[subfloat]{farskip=2pt,captionskip=1pt}
%    {\includegraphics[width=6cm,height=3.9cm]{IRTF_2MASS03300474+3023032.png}}\hfil
%    {\includegraphics[width=6cm]{Full_2MASS03300474+3023032.png}}\hfil 
%    {\includegraphics[width=6cm]{OHCH3OH_2MASS03300474+3023032.png}} 
%    
%    {\includegraphics[width=6cm,height=3.9cm]{OH_2MASS03300474+3023032.png}}\hfil 
%    {\includegraphics[width=6cm]{Si_2MASS03300474+3023032.png}}\hfil 
%    {\includegraphics[width=6cm,height=3.9cm]{Tau_2MASS03300474+3023032.png}}
%
%    {\includegraphics[width=6cm,height=3.9cm]{IRTF_2MASS03301239+3144408.png}}\hfil
%    {\includegraphics[width=6cm]{Full_2MASS03301239+3144408.png}}\hfil 
%    {\includegraphics[width=6cm]{OHCH3OH_2MASS03301239+3144408.png}}
%    
%    {\includegraphics[width=6cm,height=3.9cm]{OH_2MASS03301239+3144408.png}}\hfil 
%    {\includegraphics[width=6cm]{Si_2MASS03301239+3144408.png}}\hfil 
%    {\includegraphics[width=6cm,height=3.9cm]{Tau_2MASS03301239+3144408.png}}
%
%    {\includegraphics[width=6cm,height=3.9cm]{IRTF_2MASS03303022+3027087.png}}\hfil
%    {\includegraphics[width=6cm]{Full_2MASS03303022+3027087.png}}\hfil 
%    {\includegraphics[width=6cm]{OHCH3OH_2MASS03303022+3027087.png}}
%    
%    {\includegraphics[width=6cm,height=4.2cm]{OH_2MASS03303022+3027087.png}}\hfil 
%    {\includegraphics[width=6cm]{Si_2MASS03303022+3027087.png}}\hfil 
%    {\includegraphics[width=6cm,height=4.2cm]{Tau_2MASS03303022+3027087.png}}

  \includegraphics[width=18cm]{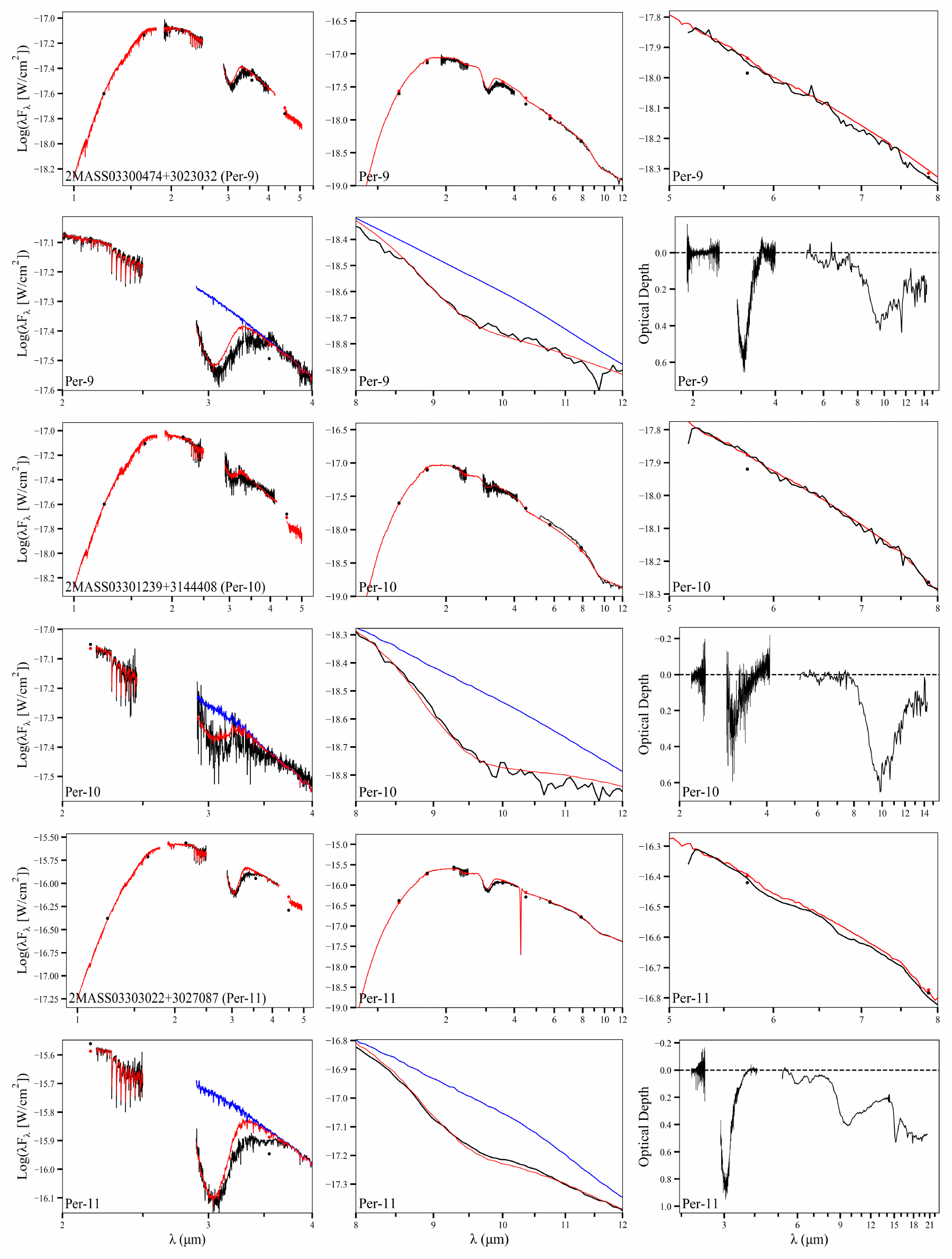}  
  \caption{(Continuation)}
\end{figure*}

\setcounter{figure}{14}

\begin{figure*}
%%    \captionsetup[subfloat]{farskip=2pt,captionskip=1pt}
%    {\includegraphics[width=6cm,height=3.9cm]{IRTF_2MASS03322030+3050485.png}}\hfil
%    {\includegraphics[width=6cm]{Full_2MASS03322030+3050485.png}}\hfil 
%    {\includegraphics[width=6cm]{OHCH3OH_2MASS03322030+3050485.png}}
%    
%    {\includegraphics[width=6cm,height=3.9cm]{OH_2MASS03322030+3050485.png}}\hfil 
%    {\includegraphics[width=6cm]{Si_2MASS03322030+3050485.png}}\hfil 
%    {\includegraphics[width=6cm,height=3.9cm]{Tau_2MASS03322030+3050485.png}}
%
%    {\includegraphics[width=6cm,height=3.9cm]{IRTF_2MASS03331023+3050177.png}}\hfil
%    {\includegraphics[width=6cm]{Full_2MASS03331023+3050177.png}}\hfil 
%    {\includegraphics[width=6cm]{OHCH3OH_2MASS03331023+3050177.png}}
%    
%    {\includegraphics[width=6cm,height=3.9cm]{OH_2MASS03331023+3050177.png}}\hfil 
%    {\includegraphics[width=6cm]{Si_2MASS03331023+3050177.png}}\hfil 
%    {\includegraphics[width=6cm,height=3.9cm]{Tau_2MASS03331023+3050177.png}}
%
%    {\includegraphics[width=6cm,height=3.9cm]{IRTF_2MASS03332416+3117470.png}}\hfil
%    {\includegraphics[width=6cm]{Full_2MASS03332416+3117470.png}}\hfil 
%    {\includegraphics[width=6cm]{OHCH3OH_2MASS03332416+3117470.png}}
%    
%    {\includegraphics[width=6cm,height=4.2cm]{OH_2MASS03332416+3117470.png}}\hfil  
%    {\includegraphics[width=6cm]{Si_2MASS03332416+3117470.png}}\hfil 
%    {\includegraphics[width=6cm,height=4.2cm]{Tau_2MASS03332416+3117470.png}}

  \includegraphics[width=18cm]{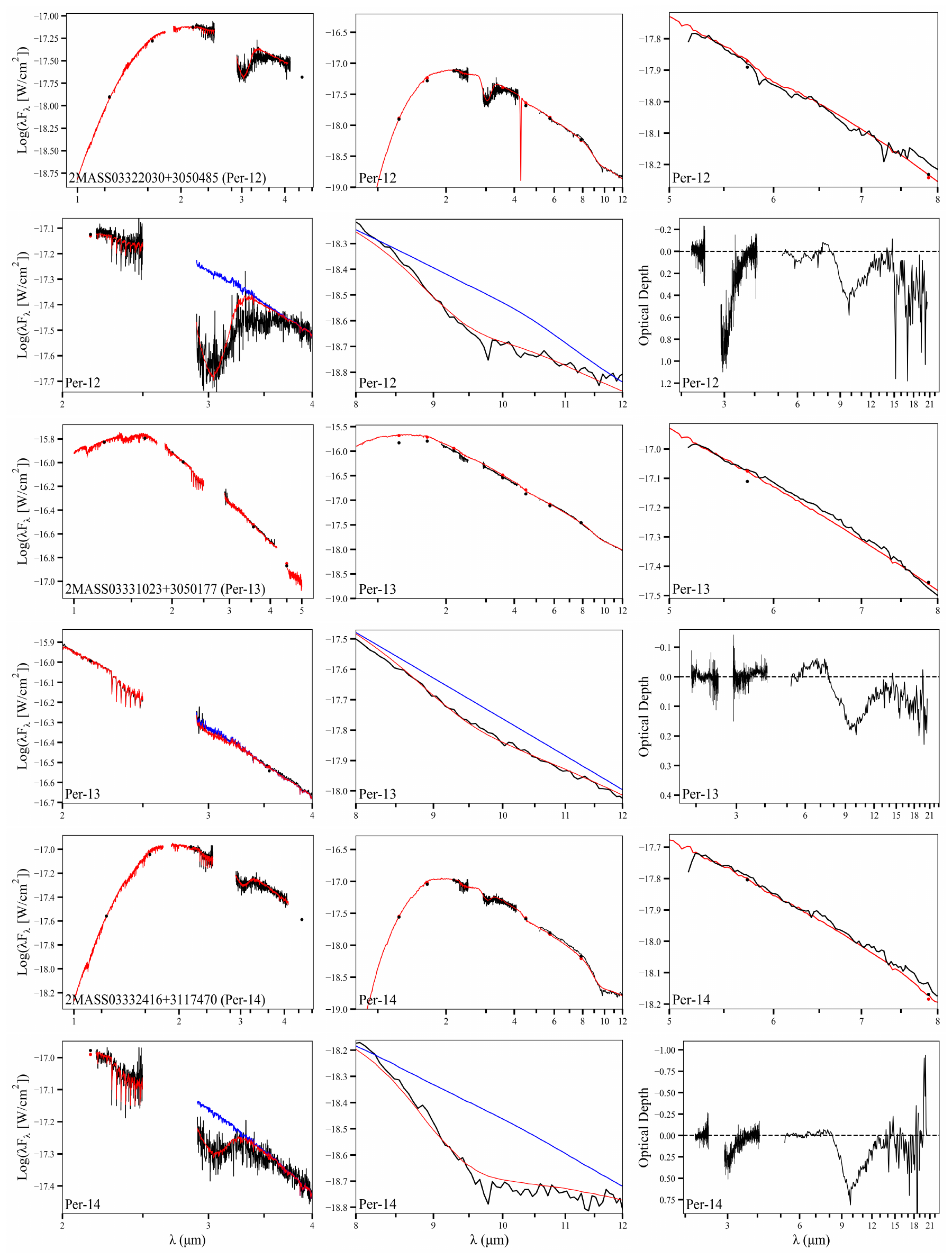}  
  \caption{(Continuation)}
\end{figure*}

\setcounter{figure}{14}

\begin{figure*}
%%    \captionsetup[subfloat]{farskip=2pt,captionskip=1pt}
%    {\includegraphics[width=6cm,height=3.9cm]{IRTF_2MASS03334078+3125007.png}}\hfil
%    {\includegraphics[width=6cm]{Full_2MASS03334078+3125007.png}}\hfil 
%    {\includegraphics[width=6cm]{OHCH3OH_2MASS03334078+3125007.png}}
%    
%    {\includegraphics[width=6cm,height=3.9cm]{OH_2MASS03334078+3125007.png}}\hfil 
%    {\includegraphics[width=6cm]{Si_2MASS03334078+3125007.png}}\hfil 
%    {\includegraphics[width=6cm,height=3.9cm]{Tau_2MASS03334078+3125007.png}}
%
%    {\includegraphics[width=6cm,height=3.9cm]{IRTF_2MASS03360868+3118398.png}}\hfil
%    {\includegraphics[width=6cm]{Full_2MASS03360868+3118398.png}}\hfil 
%    {\includegraphics[width=6cm]{OHCH3OH_2MASS03360868+3118398.png}}
%    
%    {\includegraphics[width=6cm,height=3.9cm]{OH_2MASS03360868+3118398.png}}\hfil 
%    {\includegraphics[width=6cm]{Si_2MASS03360868+3118398.png}}\hfil 
%    {\includegraphics[width=6cm,height=3.9cm]{Tau_2MASS03360868+3118398.png}}
%
%    {\includegraphics[width=6cm,height=3.9cm]{IRTF_2MASS03384753+3127345.png}}\hfil
%    {\includegraphics[width=6cm]{Full_2MASS03384753+3127345.png}}\hfil 
%    {\includegraphics[width=6cm]{OHCH3OH_2MASS03384753+3127345.png}}
%    
%    {\includegraphics[width=6cm,height=4.2cm]{OH_2MASS03384753+3127345.png}}\hfil 
%    {\includegraphics[width=6cm]{Si_2MASS03384753+3127345.png}}\hfil 
%    {\includegraphics[width=6cm,height=4.2cm]{Tau_2MASS03384753+3127345.png}}

  \includegraphics[width=18cm]{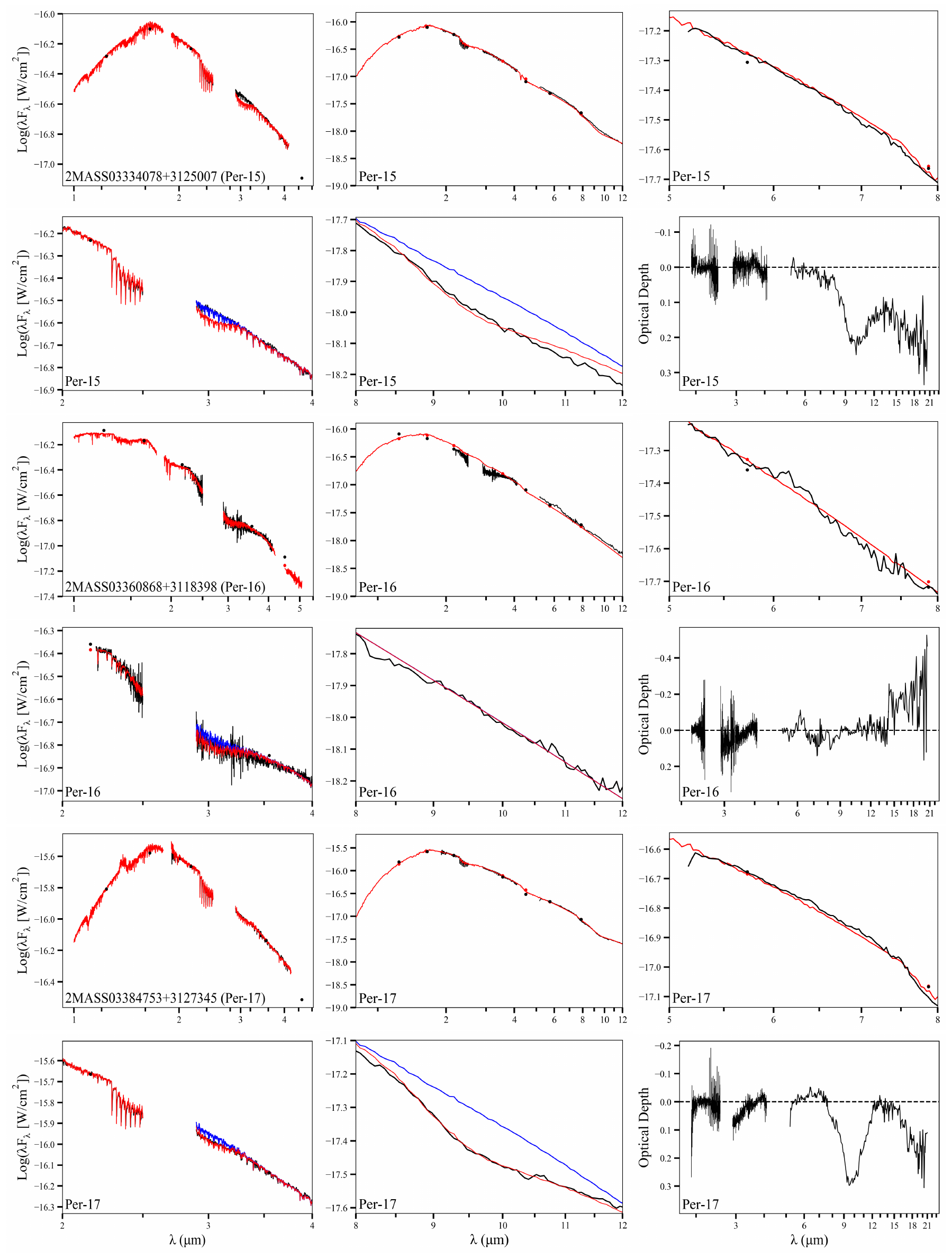}  
  \caption{(Continuation)}
\end{figure*}

\setcounter{figure}{14}

\begin{figure*}
%%    \captionsetup[subfloat]{farskip=2pt,captionskip=1pt}
%    {\includegraphics[width=6cm,height=3.9cm]{IRTF_2MASS03384901+3130173.png}}\hfil
%    {\includegraphics[width=6cm]{Full_2MASS03384901+3130173.png}}\hfil 
%    {\includegraphics[width=6cm]{OHCH3OH_2MASS03384901+3130173.png}}
%    
%    {\includegraphics[width=6cm,height=3.9cm]{OH_2MASS03384901+3130173.png}}\hfil  
%    {\includegraphics[width=6cm]{Si_2MASS03384901+3130173.png}}\hfil 
%    {\includegraphics[width=6cm,height=3.9cm]{Tau_2MASS03384901+3130173.png}}
%
%    {\includegraphics[width=6cm,height=3.9cm]{IRTF_2MASS03420993+3144139.png}}\hfil
%    {\includegraphics[width=6cm]{Full_2MASS03420993+3144139.png}}\hfil 
%    {\includegraphics[width=6cm]{OHCH3OH_2MASS03420993+3144139.png}}
%    
%    {\includegraphics[width=6cm,height=3.9cm]{OH_2MASS03420993+3144139.png}}\hfil 
%    {\includegraphics[width=6cm]{Si_2MASS03420993+3144139.png}}\hfil 
%    {\includegraphics[width=6cm,height=3.9cm]{Tau_2MASS03420993+3144139.png}}
%
%    {\includegraphics[width=6cm,height=3.9cm]{IRTF_2MASS03431627+3155097.png}}\hfil
%    {\includegraphics[width=6cm]{Full_2MASS03431627+3155097.png}}\hfil 
%    {\includegraphics[width=6cm]{OHCH3OH_2MASS03431627+3155097.png}}
%    
%    {\includegraphics[width=6cm,height=4.2cm]{OH_2MASS03431627+3155097.png}}\hfil 
%    {\includegraphics[width=6cm]{Si_2MASS03431627+3155097.png}}\hfil 
%    {\includegraphics[width=6cm,height=4.2cm]{Tau_2MASS03431627+3155097.png}}

  \includegraphics[width=18cm]{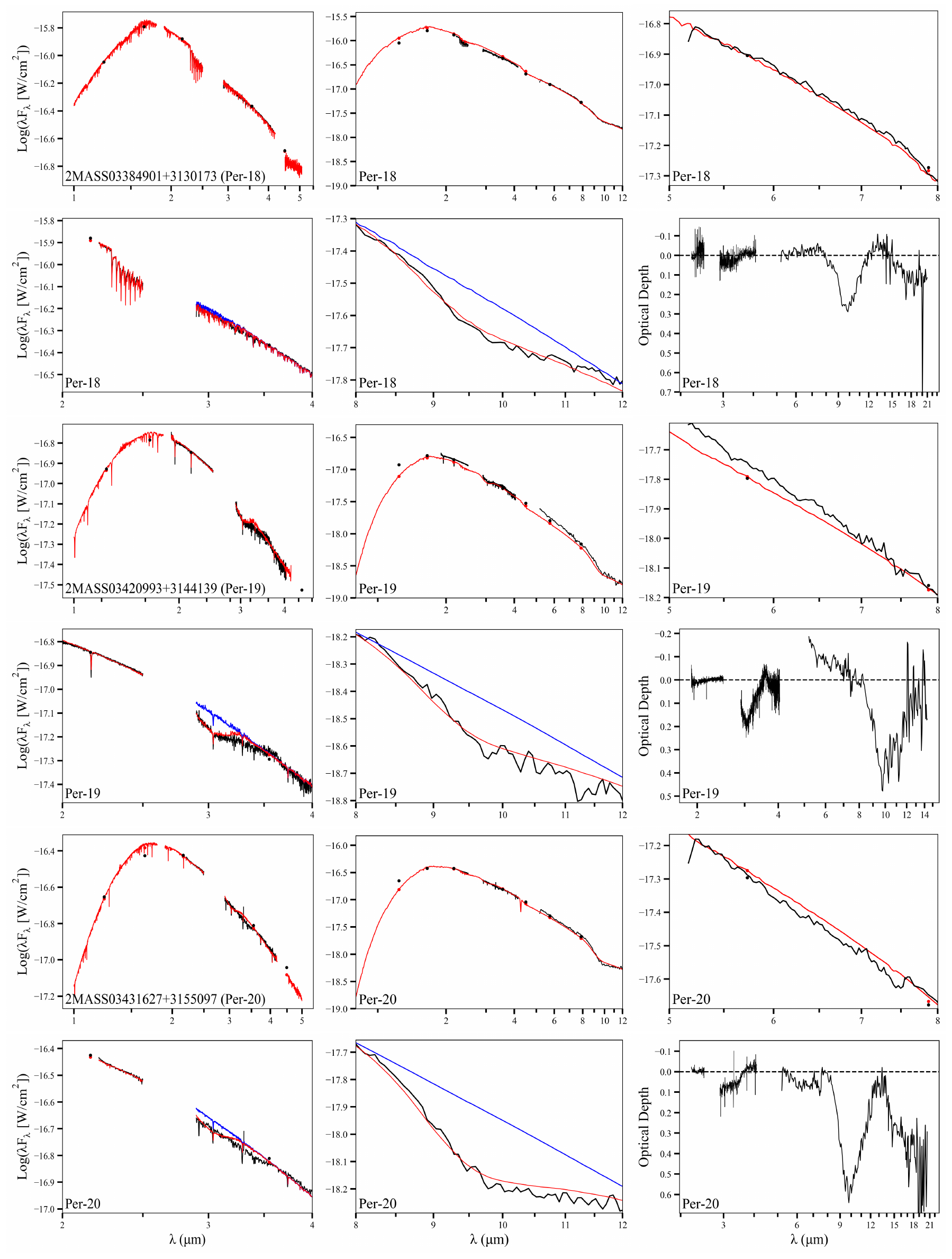}  
  \caption{(Continuation)}
\end{figure*}

\setcounter{figure}{14}

\begin{figure*}
%%    \captionsetup[subfloat]{farskip=2pt,captionskip=1pt}
%    {\includegraphics[width=6cm,height=3.9cm]{IRTF_2MASS03432386+3146110.png}}\hfil
%    {\includegraphics[width=6cm]{Full_2MASS03432386+3146110.png}}\hfil 
%    {\includegraphics[width=6cm]{OHCH3OH_2MASS03432386+3146110_v4.png}}
%    
%    {\includegraphics[width=6cm,height=3.9cm]{OH_2MASS03432386+3146110.png}}\hfil 
%    {\includegraphics[width=6cm]{Si_2MASS03432386+3146110_v4.png}}\hfil 
%    {\includegraphics[width=6cm,height=3.9cm]{Tau_2MASS03432386+3146110_v4.png}}
%
%    {\includegraphics[width=6cm,height=3.9cm]{IRTF_2MASS03434808+3151030.png}}\hfil
%    {\includegraphics[width=6cm]{Full_2MASS03434808+3151030.png}}\hfil 
%    {\includegraphics[width=6cm]{OHCH3OH_2MASS03434808+3151030.png}}
%    
%    {\includegraphics[width=6cm,height=3.9cm]{OH_2MASS03434808+3151030.png}}\hfil 
%    {\includegraphics[width=6cm]{Si_2MASS03434808+3151030.png}}\hfil 
%    {\includegraphics[width=6cm,height=3.9cm]{Tau_2MASS03434808+3151030.png}}
%
%    {\includegraphics[width=6cm,height=3.9cm]{IRTF_2MASS03450207+3141196.png}}\hfil
%    {\includegraphics[width=6cm]{Full_2MASS03450207+3141196.png}}\hfil 
%    {\includegraphics[width=6cm]{OHCH3OH_2MASS03450207+3141196.png}}
%    
%    {\includegraphics[width=6cm,height=4.2cm]{OH_2MASS03450207+3141196.png}}\hfil 
%    {\includegraphics[width=6cm]{Si_2MASS03450207+3141196.png}}\hfil 
%    {\includegraphics[width=6cm,height=4.2cm]{Tau_2MASS03450207+3141196.png}}

  \includegraphics[width=18cm]{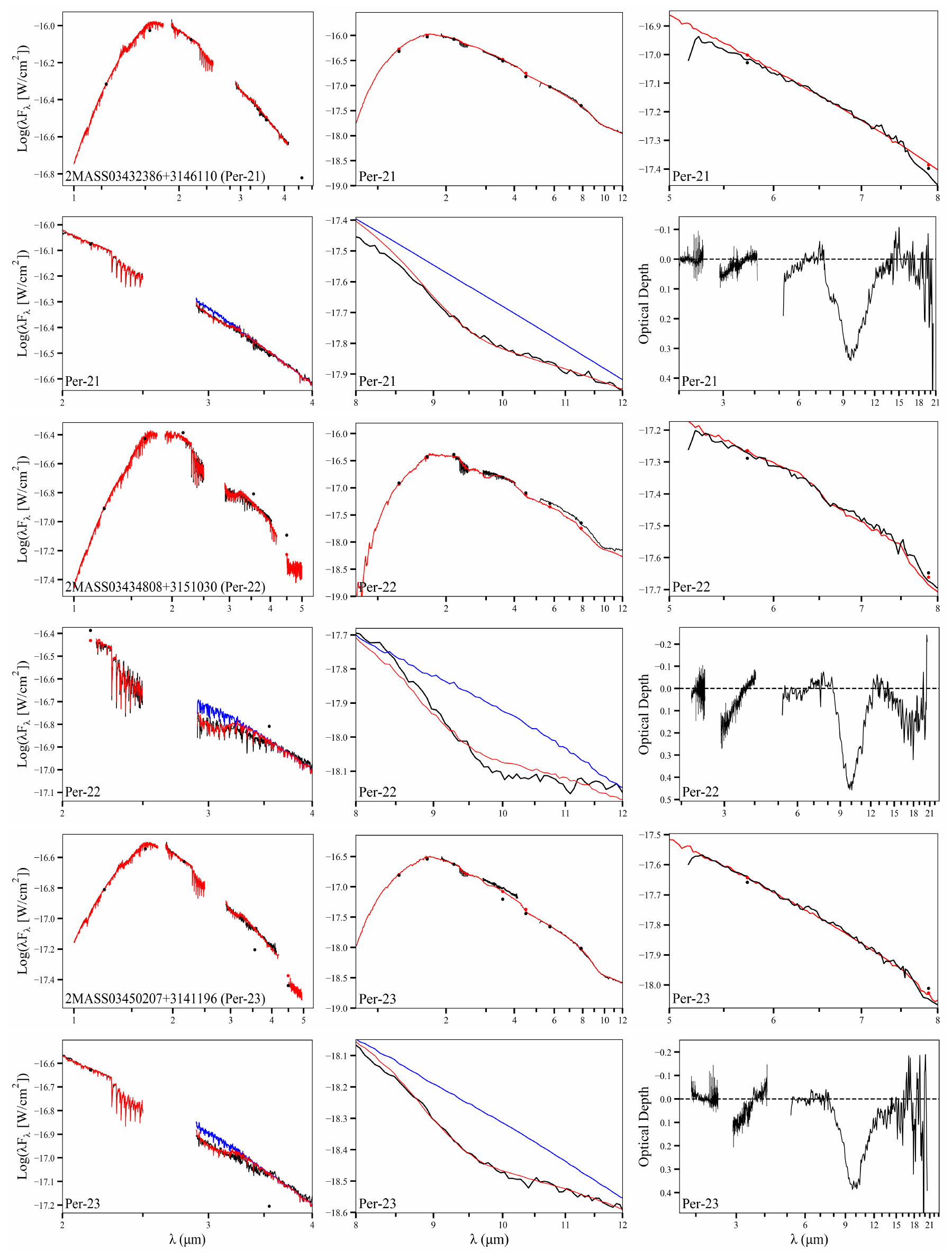}  
  \caption{(Continuation)}
\end{figure*}

\setcounter{figure}{14}

\begin{figure*}
%%    \captionsetup[subfloat]{farskip=2pt,captionskip=1pt}
%    {\includegraphics[width=6cm,height=3.9cm]{IRTF_2MASS03450796+3204018.png}}\hfil
%    {\includegraphics[width=6cm]{Full_2MASS03450796+3204018.png}}\hfil 
%    {\includegraphics[width=6cm]{OHCH3OH_2MASS03450796+3204018.png}}
%    
%    {\includegraphics[width=6cm,height=3.9cm]{OH_2MASS03450796+3204018.png}}\hfil 
%    {\includegraphics[width=6cm]{Si_2MASS03450796+3204018.png}}\hfil 
%    {\includegraphics[width=6cm,height=3.9cm]{Tau_2MASS03450796+3204018.png}}
%
%    {\includegraphics[width=6cm,height=3.9cm]{IRTF_2MASS03450839+3234202.png}}\hfil
%    {\includegraphics[width=6cm]{Full_2MASS03450839+3234202.png}}\hfil 
%    {\includegraphics[width=6cm]{OHCH3OH_2MASS03450839+3234202.png}}
%    
%    {\includegraphics[width=6cm,height=3.9cm]{OH_2MASS03450839+3234202.png}}\hfil 
%    {\includegraphics[width=6cm]{Si_2MASS03450839+3234202.png}}\hfil 
%    {\includegraphics[width=6cm,height=3.9cm]{Tau_2MASS03450839+3234202.png}}
%
%    {\includegraphics[width=6cm,height=3.9cm]{IRTF_2MASS03452349+3158573.png}}\hfil
%    {\includegraphics[width=6cm]{Full_2MASS03452349+3158573.png}}\hfil 
%    {\includegraphics[width=6cm]{OHCH3OH_2MASS03452349+3158573.png}}
%    
%    {\includegraphics[width=6cm,height=4.2cm]{OH_2MASS03452349+3158573.png}}\hfil  
%    {\includegraphics[width=6cm]{Si_2MASS03452349+3158573.png}}\hfil 
%    {\includegraphics[width=6cm,height=4.2cm]{Tau_2MASS03452349+3158573.png}}

  \includegraphics[width=18cm]{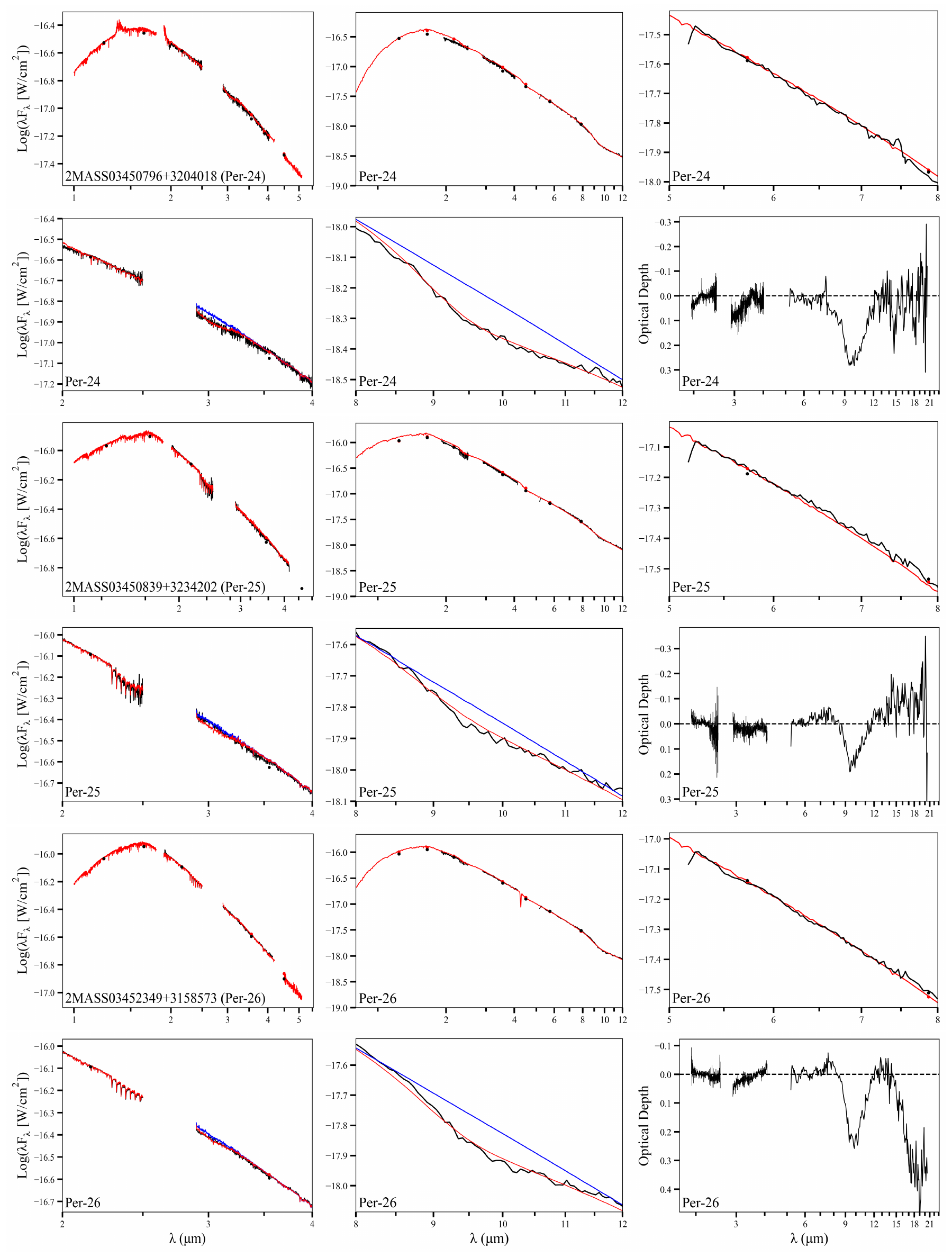}  
  \caption{(Continuation)}
\end{figure*}

\setcounter{figure}{14}

\begin{figure*}
%%    \captionsetup[subfloat]{farskip=2pt,captionskip=1pt}
%    {\includegraphics[width=6cm,height=3.9cm]{IRTF_2MASS03465115+3238494.png}}\hfil
%    {\includegraphics[width=6cm]{Full_2MASS03465115+3238494.png}}\hfil 
%    {\includegraphics[width=6cm]{OHCH3OH_2MASS03465115+3238494.png}}
%    
%    {\includegraphics[width=6cm,height=3.9cm]{OH_2MASS03465115+3238494.png}}\hfil 
%    {\includegraphics[width=6cm]{Si_2MASS03465115+3238494.png}}\hfil 
%    {\includegraphics[width=6cm,height=3.9cm]{Tau_2MASS03465115+3238494.png}}
%
%    {\includegraphics[width=6cm,height=3.9cm]{IRTF_2MASS03481723+3250595.png}}\hfil
%    {\includegraphics[width=6cm]{Full_2MASS03481723+3250595.png}}\hfil 
%    {\includegraphics[width=6cm]{OHCH3OH_2MASS03481723+3250595.png}}
%    
%    {\includegraphics[width=6cm,height=3.9cm]{OH_2MASS03481723+3250595.png}}\hfil 
%    {\includegraphics[width=6cm]{Si_2MASS03481723+3250595.png}}\hfil 
%    {\includegraphics[width=6cm,height=3.9cm]{Tau_2MASS03481723+3250595.png}}
%
%    {\includegraphics[width=6cm,height=3.9cm]{IRTF_2MASS18275901+0002337.png}}\hfil
%    {\includegraphics[width=6cm]{Full_2MASS18275901+0002337.png}}\hfil 
%    {\includegraphics[width=6cm]{OHCH3OH_2MASS18275901+0002337.png}}
%    
%    {\includegraphics[width=6cm,height=4.2cm]{OH_2MASS18275901+0002337.png}}\hfil 
%    {\includegraphics[width=6cm]{Si_2MASS18275901+0002337.png}}\hfil 
%    {\includegraphics[width=6cm,height=4.2cm]{Tau_2MASS18275901+0002337.png}}

  \includegraphics[width=18cm]{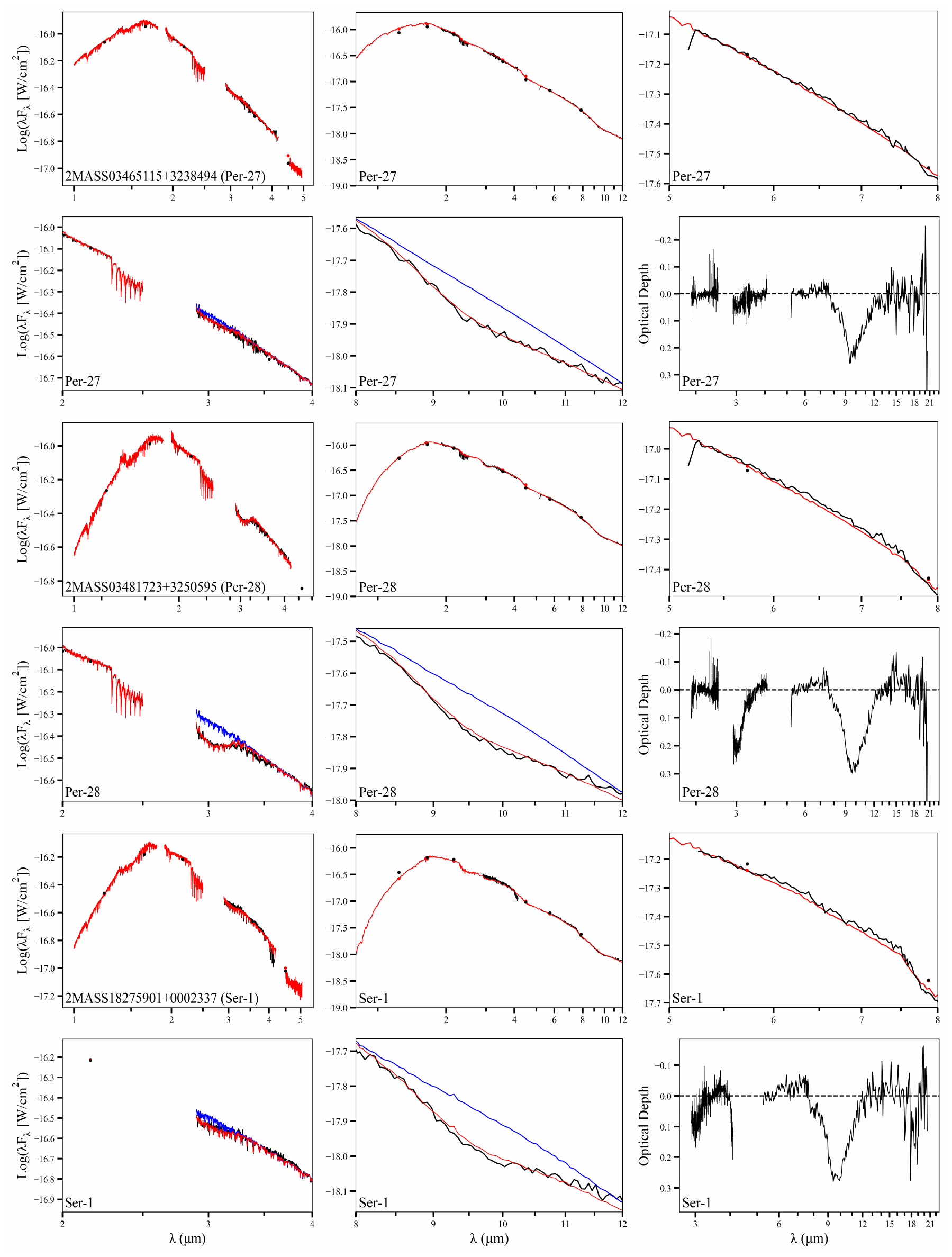}  
  \caption{(Continuation)}
\end{figure*}

\setcounter{figure}{14}

\begin{figure*}
%%    \captionsetup[subfloat]{farskip=2pt,captionskip=1pt}
%    {\includegraphics[width=6cm,height=3.9cm]{IRTF_2MASS18282010+0029141.png}}\hfil
%    {\includegraphics[width=6cm]{Full_2MASS18282010+0029141.png}}\hfil 
%    {\includegraphics[width=6cm]{OHCH3OH_2MASS18282010+0029141.png}}
%    
%    {\includegraphics[width=6cm,height=3.9cm]{OH_2MASS18282010+0029141.png}}\hfil 
%    {\includegraphics[width=6cm]{Si_2MASS18282010+0029141.png}}\hfil 
%    {\includegraphics[width=6cm,height=3.9cm]{Tau_2MASS18282010+0029141.png}}
%
%    {\includegraphics[width=6cm,height=3.9cm]{IRTF_2MASS18282631+0052133.png}}\hfil
%    {\includegraphics[width=6cm]{Full_2MASS18282631+0052133.png}}\hfil 
%    {\includegraphics[width=6cm]{OHCH3OH_2MASS18282631+0052133.png}}
%    
%    {\includegraphics[width=6cm,height=3.9cm]{OH_2MASS18282631+0052133.png}}\hfil 
%    {\includegraphics[width=6cm]{Si_2MASS18282631+0052133.png}}\hfil 
%    {\includegraphics[width=6cm,height=3.9cm]{Tau_2MASS18282631+0052133.png}}
%
%    {\includegraphics[width=6cm,height=3.9cm]{IRTF_2MASS18284038+0044503.png}}\hfil
%    {\includegraphics[width=6cm]{Full_2MASS18284038+0044503.png}}\hfil 
%    {\includegraphics[width=6cm]{OHCH3OH_2MASS18284038+0044503.png}}
%    
%    {\includegraphics[width=6cm,height=4.2cm]{OH_2MASS18284038+0044503.png}}\hfil 
%    {\includegraphics[width=6cm]{Si_2MASS18284038+0044503.png}}\hfil 
%    {\includegraphics[width=6cm,height=4.2cm]{Tau_2MASS18284038+0044503.png}}

  \includegraphics[width=18cm]{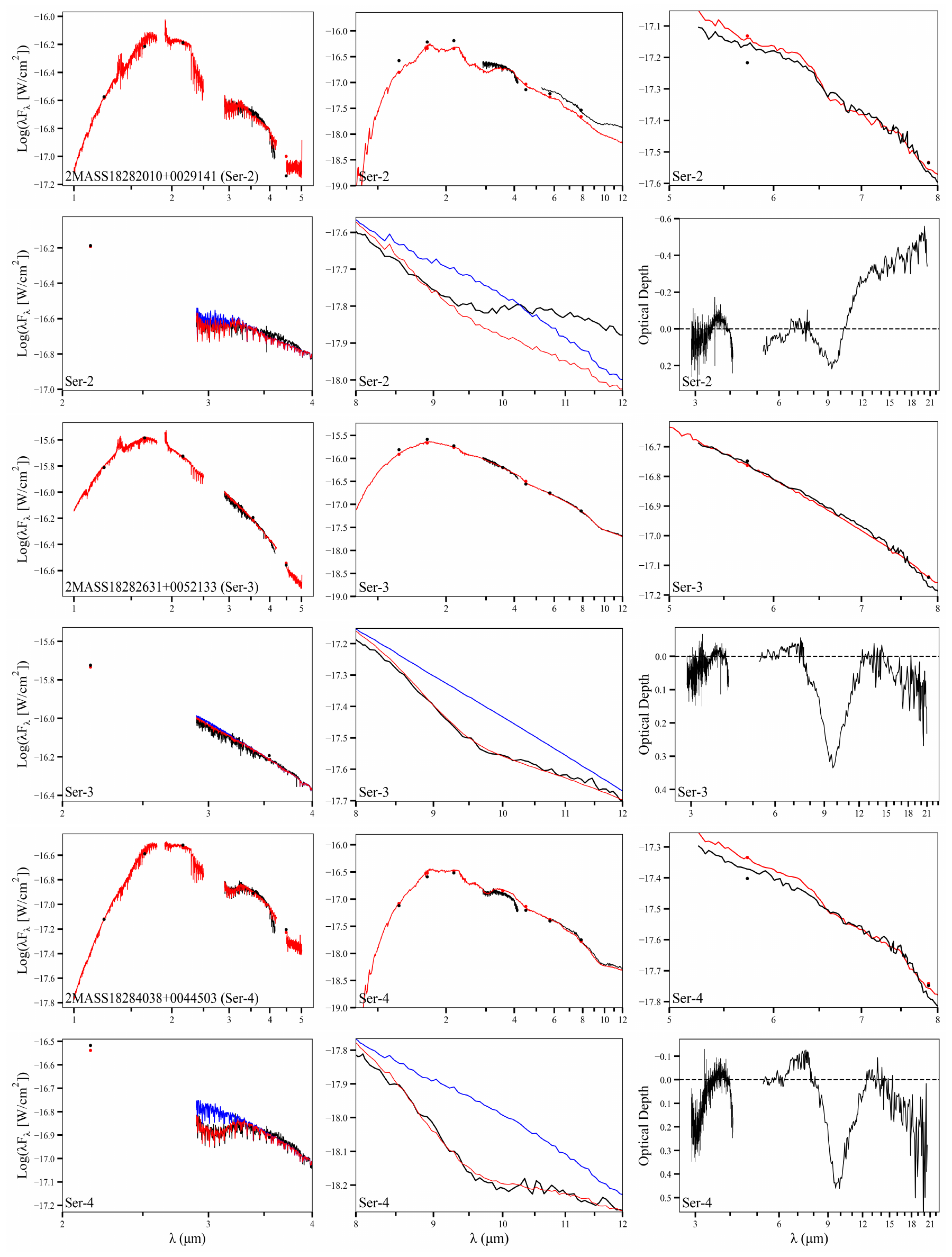}  
  \caption{(Continuation)}
\end{figure*}

\setcounter{figure}{14}

\begin{figure*}

  \includegraphics[width=18cm]{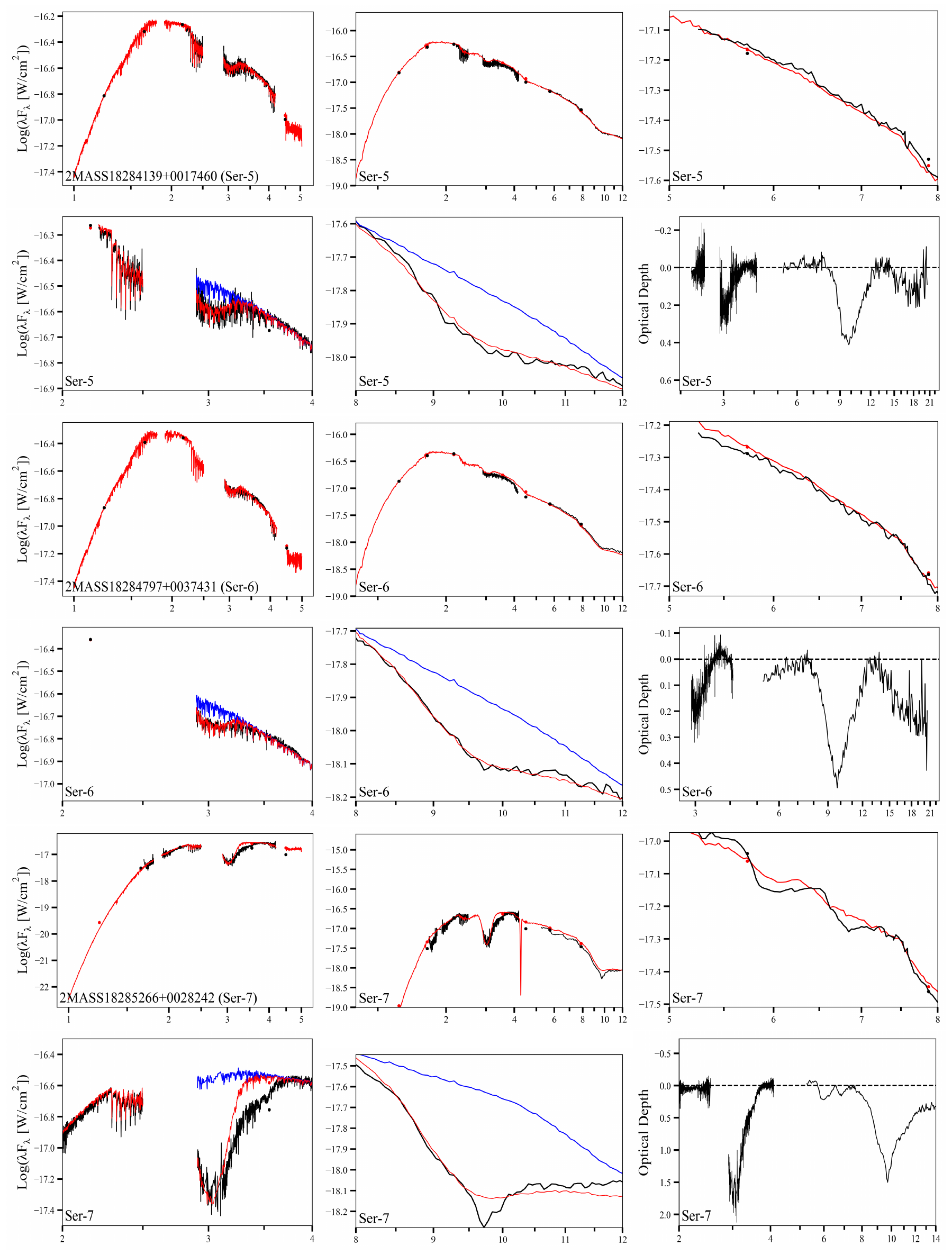}
  \caption{(Continuation)}
\end{figure*}

\setcounter{figure}{14}

\begin{figure*}
%%    \captionsetup[subfloat]{farskip=2pt,captionskip=1pt}
%    {\includegraphics[width=6cm,height=3.9cm]{IRTF_2MASS18290316+0023090.png}}\hfil
%    {\includegraphics[width=6cm]{Full_2MASS18290316+0023090.png}}\hfil 
%    {\includegraphics[width=6cm]{OHCH3OH_2MASS18290316+0023090.png}}
%    
%    {\includegraphics[width=6cm,height=3.9cm]{OH_2MASS18290316+0023090.png}}\hfil 
%    {\includegraphics[width=6cm]{Si_2MASS18290316+0023090.png}}\hfil 
%    {\includegraphics[width=6cm,height=3.9cm]{Tau_2MASS18290316+0023090.png}}
%
%    {\includegraphics[width=6cm,height=3.9cm]{IRTF_2MASS18290436+0116207.png}}\hfil
%    {\includegraphics[width=6cm]{Full_2MASS18290436+0116207.png}}\hfil 
%    {\includegraphics[width=6cm]{OHCH3OH_2MASS18290436+0116207.png}}
%    
%    {\includegraphics[width=6cm,height=3.9cm]{OH_2MASS18290436+0116207.png}}\hfil 
%    {\includegraphics[width=6cm]{Si_2MASS18290436+0116207.png}}\hfil 
%    {\includegraphics[width=6cm,height=3.9cm]{Tau_2MASS18290436+0116207.png}}
%
%    {\includegraphics[width=6cm,height=3.9cm]{IRTF_2MASS18290479-0001301.png}}\hfil
%    {\includegraphics[width=6cm]{Full_2MASS18290479-0001301.png}}\hfil 
%    {\includegraphics[width=6cm]{OHCH3OH_2MASS18290479-0001301.png}}
%    
%    {\includegraphics[width=6cm,height=4.2cm]{OH_2MASS18290479-0001301.png}}\hfil 
%    {\includegraphics[width=6cm]{Si_2MASS18290479-0001301.png}}\hfil 
%    {\includegraphics[width=6cm,height=4.2cm]{Tau_2MASS18290479-0001301.png}}

  \includegraphics[width=18cm]{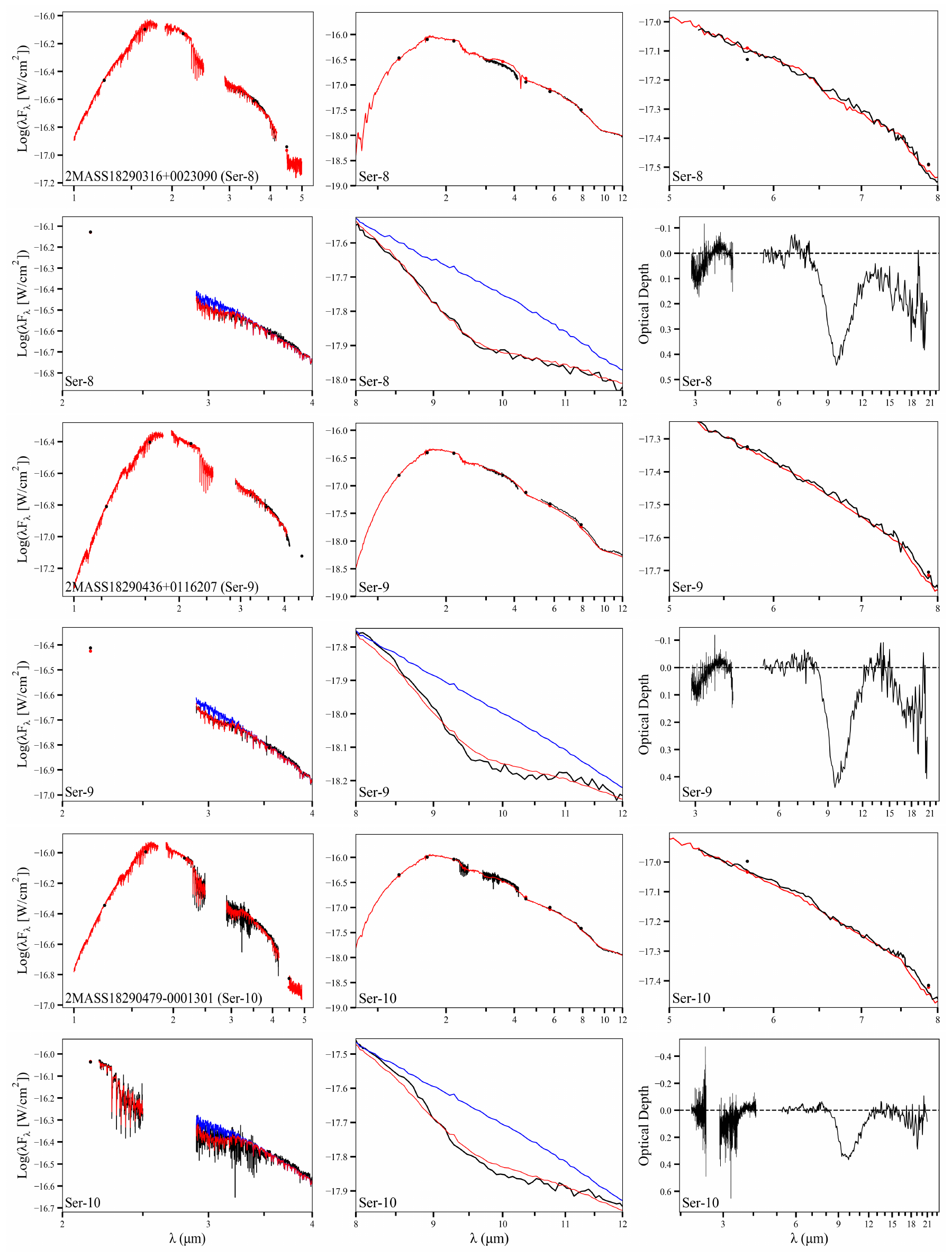}  
  \caption{(Continuation)}
\end{figure*}

\setcounter{figure}{14}

\begin{figure*}
%%    \captionsetup[subfloat]{farskip=2pt,captionskip=1pt}
%    {\includegraphics[width=6cm,height=3.9cm]{IRTF_2MASS18291546+0016422.png}}\hfil
%    {\includegraphics[width=6cm]{Full_2MASS18291546+0016422.png}}\hfil 
%    {\includegraphics[width=6cm]{OHCH3OH_2MASS18291546+0016422.png}}
%    
%    {\includegraphics[width=6cm,height=3.9cm]{OH_2MASS18291546+0016422.png}}\hfil 
%    {\includegraphics[width=6cm]{Si_2MASS18291546+0016422.png}}\hfil 
%    {\includegraphics[width=6cm,height=3.9cm]{Tau_2MASS18291546+0016422.png}}
%
%    {\includegraphics[width=6cm,height=3.9cm]{IRTF_2MASS18291600+0109382.png}}\hfil
%    {\includegraphics[width=6cm]{Full_2MASS18291600+0109382.png}}\hfil 
%    {\includegraphics[width=6cm]{OHCH3OH_2MASS18291600+0109382.png}}
%    
%    {\includegraphics[width=6cm,height=3.9cm]{OH_2MASS18291600+0109382.png}}\hfil 
%    {\includegraphics[width=6cm]{Si_2MASS18291600+0109382.png}}\hfil 
%    {\includegraphics[width=6cm,height=3.9cm]{Tau_2MASS18291600+0109382.png}}
%
%    {\includegraphics[width=6cm,height=3.9cm]{IRTF_2MASS18291619+0045143.png}}\hfil
%    {\includegraphics[width=6cm]{Full_2MASS18291619+0045143.png}}\hfil 
%    {\includegraphics[width=6cm]{OHCH3OH_2MASS18291619+0045143.png}}
%    
%    {\includegraphics[width=6cm,height=4.2cm]{OH_2MASS18291619+0045143.png}}\hfil 
%    {\includegraphics[width=6cm]{Si_2MASS18291619+0045143.png}}\hfil 
%    {\includegraphics[width=6cm,height=4.2cm]{Tau_2MASS18291619+0045143.png}}

  \includegraphics[width=18cm]{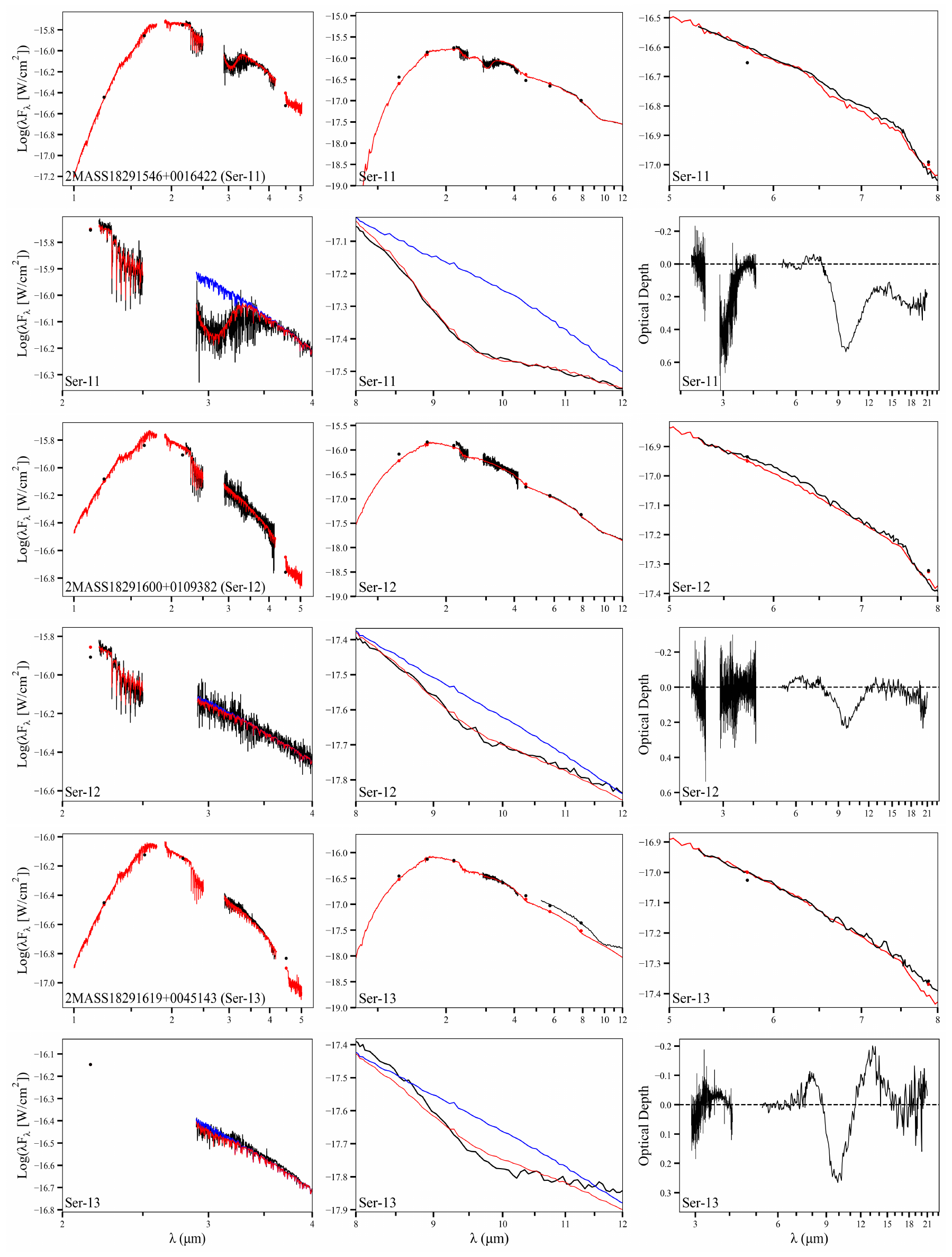}  
  \caption{(Continuation)}
\end{figure*}

\setcounter{figure}{14}

\begin{figure*}
%%    \captionsetup[subfloat]{farskip=2pt,captionskip=1pt}
%    {\includegraphics[width=6cm,height=3.9cm]{IRTF_2MASS18291699+0037191.png}}\hfil
%    {\includegraphics[width=6cm]{Full_2MASS18291699+0037191.png}}\hfil 
%    {\includegraphics[width=6cm]{OHCH3OH_2MASS18291699+0037191.png}}
%    
%    {\includegraphics[width=6cm,height=3.9cm]{OH_2MASS18291699+0037191.png}}\hfil 
%    {\includegraphics[width=6cm]{Si_2MASS18291699+0037191.png}}\hfil 
%    {\includegraphics[width=6cm,height=3.9cm]{Tau_2MASS18291699+0037191.png}}
%
%    {\includegraphics[width=6cm,height=3.9cm]{IRTF_2MASS18292528+0003141.png}}\hfil
%    {\includegraphics[width=6cm]{Full_2MASS18292528+0003141.png}}\hfil 
%    {\includegraphics[width=6cm]{OHCH3OH_2MASS18292528+0003141.png}}
%    
%    {\includegraphics[width=6cm,height=3.9cm]{OH_2MASS18292528+0003141.png}}\hfil 
%    {\includegraphics[width=6cm,height=3.9cm]{Si_2MASS18292528+0003141.png}}\hfil
%    {\includegraphics[width=6cm]{Tau_2MASS18292528+0003141.png}}
%
%    {\includegraphics[width=6cm,height=3.9cm]{IRTF_2MASS18294108+0127449.png}}\hfil
%    {\includegraphics[width=6cm]{Full_2MASS18294108+0127449.png}}\hfil 
%    {\includegraphics[width=6cm]{OHCH3OH_2MASS18294108+0127449.png}}
%    
%    {\includegraphics[width=6cm,height=4.2cm]{OH_2MASS18294108+0127449.png}}\hfil 
%    {\includegraphics[width=6cm]{Si_2MASS18294108+0127449.png}}\hfil 
%    {\includegraphics[width=6cm,height=4.2cm]{Tau_2MASS18294108+0127449.png}}

  \includegraphics[width=18cm]{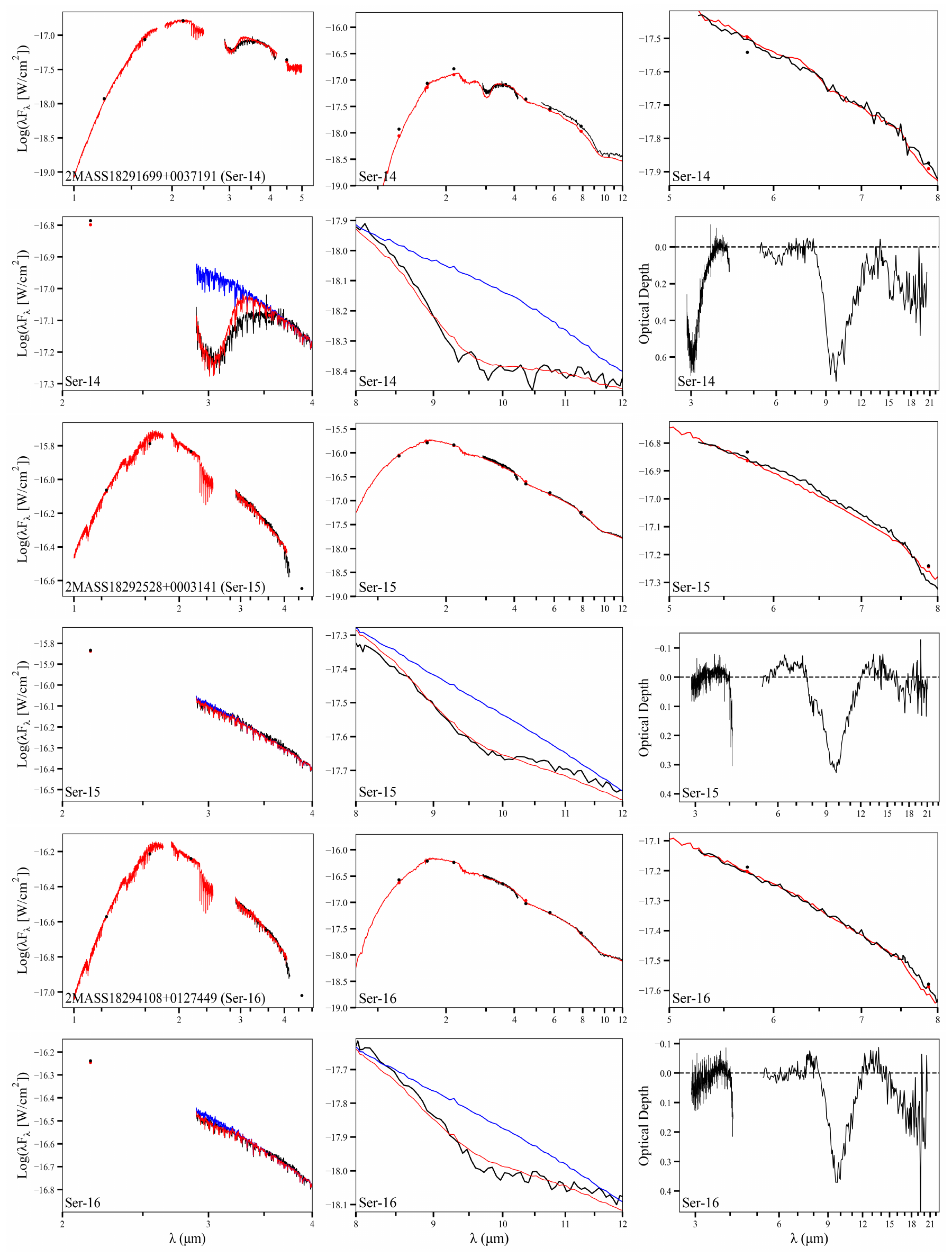}  
  \caption{(Continuation)}
\end{figure*}

\setcounter{figure}{14}

\begin{figure*}
%%    \captionsetup[subfloat]{farskip=2pt,captionskip=1pt}
%    {\includegraphics[width=6cm,height=3.9cm]{IRTF_2MASS18295604+0104146.png}}\hfil
%    {\includegraphics[width=6cm]{Full_2MASS18295604+0104146.png}}\hfil 
%    {\includegraphics[width=6cm]{OHCH3OH_2MASS18295604+0104146.png}}
%    
%    {\includegraphics[width=6cm,height=3.9cm]{OH_2MASS18295604+0104146.png}}\hfil 
%    {\includegraphics[width=6cm]{Si_2MASS18295604+0104146.png}}\hfil 
%    {\includegraphics[width=6cm,height=3.9cm]{Tau_2MASS18295604+0104146.png}}
%
%    {\includegraphics[width=6cm,height=3.9cm]{IRTF_2MASS18295940+0041007.png}}\hfil
%    {\includegraphics[width=6cm]{Full_2MASS18295940+0041007.png}}\hfil 
%    {\includegraphics[width=6cm]{OHCH3OH_2MASS18295940+0041007.png}}
%    
%    {\includegraphics[width=6cm,height=3.9cm]{OH_2MASS18295940+0041007.png}}\hfil 
%    {\includegraphics[width=6cm]{Si_2MASS18295940+0041007.png}}\hfil 
%    {\includegraphics[width=6cm,height=3.9cm]{Tau_2MASS18295940+0041007.png}}
%
%    {\includegraphics[width=6cm,height=3.9cm]{IRTF_2MASS18300085+0017069.png}}\hfil
%    {\includegraphics[width=6cm]{Full_2MASS18300085+0017069.png}}\hfil 
%    {\includegraphics[width=6cm]{OHCH3OH_2MASS18300085+0017069.png}}
%    
%    {\includegraphics[width=6cm,height=4.2cm]{OH_2MASS18300085+0017069.png}}\hfil 
%    {\includegraphics[width=6cm]{Si_2MASS18300085+0017069.png}}\hfil 
%    {\includegraphics[width=6cm,height=4.2cm]{Tau_2MASS18300085+0017069.png}}

  \includegraphics[width=18cm]{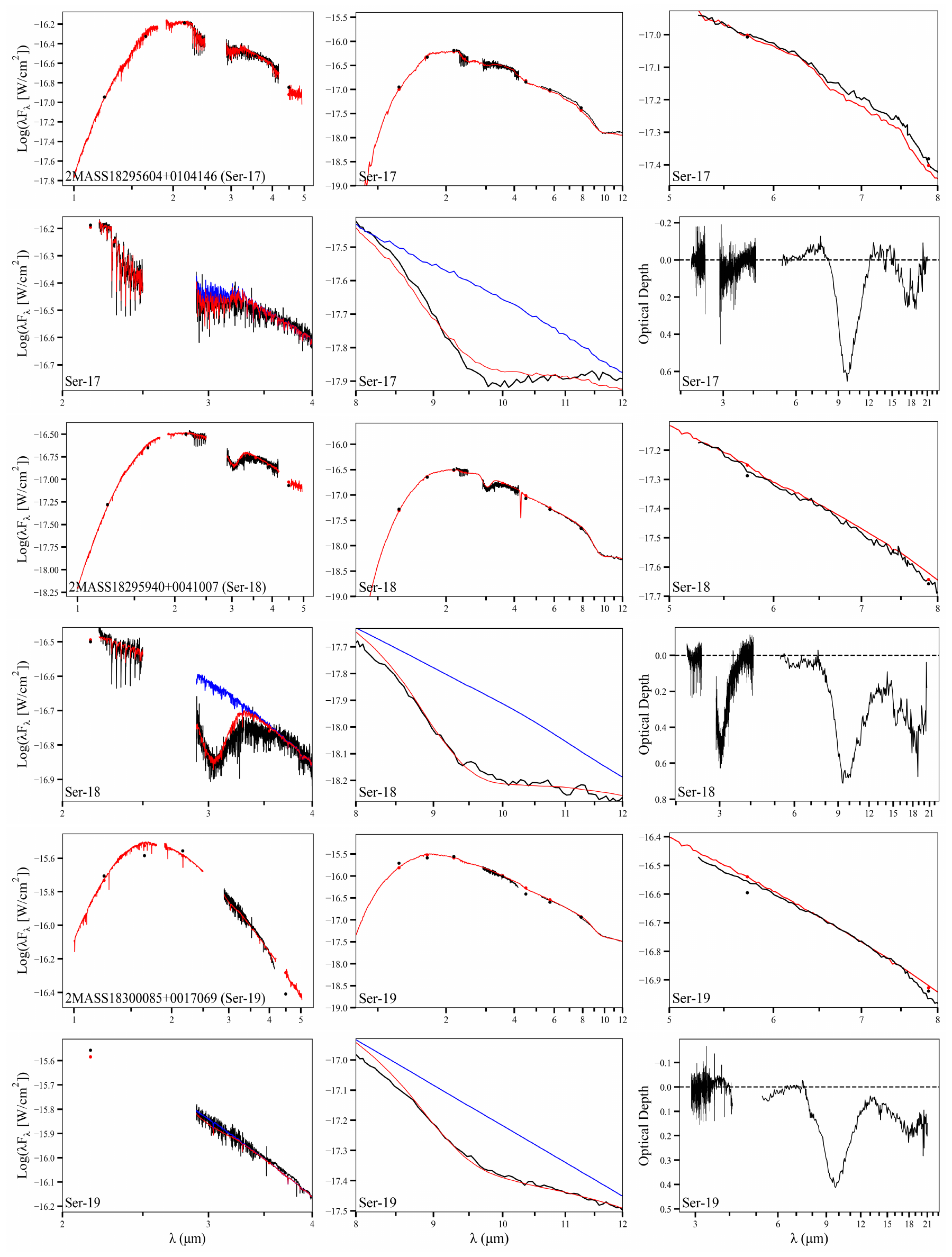}  
  \caption{(Continuation)}
\end{figure*}

\setcounter{figure}{14}

\begin{figure*}
%%    \captionsetup[subfloat]{farskip=2pt,captionskip=1pt}
%    {\includegraphics[width=6cm,height=3.9cm]{IRTF_2MASS18300896+0114441.png}}\hfil
%    {\includegraphics[width=6cm]{Full_2MASS18300896+0114441.png}}\hfil 
%    {\includegraphics[width=6cm]{OHCH3OH_2MASS18300896+0114441.png}}
%    
%    {\includegraphics[width=6cm,height=3.9cm]{OH_2MASS18300896+0114441.png}}\hfil 
%    {\includegraphics[width=6cm]{Si_2MASS18300896+0114441.png}}\hfil 
%    {\includegraphics[width=6cm,height=3.9cm]{Tau_2MASS18300896+0114441.png}}
%
%    {\includegraphics[width=6cm,height=3.9cm]{IRTF_2MASS18301220+0115341.png}}\hfil
%    {\includegraphics[width=6cm]{Full_2MASS18301220+0115341.png}}\hfil 
%    {\includegraphics[width=6cm]{OHCH3OH_2MASS18301220+0115341.png}}
%    
%    {\includegraphics[width=6cm,height=4.2cm]{OH_2MASS18301220+0115341.png}}\hfil 
%    {\includegraphics[width=6cm]{Si_2MASS18301220+0115341.png}}\hfil 
%    {\includegraphics[width=6cm,height=4.2cm]{Tau_2MASS18301220+0115341.png}}

  \includegraphics[width=18cm]{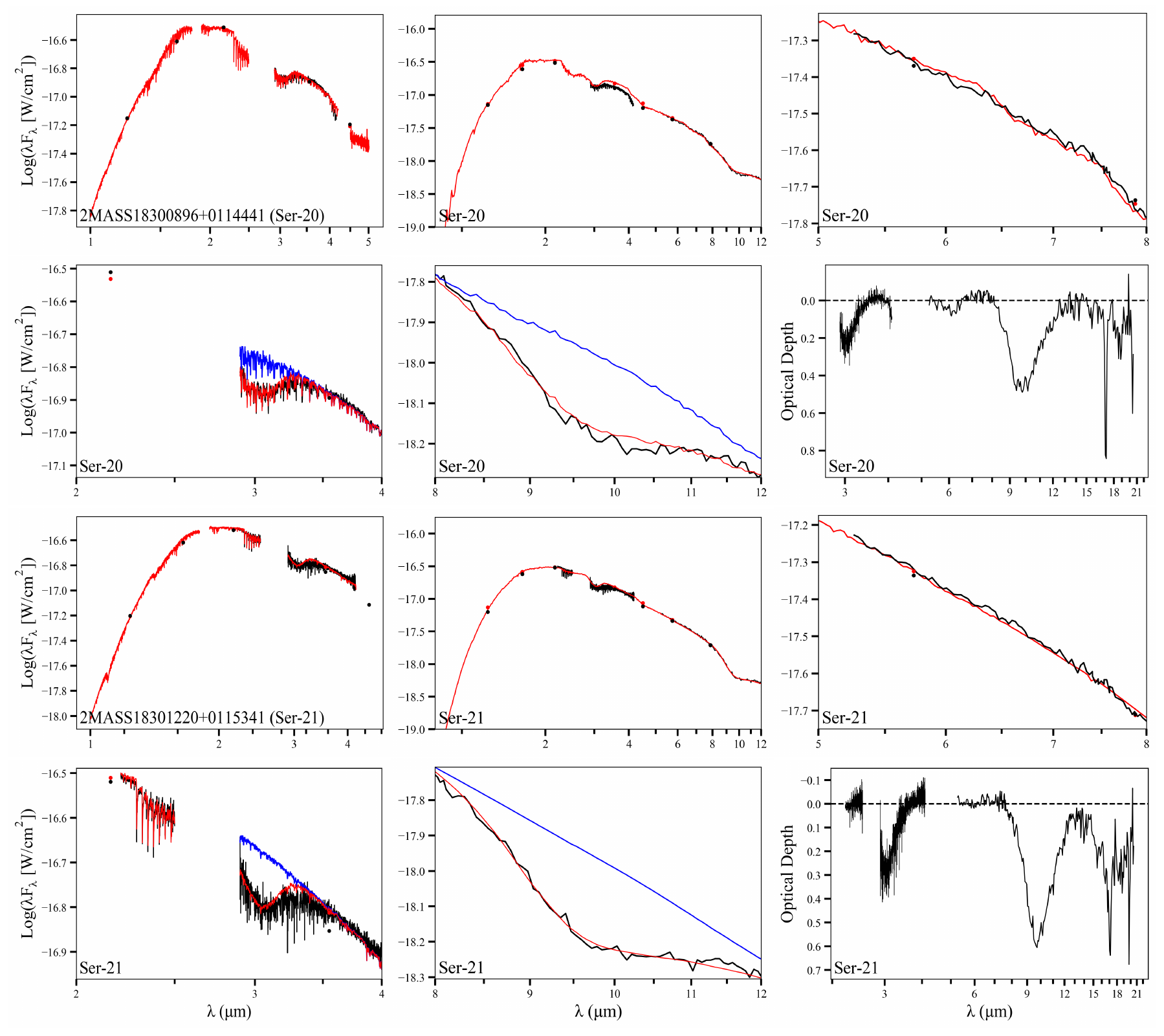}  
  \caption{(Continuation)}
\end{figure*}

\end{document}